\documentclass{aa}

\usepackage{graphicx}
\usepackage{txfonts}
\usepackage{color}
\usepackage{xspace}
\usepackage{siunitx}
\usepackage{amsmath, amsfonts}
\usepackage{amssymb}
\usepackage{xcolor}
\usepackage{dashbox}
\usepackage{framed}
\usepackage{lipsum}
\usepackage{placeins}
\usepackage{stfloats}
\usepackage{picture}
\usepackage{multirow}

\usepackage{hyperref}

\hypersetup{
        colorlinks=true, 
        breaklinks=true,
        linkcolor=blue, 
        citecolor=blue, 
        filecolor=blue, 
        urlcolor=blue,
        unicode=false, 
        pdftoolbar=true, 
        pdfmenubar=true, 
        pdffitwindow=false, 
        pdfstartview={Fit}, 
        pdftitle={}, 
        pdfauthor={Alena Rottensteiner}, 
        pdfsubject={},
        pdfcreator={Alena Rottensteiner}, 
        pdfkeywords={},
        pdfnewwindow=true, 
        pdfdisplaydoctitle=true 
}

\makeatletter
\renewcommand*\aa@pageof{, page \thepage{} of \pageref*{LastPage}}
\makeatother

\begin{document}

\defcitealias{Cantat-Gaudin2020a}{Catalog~I}
\defcitealias{meingast2020}{Catalog~II}
\defcitealias{2022Ratzenbock_Sigma}{Catalog~III}

\title{An empirical isochrone archive for nearby open clusters}

\author{Alena Rottensteiner\inst{1} \and Stefan Meingast\inst{1}}

\institute{Department of Astrophysics, University of Vienna, T\"urkenschanzstrasse 17, 1180 Wien, Austria 
\\ \email{alena.kristina.rottensteiner@univie.ac.at}
}

\date{Received 10 August 2023 / Accepted 3 June 2024}

\abstract
{The ages of star clusters and co-moving stellar groups contain essential information about the Milky Way. Their special properties and placement throughout the galactic disk make them excellent tracers of galactic structure and key components to unlocking its star formation history. Yet, even though the importance of stellar population ages has been widely recognized, their determination remains a challenging task often associated with highly model-dependent and uncertain results.}
{We propose a new approach to this long-standing problem, which relies on empirical isochrones of known clusters extracted from high-quality observational data. These purely observation-based data products open up the possibility of relative age determination, free of stellar evolution model assumptions.}
{For the derivation of the empirical isochrones, we used a combination of the statistical analysis tool principal component analysis for preprocessing and the supervised machine learning method support vector regression for curve extraction. To improve the statistical reliability of our result, we defined the empirical isochrone of a color-magnitude diagram (CMD) of a cluster as the median calculated from a set of $n_{\mathrm{boot}}=1000$ curves derived from bootstrapped data. The algorithm requires no physical priors, is computationally fast, and can easily be generalized over a large range of CMD combinations and evolutionary stages of clusters.}
{We provide empirical isochrones in all \emph{Gaia} DR2 and DR3 color combinations for 83 nearby clusters ($d < 500$ pc), which cover an estimated age range of 7 Myr to 3 Gyr. In doing so, we pave the way for a relative comparison between individual stellar populations based on an age-scaling ladder of empirical isochrones of known clusters. Furthermore, due to the exceptional precision of the available observational data, we report accurate lower main sequence empirical isochrones for many clusters in our sample, which are of special interest as this region is known to be especially complex to model. We validate our method and results by comparing the extracted empirical isochrones to cluster ages in the literature. We also investigate the added information that empirical isochrones covering the lower main sequence can provide on case studies of the IC~4665 cluster and the Meingast~1 stream.}
{The archive of empirical isochrones offers a novel approach to validating age estimates and can be used as an age-scaling ladder or age brackets for new populations and serve as calibration data for further constraining stellar evolution models.}

\keywords{Stars: Hertzsprung-Russel and C-M diagrams - Methods: data analysis - Methods: statistical - Open clusters and associations: general}

\maketitle

\section{Introduction}
\label{sec:Introduction}

Among the wide variety of sources that can be observed in the plane of the night sky, star clusters have long been held in a particularly prominent regard. This fascination is at least partly due to the fact that these complex systems, comprising tens, hundreds, or even thousands of stars that formed in the same molecular cloud, often map spectacular constellations onto the firmament. But also besides their picturesque appearances, star clusters have long been established as objects of dedicated study in various fields, among them the research into the formation and evolution of galactic structure across all scales. They are exceptional laboratories for studying characteristics of a stellar sample with largely homogeneous initial conditions and formation history, and especially for open clusters, their predominant spatial occurrence within the galactic plane identifies them as strategic targets for probing different galactic properties. Their generally young ages and formation history forge a deep connection between clusters and past or still active star formation sites.

The advent of extensive space-based, all-sky astrometric surveys, such as \textsc{Hipparcos} \citep{1997Hipparcos} and later the \emph{Gaia} mission with its various data releases \citep{Gaia_mission_2016, 2017Gaia_DR1, Gaia_DR2_Summary_2018, 2022DR3} and unprecedented observational accuracy and quantity, has heralded new insights not only into the shape and structure of clusters themselves \citep[e.g.,][cluster coronae]{meingast2020} but also regarding their mean properties \citep[e.g.,][]{2019Bossini, 2020b_Cantat-Gaudin_ages, 2021Dias}. Moreover, access to parallaxes, proper motions, and radial velocities has caused the scientific community to move away from studying only the clusters visible in positional space and instead focus on exploring the combined spatial and kinematic phase space. As a result, an ever-growing number of clusters and stellar groups is being discovered using various methods measuring overdensities within this space \citep[e.g.,][]{2018b_Cantat-Gaudin_Dr2, Cantat-Gaudin2020a, 2021SPYGLASSI,  2022_SPYGLASSIII, SPYGLASSII, 2022Castro-Ginard, 2022Ratzenbock_Sigma, 2023Hunt}, while at the same time conglomerates formerly regarded as clusters have been revealed to be only asterisms \citep[e.g.,][]{ 2018b_Cantat-Gaudin_Dr2, Cantat-Gaudin2020a}. Both findings are crucial for advancing the cluster and association census in the Milky Way. This, in turn, has consequences for our understanding of spatial structures on a galactic scale, as shown, for example, by \cite{2021Castro-Ginard}, who successfully traced the spiral arm structure of the Milky Way using open clusters.

Open clusters also greatly aid the accessibility of our galactic past due to their close relation to star-forming regions and as direct representations of star formation history. However, to fully access the temporal axis of structure formation and evolution on larger-scale objects, one requires knowledge about a crucial yet elusive physical parameter – the age of a cluster or a stellar population. The age parameter plays a critical role in developing timescales, for instance, for formation processes of large-scale star formation regions. For example, \cite{2023Ratzenbock_Sco-Cen_ages} recently mapped a possible evolutionary scenario for the nearest active star-forming site, Scorpius-Centaurus (Sco-Cen), using only the information deduced from the over thirty different star clusters they isolated in the 6D phase space. Cluster ages are also a vital piece for enhancing our understanding of molecular cloud formation and dispersal \citep[e.g.,][]{2011Murray, 2013Adamo, 2020SSChevance}, or even for constraining possible windows for exoplanet formation \citep{2015David_Pleiades_Benchmark}.

However, ages are not among the list of directly accessible parameters of stars, which means that, despite their importance across many different physical scales, their determination remains a notoriously difficult task to this day. This is especially true for stellar population ages. Various age estimation techniques have been developed for these objects, but their results are seldom unanimous. Even for well-studied, local clusters, as for example the Pleiades (Melotte~22), for which high-quality observational data are available, the age estimation process is often characterized by the need for prior assumptions and numerical models, leading to, at times, vastly deviating age estimates (see Fig.~\ref{fig:01-Blindspot-Pleiades}). The heavy model dependence of population age estimation also holds for empirical methods to some extent, as they need to be calibrated using benchmark clusters and, therefore, partially inherit their associated uncertainties. Notably, the Pleiades cluster seen in Fig.~\ref{fig:01-Blindspot-Pleiades} with its evident discrepancies in age estimates, is among the most commonly used benchmarks \citep[e.g.,][]{2012Bell, 2018Pleiades_DANCe, 2022Messina, 2023Brandner}. Moreover, uncertainties or gaps in current stellar evolution theories, computational limitations in the implementation of stellar interior physics, as well as measurement errors in observational data cause further complications and together preclude the determination of absolute ages with universal, unbiased, and academically undisputed methods \citep[see, e.g.,][ and references therein]{Soderblom2010}.

To circumvent many of the drawbacks associated with the age determination of stellar populations, in particular regarding stellar evolutionary model dependencies, we have developed an innovative approach to age estimation: Using only the information from observational Hertzsprung-Russel diagrams (HRDs) of stellar populations, we extract empirical isochrones for open clusters and thus pave the way for the derivation of relative cluster ages. Their precise representation of the source distribution can add vital information about the lower mass end of a stellar population and could be valuable for better constraining the physical models of stellar interiors in the low-mass regime. Based exclusively on widely available observational data, the distinguishing property of our approach, compared to other age estimation methods, is its purely factual nature. By providing empirical isochrones for a comprehensive, representative, and relevant selection of nearby open clusters, we establish a way of quantifying ages via a relative and homogeneous comparison between individual stellar populations.

The remainder of this work is structured as follows: We start by briefly explaining the current state of the art of stellar population age determination and the complications arising from different methods, that inspired the creation of the empirical isochrone archive in Sect.~\ref{sec:Ages}. We describe the data and cluster memberships we used in Sect.~\ref{sec:Data} before reviewing our methods and the isochrone extraction algorithm in Sect.~\ref{sec:Methods}. The results are presented in Section~\ref{sec:Results} and validated against age estimates found in contemporary literature. We follow up with a discussion of the isochrone quality and significance, especially for the region of interest in the lower main sequence, with the aid of case studies in Section~\ref{sec:Discussion}. Lastly, we summarize our findings in Sect.~\ref{sec:Conclusion}.

\section{The current state of population age estimation}
\label{sec:Ages}

Since there is no measurable parameter that enables unequivocal age determination, various methods have been developed to estimate stellar population ages using different indicators. These include asteroseismology \citep[e.g.,][]{2022Pamos-Ortega}, evaporation ages \citep{2024_Pelkonen}, gyrochronology \citep[e.g.,][]{2007Barnes, 2019Curtis_Meingast1}, kinematic and traceback ages \citep[e.g.,][]{2020Miret_roig_beta-pic, 2022_SPYGLASSIII}, lithium depletion (boundary) measurements \citep[e.g.,][]{1996_LD_Pleiades, 2014_LDB_beta-pic, 2018Martin}, or lithium equivalent widths \citep[e.g.,][]{2023Jeffries}. However, these methods all share a ``specialist'' limitation, in that they may only be suited for specific age, temperature, or mass ranges, require precise, work-intensive observed parameters, or can only be applied to either individual stars or ensembles. \citep[see, e.g., Table~1 in][]{Soderblom2010}.

Contrasting these rather specialized methods with narrow application ranges is an approach often praised for its generalistic nature -- isochrone fitting. Based on variants of the Hertzsprung-Russel diagram, particularly the color-absolute magnitude diagram (CMD), it is a popular estimation technique with only minimal observational data requirements. As a result of \emph{Gaia} providing photometry and astrometry for billions of sources, isochrone fitting has become perhaps the most used approach to age determination to date. Because of its connection to the HRD, the method is widely applicable across stellar age and mass ranges, for example from 0.5 Myr to 10 Gyr for $M_* = 0.01 - 1.4~M_{\odot}$ \citep{BHAC15} or 10 Myr to 12.5 Gyr for $M_* = 0.09 - 14~M_{\odot}$ \citep{2022Parsec_V2}. There are two major drawbacks to fitting isochrones, though: On the one hand, it heavily relies on the specific physics considered when modeling the theoretical isochrones. As the different physical processes of stellar evolution are highly complex and vary between different stellar ages and masses \citep{deBoerSeggewiss+2008}, this translates into a large number of free parameters, which can produce systematic errors in age estimates across different stellar evolution models and prior assumptions. Given the many different available stellar evolution codes for model isochrones, such as BHAC15 \citep{BHAC15}, DSEP-magnetic \citep{2016_DSEP-Magnetic}, BT-Settl \citep{2014Allard}, PARSEC \citep{2012PARSEC,2017marigo, 2022Parsec_V2}, MIST \citep{2016MIST1, 2016MIST2} and BaSTI \citep{2018Basti1, 2021Basti2, 2022Basti3}, this dependence has a profound impact on age estimates found in literature. On the other hand, the fitting technique of the isochrones to the CMD data itself can strongly impact the results. From the observational side, measurement errors and incomplete or contaminated cluster member data are drivers of age uncertainties \citep{deBoerSeggewiss+2008}. 

Additionally, observational cluster CMDs include an unknown amount of unresolved binary stars. The extent of this contaminating population has been estimated to range from about 29 to 50 \% \citep{MiMO}, and equal mass unresolved binaries reside at the maximum possible displacement of \SI{-0.753}{mag} above the main sequence. Additionally, reddening effects of non-equal mass unresolved binaries have been found to cause a shift of the corresponding data points closer, or right onto, the equal mass binary sequence \citep[e.g.,][]{1998Hurley,MiMO_only_binaries}. Consequently, it often appears disproportionately enlarged compared to the expected number of equal mass binaries. The presence of such a well-defined locus of contaminants harbors another challenge to isochronal age determination. However, the impact of the binary sequence on the empirical isochrones presented in this work has been specifically addressed and quantified (see Appendix \ref{appendix:Uncertainty-quantification}.

Consequently, it is challenging to verify age estimates obtained through different age estimation methods or to establish a uniform age for an object using the literature \citep[see, e.g., the comparison tables in][]{2014Mamajek_beta-pic, 2020Miret_roig_beta-pic, SPYGLASSII}. In the following paragraphs, we examine this situation's implications, focusing on isochrone fitting. For a more detailed explanation of the age determination methods discussed here, readers can refer to the review works of \cite{Soderblom2010, Soderblom2014}.

\subsection{Age ambiguity}
\label{sec:Ages-Ambiguity}

The apparent ambiguity of stellar ages derived with different methods is at least in part caused by their use of different age indicators. 
For instance, dynamical traceback ages, which measure the time since a group of stars was most concentrated in space, seem to systematically underestimate the ages compared to other techniques, such as isochrone fitting or lithium depletion boundary measurements, of young stellar agglomerates ($\lesssim 50$ Myr). A possible explanation for this observation could be that they already existed for a few Myr without significant expansion, which would not be reflected in the kinematic age indicator \citep{2024_Miret-Roig}. Another example concerns very young clusters, whose lithium depletion boundary ages have been reported to systematically be about 50 \% greater than those determined from their main sequence turn-off points \citep[][and references therein]{Soderblom2010}.

In contrast to methods based on different age indicators, the strong dependence of the results on the individual model assumptions causes isochrone fitting to produce significant age differences even though the overarching determination technique remains the same. This seriously impacts the reliability and the reproducibility of isochronal ages, as highlighted by \cite{2022_SPYGLASSIII, SPYGLASSII}. They compare isochronal ages from three stellar evolution codes, namely PARSEC \citep{PARSEC_CHEN2015}, BHAC15 \citep{BHAC15}, and DSEP-Magnetic \citep{2016_DSEP-Magnetic}, on different stellar groups and find age disparities of up to a factor of $\sim 2.6$ between the different models.

\begin{figure}[t]
        \centering
        \resizebox{\hsize}{!}{\includegraphics[]{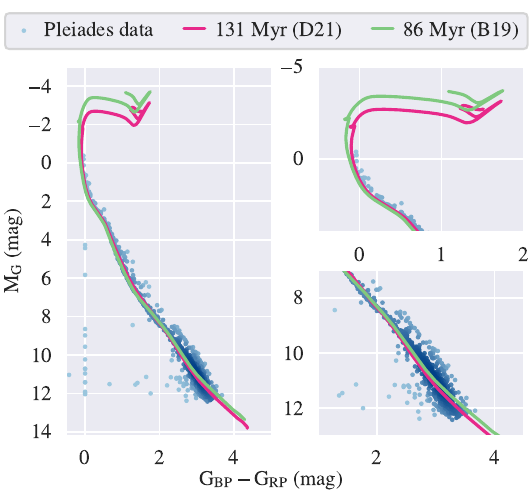}}
    \caption{Comparison between two different isochronal ages for the Pleiades cluster. The data points correspond to the \emph{Gaia} DR2 cluster selection of \cite{Cantat-Gaudin2020a} , 
 and the isochrones were produced with PARSEC, using the parameters calculated in the respective works \citep[][]{2019Bossini, 2021Dias}, indicated by B19 and D21 in the figure legend, respectively.}
    \label{fig:01-Blindspot-Pleiades}
\end{figure}

Further investigations show that even employing theoretical isochrones from the same model and the same member selection can lead to extreme age differences, as in the case of the Pleiades cluster depicted in Fig.~\ref{fig:01-Blindspot-Pleiades}. The plot shows the DR2 member selection of the cluster by \cite{Cantat-Gaudin2020a}\footnote{In the original selection of the authors, an artifact appears in the form of a vertical line of data points at the color index zero. It disappears when cross-matching their membership list with DR3 data.}, along with theoretical isochrones based on two published sets of age, metallicity, and extinction values, which were determined via interpolation of a parameter grid of PARSEC V1.2S isochrones. Both the theoretical isochrones displayed in the figure \citep[][denoted as B19 and D21]{2019Bossini, 2021Dias}\footnote{The authors used different parameter grids and \emph{Gaia} passband corrections. They report a similar cluster extinction ($A_{V,~\textrm{Dias}} = 0.168$, $A_{V,~\textrm{Bossini}} = 0.14$); however, \cite{2021Dias} treated the cluster metallicity as a free parameter (Z$_{\textrm{Dias}}$ = 0.032), while \cite{2019Bossini} assumed solar metallicity for the Pleiades.} present a good fit to the observations over almost the entire populated dynamical range between ca. 0 and 10 absolute G-band magnitudes. However,  their absolute age difference amounts to 45 Myr, which is around a 50 \% deviation of the lower age estimate. Yet, they are almost indistinguishable from each other, with the exception of the faint mass end of the sequence ($\mathrm{M}_{\mathrm{G}} \gtrsim 11$ mag), where neither isochrone presents a good fit to the data, and the the upper main sequence, which lacks observations.

The fact that such inconclusive age estimates exist even for one of the nearest, best-studied open clusters in the solar neighborhood, which is moreover a popular benchmark for empirical estimation methods, illustrates the extent of uncertainty that can be accrued when using isochrone fitting.

\subsection{Isochrone blindspot}
\label{sec:Ages-blindspot}

\begin{figure*}[ht]
        \centering
        \resizebox{\hsize}{!}{\includegraphics{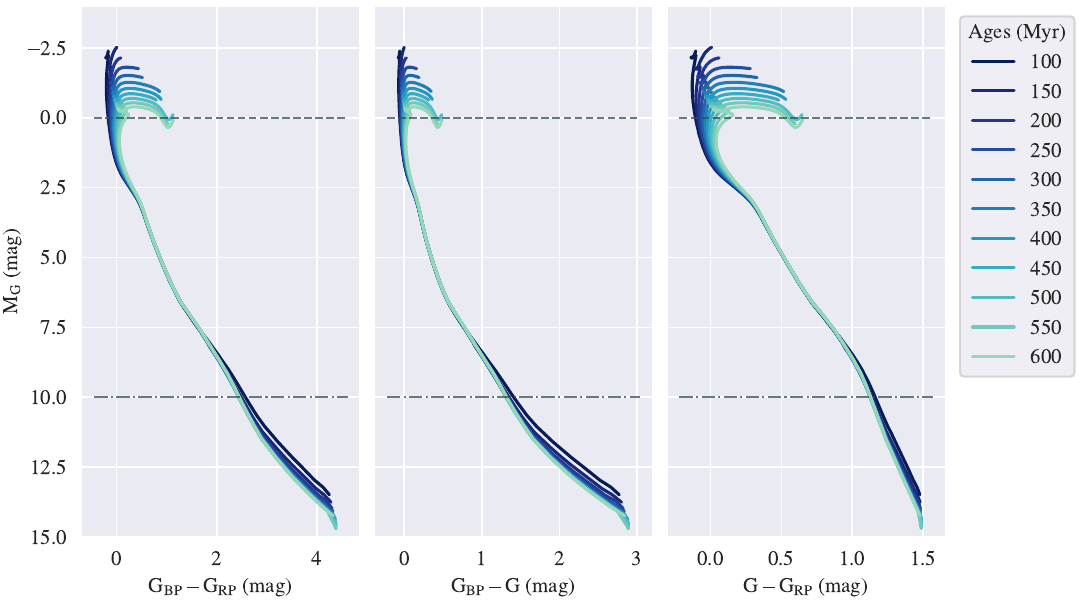}}
        \caption{Depiction of the largely featureless region, dubbed the ``isochrone blindspot.'' It covers an age span of approximately $100-500$ Myr and roughly $0-12$ G magnitudes in the \emph{Gaia} passbands. The plot features PARSEC isochrones \citep{2012PARSEC} with default parameters. The \emph{dashed line} indicates the upper main sequence region where observed sources are rare, while the \emph{dash-dotted line} delineates the lower main sequence, which is a region of interest regarding the information gained through empirical isochrones.}
        \label{fig:01-Blindspot}
\end{figure*}

As shown in the example of the Pleiades cluster, theoretical isochrones can appear almost indiscernible over a large portion of the main sequence, despite being associated with significantly deviating ages \citep[see also, e.g.,][]{deBoerSeggewiss+2008, Madys}. This effect is particularly evident for clusters in an age range around $\sim 100-500$ Myr and is prevalent across several photometric bands, not only for the specific set of \emph{Gaia} passbands. Moreover, it is not even restricted to the optical wavelength range, but can for example also be observed in the infrared $J$, $H$ and $K_S$ filters. It should be noted, that the dynamical range affected by the blindspot varies depending on the passbands used in the CMD.

Regarding the \emph{Gaia} photometric system, the overlapping of the theoretical isochrones commences as soon as their main sequences\footnote{In this context we refer to the lower main sequence as roughly $\mathrm{M}_{\mathrm{G}} \gtrsim 10$ mag, and to the upper main sequence when $\mathrm{M}_{\mathrm{G}} \lesssim 2$ mag.} do no longer appear flatter and shifted to redder colors like those of very young clusters ($\lesssim 30$ Myr). To illustrate this issue, Fig.~\ref{fig:01-Blindspot} shows a set of PARSEC isochrones calculated with default metallicity and extinction values for this age range in the three \emph{Gaia} DR3 color variations. The ambiguity of the main sequence across large parts of the CMD requires the isochronal age determination to hinge on available sources located in the upper main sequence and beyond (\emph{dashed line}), where the isochrones start to deviate again. However, the main sequence turn-off and evolved star regions only begin to get well populated for very evolved clusters, meaning no reliable information for the depicted age range is usually available for isochrone fitting. In addition, according to simulations and observational trends, massive stars are more likely part of binary or multiple systems \citep[e.g.,][and references therein]{2014deRosa}. If such a system was observationally unresolved, the age determination could be seriously influenced. The same holds for possible field star contaminants around the turn-off region, simply because of the typically low number of observations in this crucial region for isochrone fitting.

 In short, from the point at which young clusters are no longer separated from older populations via their shifted, flatter main sequence until the time at which the main sequence turn-off has shifted to lower stellar masses and becomes more populated for evolved clusters, there are no decisive features in the shapes of theoretical isochrones over a sizeable dynamical range -- resulting in the appearance of an ``isochrone blindspot.'' Physically, this blindspot is a consequence of the very stable and long-lasting burning processes occurring during the main sequence phase of lower mass stars \citep[e.g.][]{2007Barnes, Madys}, that results in only minute variations of the luminosity and temperature, or color, over their main sequence lifetimes. What is more, the blindspot does not only exist for theoretical isochrones but also partly occurs when considering other age determination methods for stellar populations. For instance, gyrochronology only works for single, solar-type stars, whereas lithium depletion boundary measurements or kinematic modeling apply only to young groups of $\lesssim 200$~Myr and $\lesssim 30$~Myr, respectively \citep{Soderblom2014}.


A possible workaround of the blindspot problem would be to consult the lower main sequences of clusters, below the \emph{dash-dotted line} in Fig. \ref{fig:01-Blindspot}. As can be seen, for this region the model isochrones are again better distinguishable; however, as depicted in the lower right panel of Fig.~\ref{fig:01-Blindspot-Pleiades} for the Pleiades, models often do not capture the observations in enough detail to permit an age estimation using only this part of the CMD with theoretical isochrone fitting. Empirical isochrones, on the other hand, very precisely match the shape of the lower main sequence of a CMD, provided there are observations of faint stars available for a given cluster. Comparing empirical lower main sequence isochrones thus provides an additional piece of information to be leveraged when deciding on an absolute age estimate, or can even be sufficient for a relative bracketing or ordering of observed stellar population ages (Sect. \ref{sec:Discussion-Case studies}).

\subsection{Empirical isochrones -- A largely untapped potential}
\label{sec:Ages-empirical-isochrones}

Despite current limitations in technology and physical knowledge concerning stellar evolution modeling, the recent rise in precision and quantity of stellar photometry and astrometry is opening up a new, data-driven approach for using isochrones as a decisive age indication tool: in an empirical form. Both empirical and theoretical isochrones have a fundamental connection to the HRD, which renders almost all aspects of stellar evolution accessible from a 2D representation. Consequently, empirical isochrones do not suffer from the problem of being too ``specialist'' compared to other age determination methods discussed in this section. However, in contrast to theoretical isochrones, the empirical curves directly originate from the CMD, meaning they are purely observation-based. Thus, in creating the empirical isochrone archive, we avoid the caveats and drawbacks associated explicitly with the model dependency of theoretical isochrones, while still taking advantage of the extensive applicability of the constructs of isochrones in general. Since the CMD comprises only three easily accessible observational parameters, empirical isochrone extraction profits from a tiny list of requirements and no physical complexity compared to the models necessary for computing theoretical isochrones.

The value of empirical isochrones has already been recognized among the scientific community, and (semi)-empirical isochrones have been extracted using different techniques and published for individual case studies or small groups of populations by, for example, \cite{2014Sarro_DANCe, 2015Bouy_7sisters, 2015Herczeg, 2018Olivares,  2019_IC4665_DANCe, 2019Dance_Ruprecht147, 2021_Li, 2021Olivares, 2023DANCe_IC348}. We now want to take this effort a step further and create an archive of empirical isochrones for a large sample of open clusters in the solar neighborhood, using purely observational data combined with a homogeneous extraction method based on statistical analysis and machine learning tools. By sacrificing the ability of absolute age quantification, we can decouple isochrones from their strong associated model dependencies and still have a relative age determination method that is entirely based on observational data. This approach is also independent of any assumptions regarding extinction or metallicity. By comparing new and known clusters, our archive can provide insights into cluster evolution and the properties of large-scale structures in the galaxy. Extracting and analyzing empirical isochrones may lead to the discovery of new features in their shapes, particularly regarding the lower main sequences, and provide calibration data for stellar evolution models, improving their quality and broadening our understanding of stellar evolution.

\section{Data}
\label{sec:Data}

In the following paragraphs, we outline the surveys and cluster member lists that we consulted for creating the empirical isochrone archive. As their specific membership determination method often influences our choice of applied quality cuts, we briefly describe the applied techniques used by the respective data source.

\subsection{Survey data}
\label{sec:Data-survey-data}

The observational information for the archive clusters is collected almost exclusively from the data products of the \emph{Gaia} mission \citep{Gaia_mission_2016}. Specifically, the extraction algorithm was built on input from the early third data release \citep[EDR3,][]{Gaia_EDR3_2021}, as well as the full DR3 \citep{2022DR3}, which contain identical astrometric and photometric measurements. To facilitate comparisons to pre-DR3 works, we additionally include isochrones based on \emph{Gaia} DR2 \citep{Gaia_DR2_Summary_2018} in our archive.\footnote{All CMDs using \emph{Gaia} passbands shown throughout this work depict \emph{Gaia} DR3 data unless explicitly indicated otherwise.}

We chose the \emph{Gaia} survey as our primary data source mainly due to its unprecedented size and precision. Its latest data releases include five-parameter astrometric solutions for 1.2 and 1.5 billion sources, respectively. Profiting from a homogeneous source of photometric and astrometric measurements for all archive clusters, we greatly reduce the need for gathering data across multiple sources \citep{2018b_Cantat-Gaudin_Dr2}, which has been reported to cause discrepancies in the results of cluster analyzes of different authors \citep[e.g.,][]{2015Netopil}. Furthermore, the low parallax uncertainties of $0.1$ \si{mas} for sources with G~$\lesssim$~18~mag, and the sub-millimagnitude photometric uncertainties in the color measurements (~$\leq$ 13 mag) provide the best possible conditions for creating sharp color-magnitude diagrams of star clusters. Finally, due to its homogeneous nature and vast source catalog, the \emph{Gaia} releases have been established among the most commonly used optical surveys for performing cluster membership and isochronal age determinations.

Alongside the \emph{Gaia} data, we also draw on data from the Two Micron All Sky Survey \citep[][2MASS]{2MASS} $J, H, K_S$ photometric system, as well as the Panoramic Survey Telescope And Rapid Response System \citep[][Pan-STARRs]{Panstarrs} $u, g, r, i, z, y$ bands for the two nearby clusters IC~4665 and the Pleiades. We do this to test the code on photometric systems other than the one it was designed on and to study empirical lower main sequences down to the mass level of brown dwarfs.

The required parameters for the extraction of empirical isochrones are constrained to apparent magnitude or flux measurements in at least two passbands for calculating the color indices, and stellar distances, to infer the absolute magnitudes via the distance modulus. The distances are computed via an inversion of the parallax, rather than for example using a Bayesian method, such as the one proposed by \cite{2015Bailer-Jones}, or the systematic parallax offset described in \cite{2018Lindegren-DR2}. To justify this choice, we analyzed the fractional parallax errors of all datasets where we performed the distance estimation ourselves, meaning the \emph{Gaia} DR2 and DR3 source catalogs. We found that only for 0.03 \% of all DR3 sources and 0.42 \% of all DR2 sources do those errors exceed 10 \%. Therefore, distance calculation via parallax inversion is a viable approach. However, we note that in case of using datasets with typical parallax errors larger than 10 \%, we recommend an alternative way of distance estimation when calculating absolute magnitudes.

\subsection{Cluster data and membership lists}
\label{sec:Data-cluster-memberships}

\begin{table*}[ht]
    \caption{Overview of the catalog contents, mean cluster parameters, and the applied quality cuts.}

    \begin{tabular*}{\linewidth}{@{\extracolsep{\fill}}lcccccc}
    
\hline \hline
Member  & Cluster(s)    & Mean distance & Member stars  & Probability                             & Age ref. & Further cleaning  \\
selection               &               & (pc)          & $N_*$         & or score                                           &        &                   \\
\hline
\multicolumn{7}{c}{\emph{Catalogs}} \\
\hline
(1)              & 64            & $135  - 498$    & $~96 - 1749$     & $p \geq 0.5$                            & (a), (b)     & Q1                \\
(2)              & 10            & $137 - 402$     & $321  - 1828$   & no                                      & (a)         & Q1                \\
(3)              & 16            & 1$06  - 177$    & $150  - 1196$   & \texttt{stability} $> 6$                & (c)         & Q2; $G_{\mathrm{BP,err}} < 0.05$ mag \\
\hline
\multicolumn{7}{c}{\emph{Single clusters}}\\
\hline
(4)              & Hyades         & 67           & 972           & no                                                  & (a)       & Q1                        \\
(5)              & Coma Berenices & 86           & 181           & no                                                  & (d)       & Q1                        \\
(6)              & Meingast~1     & 135          & 1366          & \texttt{stability} $\geq 0.24$                      & (a)      & Q1                        \\
(7)              & IC~4665 (DANCe)  & 356          & 819           & $p_{\text{\emph{Gaia}}} > 0.7$ or $p_{\text{DANCe}} > 0.5$ & (a)       & $i > 13$ mag; $r > 9$ mag                       \\
(8)              & Pleiades (DANCe) & 136          & 1483          & $p_t \geq 0.84$                                     & (a)       & $e_{\mathrm{i}} < 0.3$ mag, $z > 0$ mag \\
 \hline 
    \end{tabular*}
    \label{tab:02-Cluster-specs}
    \tablebib{\emph{Member selections.} (1) \citet{Cantat-Gaudin2020a}; (2) \citet{meingast2020}; (3) \citet{2022Ratzenbock_Sigma}; (4) Meingast, priv. comm.; (5) \citet{2019Furnkranz_ESSIII}; (6) \citet{2020ESSIV_Meingast1}; (7) \citet{2019_IC4665_DANCe}; (8) \citet{2018Pleiades_DANCe}; 
              \emph{Ages.} (a) \citet{2020b_Cantat-Gaudin_ages}; (b) \citet{2021Dias}; (c) \citet{2023Ratzenbock_Sco-Cen_ages}; (d) \citet{2019Curtis_Meingast1}}
    \tablefoot{All distances rounded to integer values. Numbers and quality criteria referring to DR3 data.\\ Quality filter 1 (Q1): $\varpi > 0$, RUWE$_{\mathrm{DR3}} \leq 1.4$.\\ Quality filter 2 (Q2): $\varpi > 0$, RUWE$_{\mathrm{DR3}} \leq 1.4$, \texttt{fidelty\_v2} $> 0.5$, $G_{\mathrm{err}} < 0.007$ mag, $G_{\mathrm{RP,err}} < 0.03$ mag \citep{2022Ratzenbock_Sigma, 2023Ratzenbock_Sco-Cen_ages}.}
\end{table*}

We collect cluster memberships from various sources found in contemporary literature, under the consideration of two global quality cuts: First, the mean cluster distance is limited to $d \leq$ \SI{500}{pc}, and secondly, the number of high fidelity cluster members is required to be $N_* \geq 100$. By ``high fidelity'' we group the probability or stability scores calculated by different methods, as can be seen in the fifth column of Table~\ref{tab:02-Cluster-specs}. 
 
 The first cut is justified from an observational point of view, as sources grow increasingly faint as a function of the cluster distance, resulting in an ever-increasing under-representation of low-mass sources in the CMDs of more distant clusters. At the same time, \emph{Gaia} parallax errors increase disproportionately when moving out to farther distances. For instance, \cite{Cantat-Gaudin2020a} state that the uncertainty in the proper motion measurements of DR2 starts to contribute significantly to the scattering of the cluster members already at distances around $\sim$ \SI{500}{pc}; hence we pick this distance as our cutoff value.

 The member number limit originates from the extraction method's perspective: Empirical isochrones should ideally represent a significant portion of a cluster's stellar initial mass function (IMF) and trace its population over a wide dynamical range. As clusters age, low-mass members may be dynamically ejected, leading to a depletion in the low-mass region of the main sequence. However, while this mass loss and its treatment in different membership determination methods are noteworthy, the general abundance of low-mass stars, owing to the shape of the IMF, ensures a well-populated lower main sequence. Consequently, the proposed empirical isochrone extraction is not significantly affected by a slight change of the mass distribution in the low-mass regime.

 Another motivation of the member number cut stems form the fact that the extractions are performed with statistical methods, which require a sufficiently large set of data points for each cluster to yield reliable results. From heuristic trial-and-error experiments, we determined that a member list of at least 100 sources provides adequate CMD populations in the overwhelming majority of the cases.\footnote{After the cross-matching and data cleaning, a cluster occasionally ends up with less than 100 members. These variations of a maximum $\pm 5~\%$ are allowed within our cut.} Depending on the sharpness of the CMD of an observed population, one could also relax this criterion, but it would come with the stipulation of having to inspect each case individually manually.

We accumulate 100 membership lists corresponding to 88 individual star clusters. The discrepancy between the number of memberships and actual clusters in the archive arises from a twofold overlap: On the one hand, the selection of 10 clusters from \cite{meingast2020} corresponds to a subset of the sample collected from \cite{Cantat-Gaudin2020a}. On the other hand, for the two clusters IC~4665 and the Pleiades, we additionally collect data from the Dynamical Analysis of Nearby Clusters (DANCe)   survey \citep{2018Pleiades_DANCe, 2019_IC4665_DANCe} along with the \emph{Gaia} observations. Concerning the member selections based on the \emph{Gaia} passbands, only the selection of \cite{2022Ratzenbock_Sigma} was performed on DR3 data, whereas all others used DR2 data. Since the third data release generally comprises more accurate measurements, we cross-match the DR2 selections with their DR3 counterparts.

Due to quality concerns regarding the empirical isochrones that the algorithm calculated in isolated cases, we excluded five clusters from our original sample (details in Sect.~\ref{sec:Data-final-cut}), bringing the number of archive clusters to 83 in total. A summary of the cuts and mean cluster parameters of our final selection, divided into their respective source catalogs, can be found in Table~\ref{tab:02-Cluster-specs}. The following provides a brief overview of the literature sources and a short description of their membership determination processes.

\subsubsection{\texorpdfstring{\cite{Cantat-Gaudin2020a} selection}{}}

 Our primary source is the extensive open cluster catalog by \cite{Cantat-Gaudin2020a}, who created membership lists for 1481 clusters in the Milky Way. After applying the global quality cuts, with high fidelity sources being defined as $p \geq 0.5$ on the probability score determined by the authors, we obtain membership lists for 67 open clusters, hereafter referred to as \citetalias{Cantat-Gaudin2020a}. Of these, 59 were first determined in \cite{2018b_Cantat-Gaudin_Dr2} with the unsupervised, two-step UPMASK procedure \citep[Unsupervised Photometric Membership Assignment in Stellar Clusters,][]{2014_UPMASK}. Their method consisted of clustering stars based on astrometric proper motion and parallax measurements ($\varpi$, $\mu_{\alpha_*}$, $\mu_{\delta}$) of \emph{Gaia} DR2, and subsequently validating each cluster by comparing its density to that of a random distribution. In doing so repeatedly with values drawn from the astrometric parameter distributions, a membership probability was determined for each star in their input sample \citep[][]{2018b_Cantat-Gaudin_Dr2}. \citetalias{Cantat-Gaudin2020a} further includes 5 UPK clusters \citep{2019UPK_clusters}, for which \cite{Cantat-Gaudin2020a} also provide membership lists calculated with UPMASK. Finally, 3 of the 67 groups are UBC clusters found by \cite{2018Castro-Ginard_UBC}, who used DBSCAN \citep{1996DBSCAN} in combination with artificial neural networks on the 5D astrometric parameter space of \emph{Gaia} DR2 data to perform their selection. It should be noted that \cite{2018Castro-Ginard_UBC} did not compute membership probabilities for their clusters, and thus the probability column provided in \cite{Cantat-Gaudin2020a} equates to one for all UBC clusters. 

\subsubsection{\texorpdfstring{\cite{meingast2020} selection}{}}

For a subset of \citetalias{Cantat-Gaudin2020a}, namely the ten open clusters $\alpha$~Per (Melotte~20), Blanco~1, IC~2602, IC~2391, NGC~2451A, NGC~2516, NGC~2547, NGC~7092, Platais~9, and the Pleiades we also collect alternative membership lists from \cite{meingast2020}, hereafter referred to as \citetalias{meingast2020}. These authors determined cluster membership based on deprojected proper motion measurements, which enabled them to not only retrieve members in the cluster cores but also in sub-field-density stellar coronae that can extend several 100 pc from the central region. As a result, comparatively more members were detected, especially down to the low-mass end of the main sequence. With their clustering method, no membership probability was assigned to the members of the clusters.

We use this subset of open clusters as main sample for testing and refining the extraction algorithm. Furthermore, as the lower main sequences of their clusters are generally better populated than those of the \citetalias{Cantat-Gaudin2020a} clusters, they are well suited for studying the information gain via empirical isochrones in the very faint regions (see Sect.~\ref{sec:Discussion-MG1}).

\subsubsection{\texorpdfstring{\cite{2022Ratzenbock_Sigma} selection}{}}

To extend the empirical isochrone archive toward young clusters, we consult the work of \cite{2022Ratzenbock_Sigma}, who recently extracted 37 co-moving star clusters in the nearby Sco-Cen star-forming region. To determine the cluster memberships, \cite{2022Ratzenbock_Sigma} employed the Significance Mode Analysis (\texttt{SiGMA}) tool to \emph{Gaia} DR3 data in a box around the molecular cloud complex. \texttt{SiGMA} works by identifying modal regions in the 5D astrometric parameter space of the observational data. It first estimates the data density and performs a graph-based gradient hill climb to identify local density peaks. The multiple resulting preliminary clusters are then iteratively merged using a modality test to measure the significance of the separating density dip. Instead of membership probabilities, the algorithm assigns each cluster member a \texttt{stability} value between zero and 100, though a given cluster may have a maximum stability of less than 100. We choose a generous threshold of \texttt{stability} $ > 6$ for high fidelity members. Furthermore, we apply the quality cuts listed in \cite{2023Ratzenbock_Sco-Cen_ages} (see Table~\ref{tab:02-Cluster-specs}) and set a stricter quality cut in the blue passband error $G_{\mathrm{BP,err}} < 0.05$~mag to remove outliers from the lower left region of the CMD. As a result, we retain 18 of their determined clusters, which we hereafter refer to as \citetalias{2022Ratzenbock_Sigma}.

\begin{figure*}[ht]
        \centering
        \resizebox{\hsize}{!}{\includegraphics{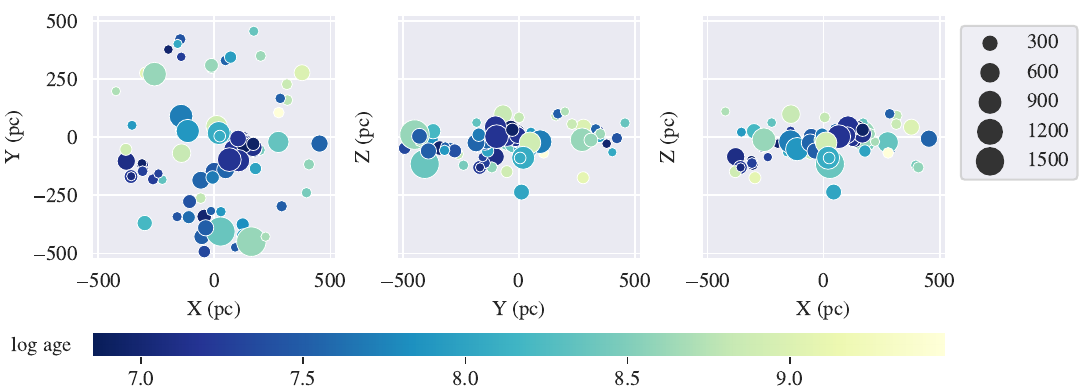}}
        \caption{Spatial distribution and relative sizes of the archive clusters, determined by the different source catalogs. The cluster positions included in \citetalias{meingast2020} are not displayed to avoid duplication of \citetalias{Cantat-Gaudin2020a} sources. The marker colors correspond to the estimated cluster ages from literature (Table~\ref{tab:02-Cluster-specs}).}
        \label{fig:02-Cluster-XYZ}
\end{figure*}

\subsubsection{Hyades and Coma Ber clusters}

To complement the clusters of \citetalias{Cantat-Gaudin2020a}, we collect membership lists for the Hyades (Melotte~25) and the Coma Berenices (Melotte~111) cluster. The two well-known OCs are absent from the works of \cite{2018b_Cantat-Gaudin_Dr2, Cantat-Gaudin2020a}, as their proximity makes those clusters incompatible with their membership selection method. Regarding Coma Ber, we use the membership selection published by \cite{2019Furnkranz_ESSIII}, who performed a wavelet decomposition to single out stellar overdensities in velocity space and subsequently employed DBSCAN to identify cluster members without the use of a probability or score measure. For the Hyades, the clustering method is the same that was employed for the clusters of \citetalias{meingast2020} (Meingast, priv. comm.). 

\subsubsection{Meingast~1 stellar stream}

The Meingast~1 stream \citep[][also referred to as Pisces-Eridanus stream]{2019Meingast_ESSII, 2019Curtis_Meingast1, 2020ESSIV_Meingast1, 2020Roeser} is not a classical open cluster, but rather a large, spread-out stellar group of co-moving objects, which traverse the solar neighborhood at roughly a distance of \SI{100}{pc}. At a length of more than \SI{400}{pc} and width of \SI{50}{pc}, it is speculated to be the remnant of a tidally disrupted cluster or OB association. The stream was initially identified via a wavelet decomposition of the 3D velocity space of stars in the solar neighborhood ($d <$ \SI{300}{pc}) by \cite{2019Meingast_ESSII}. In a later work, \cite{2020ESSIV_Meingast1} revisited the object and determined new memberships based on the five-dimensional DR2 position and velocity space ($\alpha$, $\delta$, $\varpi$, $\mu_{\alpha}$, $\mu_{\delta}$) clustering of a supervised one-class support vector machine \citep[OCSVM,][]{1995Cortes_OCSVM}. Similar to the clusters of \citetalias{2022Ratzenbock_Sigma}, possible member stars were assigned a \texttt{stability} parameter, albeit normalized and provided as percentages for the stream. Of the two filter criteria proposed by \cite{2020ESSIV_Meingast1} (\texttt{stability} $> 4$ \% or \texttt{stability} $\geq 24$ \%), we apply the stricter one.

As the stream members are still thought to be of a common origin, displaying a narrow main sequence very similar to the Pleiades \citep{2020ESSIV_Meingast1}, we also choose to test our isochrone extraction algorithm on this object. We further compare the empirical isochrones of the stream with those of OCs of a similar evolutionary state in Sect.~\ref{sec:Discussion-MG1} to analyze the impact of empirical isochrones on the lower main sequence in the isochrone blindspot region.

\subsubsection{DANCe data for IC~4665 and the Pleiades}

 The data for the clusters IC~4665 and the Pleiades was collected as a part of the DANCe survey \citep{2013Bouy_DANCe, 2018Pleiades_DANCe, 2019_IC4665_DANCe}. As such, the measurements for the clusters were gathered from different observatories and instruments and contain observations in near-infrared passbands. Since their data do not include parallaxes, these values need to be substituted. For IC~4665, \cite{2019_IC4665_DANCe} already calculated a median parallax value from the portion of cluster members visible in \emph{Gaia} DR2. Concerning the Pleiades, we adapt the distance from the cluster parallax denoted in \cite{Cantat-Gaudin2020a}.
Regarding the membership determination methods, \cite{2019_IC4665_DANCe} used an algorithm that models a field and a cluster component, respectively, based on an initial member list \citep{2014Sarro_DANCe,2019Dance_Ruprecht147}. Their approach is of Bayesian nature, and a membership probability is provided for the clustered sources. The membership determination for the Pleiades was undertaken with a similar, but more computationally expensive, Bayesian-based technique, consisting of a generative mixture model and a Markov Chain Monte Carlo method for calculating the posterior distributions of the parameters. A probability score for cluster members was created by adding the Bayesian membership probability and its sensitivity to the cluster parameter for each star in their sample. Their probability threshold was defined at $p_t = 0.84$ (see Table~\ref{tab:02-Cluster-specs}), which we adopt for the high fidelity sources.

Including the two clusters in our selection offers a convenient way to accomplish studying the shapes and added value of empirical isochrones regarding the faint, low-mass end of the main sequence of star clusters. Additionally, it permits exploring the extension of the method toward different photometric systems.

\subsection{Final cluster selection and general archive properties}
\label{sec:Data-final-cut}

After applying our method to all 88 clusters described in the previous section and visually inspecting the results and the respective cluster CMDs, we had to perform a final cut on our archive selection. This is because, for a few select sources, no physically meaningful isochrones could be extracted from their CMDs due to either a severe distribution scattering or concerns regarding their member selections. The concerned objects include the three clusters L1641~S, IC~348, and RSG~7 from \citetalias{Cantat-Gaudin2020a}, as well as the two groups $\rho$ Oph/L1688, and Lupus $1-4$ from \citetalias{2022Ratzenbock_Sigma}.

Three of the excluded clusters are estimated to be very young (L1641~S $\sim$ \SI{3}{Myr}; \citealt{2017_L1641S_age}, $\rho$ Oph/L1688 $\sim$ \SI{3.1}{Myr}, and Lupus $1-4$ $\sim$ \SI{4.5}{Myr}; \citealt{2023Ratzenbock_Sco-Cen_ages}), which is likely the reason for their scattered CMDs.

On the other hand, with an estimated age of $\sim$ \SI{12}{Myr} \citep{2020b_Cantat-Gaudin_ages}, the cluster IC~348 does not fall below the apparent lower age threshold of the archive. However, its estimated extinction of $A_V = 1.91 - 2.3$ mag exceeds that of the second highest reported extinction (UBC~17a, $A_V = 0.8$ mag) by almost a factor of two (references from Table~\ref{tab:02-Cluster-specs}). The region was for example studied by \cite{2023DANCe_IC348}, who further determined that the extinction varies significantly over the dynamical range of the cluster CMD and discussed its extinction law. We note that along with $A_V$ and $R_V$ distributions, the authors also provide empirical, extinction-free isochrones derived with a similar method for IC~348 and five other Perseus groups in their work.

In contrast to the other excluded clusters, RSG~7 is not characterized by any extreme values concerning either age or extinction. Still, following our quality cuts, its CMD displays two distinct populations along the main sequence, directly contrasting the sharp, well-defined age sequence one would expect for a young, bound cluster. As both subbranches are comparable in member numbers, it is impossible to distinguish between contaminants and actual cluster members without further investigation. Therefore, we exclude this cluster.

\begin{figure}[t]
    \resizebox{\hsize}{!}{\includegraphics{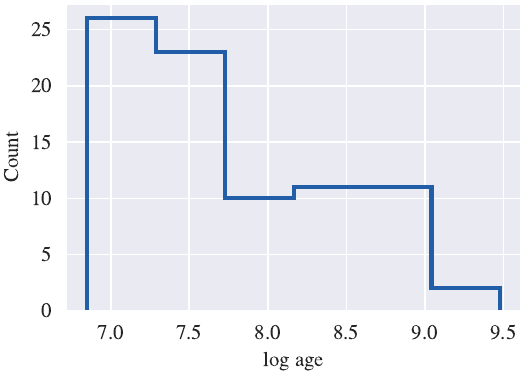}}
    \caption{Histogram depicting the logarithmic age distribution of the archive clusters, according to the literature values listed in Table~\ref{tab:02-Cluster-specs}. Due to visualization purposes, the ordinate is scaled logarithmically. The ages are binned using Knuth's rule, as is implemented in \texttt{astropy} \citep{astropy:2013,astropy:2018}.}
    \label{fig:02-Age-hist}
\end{figure}

Figure~\ref{fig:02-Cluster-XYZ} displays the spatial distribution of the final 83 clusters selected for the archive in a heliocentric Cartesian coordinate system. The number of cluster members, determined by the respective source catalog, is reflected in the relative size of the position markers. The clusters are color-coded according to the estimated ages found in the literature. The references for the cluster ages are provided in the rightmost column of Table~\ref{tab:02-Cluster-specs}. For the cases of \citetalias{Cantat-Gaudin2020a}, we preferably use the ages from \cite{2020b_Cantat-Gaudin_ages} and only consult the work of \cite{2021Dias} as a secondary source for clusters not covered by the former. A histogram showing the age distribution of the archive clusters is displayed in Fig.~\ref{fig:02-Age-hist}. As can be seen, the bulk of the archive clusters are younger than \SI{100}{Myr}, and only very few clusters are older than one Gyr, whereas the lower age limit is around 7 Myr. The displayed age caps indicate a reliable age range for a representative cluster catalog and correspond to natural limits in open cluster observations. After all, very young clusters ($\lesssim$ \SI{5}{Myr}) are at least partly obscured and reddened by their parental dust cloud, rendering them invisible to \emph{Gaia} or badly scattered, as discussed above. Very old clusters ($\gtrsim 1$ Gyr), on the other hand, are naturally rare due to infant mortality \citep[][and references therein]{2003Lada&Lada}.

\section{Methods}
\label{sec:Methods}

In search of an isochrone extraction routine, we tested several methods, some of which have their roots in data analysis, while others originate from the research field of image processing. In addition, we investigated various preprocessing strategies. After evaluating their respective performance on the clusters of \citetalias{meingast2020}, we decided on a combination of Support Vector Regression (SVR) and Principal Component Analysis (PCA) as the final procedure. Not only did this combination yield the best results, but it also conforms to an important applicability aspect, as it is very adaptive toward the many different CMD shapes in our sample. Thus, it can be easily applied to CMD distributions reflecting various evolutionary states of the corresponding clusters and diagrams created from different passband combinations or photometric systems, without requiring extensive manual parameter tuning. We employ PCA as preprocessing step to provide continued functional dependency of the cluster data, particularly concerning evolved clusters. SVR is then used to perform the actual extraction of the empirical isochrones. While other authors have also started providing empirical isochrones for individual case studies of open clusters, to the best of our knowledge, we are the first to design an approach utilizing SVR and to derive empirical curves for a large, age-representative sample of clusters in a homogeneous manner. For comparison, \cite{2015Bouy_7sisters, 2019_IC4665_DANCe, 2019Dance_Ruprecht147} for instance employ a version of the Principal Curves technique first described by \cite{PC_Hastie}, which we also tested but ultimately discarded due to the loss of some key characteristics of the isochrones. Both PCA and SVR are implemented in the \texttt{Python} library \texttt{scikit-learn} \citep{scikit-learn} in their nonlinear extensions, which we use for our algorithm.


\subsection{Requirements and possible pitfalls}
\label{sec:Methods-pitfalls}

Despite the major benefit of model independence, one should be aware of some potential pitfalls when relying on observational data to produce isochrones. For instance, a low signal-to-noise ratio in the measurements can lead to a large scatter in the star positions in the CMD. This makes it difficult to derive the finer characteristics of the isochronal curve empirically. Another common occurrence in observational data concerns outliers. In cluster CMDs, they often appear in the form of unresolved  main sequence binaries (see Sect. \ref{sec:Ages}). Unresolved binary pairs of main sequence stars with evolved objects such as red giants or white dwarfs can induce further scatter both above and to the right, as well as below and to the left of the main sequence in a CMD \citep{Gaia_DR2_HRDs_2018}. Field stars can cause additional contamination of the member selection, either assuming random positions in the CMD or forming a distinct sequence (e.g., RSG~8 in Fig.~\ref{fig:04-Summary_Matrix_1}). Furthermore, all observational data are associated with measurement errors, which in the case of \emph{Gaia} DR3 are readily available and should be considered in the isochrone extraction. A final caveat arises from the fact that the extraction technique is based on data analysis, which generally requires a functional dependency between the variables one is interested in. Common isochrone characteristics such as the main sequence turn-off, white dwarf regions, or even a relatively vertical distribution of upper main sequence stars directly counteract such a functional dependency in the original CMD parameter space and must be accounted for in the algorithm design.

\subsection{The algorithm}
\label{sec:Methods-algorithm}

To counteract the described possible caveats, the cluster data need to be preprocessed in a way that preserves information while simultaneously ensuring a continued functional relationship between the input variables. Furthermore, the isochrone extraction needs to be robust against outliers assuming various forms and needs to be able to incorporate measurement errors. Hence, the workflow of creating a robust empirical isochrone from observational data with our developed method consists of the following main steps:
\begin{enumerate}
    \item Preprocessing via PCA transformation of the cluster data.
    \item SVR hyperparameter tuning via gridsearch and cross-validation.
    \item Extraction of a preliminary empirical isochrone with the tuned SVR model and the weights calculated from measurement errors.
    \item Generation of a set of resampled isochrones via iterative bootstrapping ($n_{\text{boot}} = 1000$).
    \item Calculation of the median, the 5$^{\text{th}}$ and 95$^{\text{th}}$ percentile values from the collection of resampled and reverse-transformed isochrones.
\end{enumerate}
In the following paragraphs, each step of the workflow is discussed in more detail.

\subsubsection{Preprocessing: Principal component analysis}

Principal component analysis is an adaptive data analysis technique \citep{2016Jolliffe}, whose variables are always defined by the specific input dataset instead of a priori. Thus, it is well suited for an isochrone extraction algorithm that needs to perform on a versatile input cluster CMDs. 
Briefly put, PCA transforms the input parameters, in this case the color index and absolute magnitude of a cluster CMD, into new, uncorrelated variables that successively maximize the variance in the data. It can be shown that this maximization can be reached by defining the new variables as a set of mutually orthogonal linear combinations of the input variables, together with the eigenvectors corresponding to the largest eigenvalues of the data. These combinations are commonly referred to as ``principal components.'' For an introduction to the basic principle of the method, we refer readers to works such as \citet{PCA_Bishop, 2016Jolliffe}. A detailed description of the application of the technique to the problem at hand can be found in Appendix~\ref{Appendix:PCA}.

\begin{figure}[t]
    \centering
\resizebox{\hsize}{!}{\includegraphics{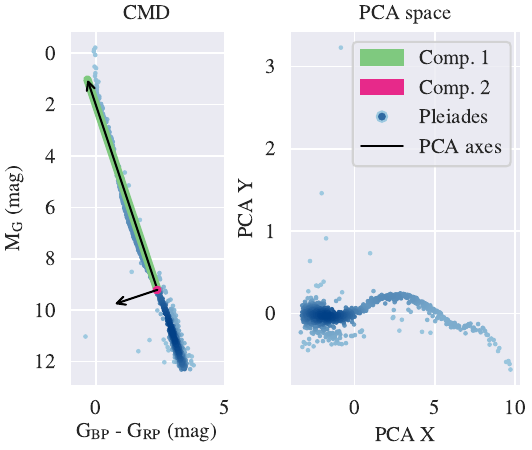}}
    \caption{Illustration of a PCA transformation of CMD data on the example of the Pleiades cluster. \emph{Left:} Extent and orientation of the principal components. The first component (\emph{green}) is much larger than the second one (\emph{magenta}), reflecting the different orders of magnitude of variance in the observation matrix. The directions of the PCA space axes are denoted by black arrows, with the second one magnified by a factor of 50 for better visibility. \emph{Right:} Transformed cluster data in PCA space, with the new axes pointing in the direction of the two principal components.}
    \label{fig:03-PCA-components}
\end{figure}

We use PCA to rotate the reference frame into the direction of the principal components spanned by the new variables (see Eq. \ref{eq:03-Linear-combis}). The purpose of this data transformation is illustrated in Fig.~\ref{fig:03-PCA-components} on the example of the Pleiades cluster. The left panel shows a regular CMD, overlaid by two colored bars indicating the extent of the principal components, which have been scaled by their corresponding variance. The arrows represent the direction of the eigenvector associated with the respective component and therefore span the axes of the PCA space. Due to the source distribution in most CMD configurations, which mostly comprises stars populating the diagonally or vertically inclined main sequence region, the variance in the direction of the absolute magnitude axis is much larger than the one for the color axis. This physical fact is accurately represented in the ratio of the principal components; it can be seen that the second principal component is much smaller than the first one, and the superimposed arrow denoting the direction of its eigenvector has been magnified by a factor of 50 for better visibility. The PCA-transformed cluster data can be seen in the right panel of Fig.~\ref{fig:03-PCA-components}. A look at the $y-$axis of the PCA space shows that the absolute data scatter is much smaller in that direction than for the CMD. In preprocessing the cluster data with PCA, we circumvent two of the pitfalls of observational data discussed in Sect.~\ref{sec:Methods-pitfalls}: Firstly, projecting the data into PCA space concentrates the variance on the first component instead of distributing it between the two input variables, which is beneficial for the isochrone extraction. Secondly, the functional dependency between the two variables is preserved for all shapes and evolutionary stages of cluster CMDs.

\subsubsection{Curve extraction: Hyperparameter tuning and weighted support vector regression}

The central component of the algorithm, meaning the method governing the generation of empirical isochrones, is based on the machine learning technique of support vector regression. The sparse, memory-efficient approach originates in statistical learning theory \citep{SVR_smola2003} and was, for example described by \cite{SVR_Bishop2006}, to whom interested readers are referred for a more detailed description. The procedure's working principles are outlined in Appendix~\ref{Appendix:SVR}, based on its implementation in \texttt{sklearn}. 

The general benefits of the method include its straightforward implementation, low computational cost, and small number of hyperparameters, meaning variables not subjected to analytical optimization. Instead, they can be defined somewhat arbitrarily and impact the resulting regression to varying degrees. The hyperparameters of SVR originate at different points of the formal derivation of the method (see Appendix~\ref{Appendix:SVR} for a detailed discussion). The first two, \texttt{epsilon} and \texttt{C}, are defined via the characteristic loss function of the regression (Eqs.~\ref{eq:03-epsilon-loss-function}$ - $\ref{eq:03-Primal-problem}). The former determines the number of support vectors, meaning the size of the subset of training data at which the kernel function is evaluated. By adjusting its value we can enhance the method's sensitivity for outliers, particularly regarding the unresolved binary sequence. The latter parameter, also called the penalty, determines the leniency toward considering points far away from the main distribution in the regression. Lastly, for our algorithm we use a radial basis function as kernel, adding a final hyperparameter called \texttt{gamma}, which is connected to the variance of the input data. 
Due to the many possible CMD shapes, the hyperparameters must be tuned once for a given CMD variation of a cluster at the beginning of a regression. However, they do not need to be re-established after the isochrone has already been calculated for a specific CMD or group of similar CMDs. We tune the hyperparameters on a predefined parameter grid using a gridsearch and 5-fold cross-validation within the \texttt{sklearn} package. For a detailed analysis of the influence of each hyperparameter and a description of the tuning process, readers are referred to Appendix~\ref{appendix:SVR-hyperparameters}.

Once the hyperparameters for a given CMD configuration have been determined, the tuned SVR model is fitted to the entire transformed cluster data, corresponding to the whole set of observations. This is done as we are not predicting further points but instead want to generate a regression curve representing the observed data as accurately as possible. Using the trained model to predict the response variable, meaning the PCA Y data, yields a regression curve that can then be easily transformed back into the original parameter space, where it corresponds to an empirical isochrone, akin to the data transformation shown in Fig.~\ref{fig:03-PCA-components}. A benefit of the SVR implementation of \texttt{scikit-learn} \citep{scikit-learn} is its ability to incorporate a weight parameter for each data point, which is subsequently used to multiply the penalty parameter \texttt{C}. Large weights urge the SVR model to include a data point in the regression, whereas small weights allow more lenience. Hence, we can solve the problem of incorporating measurement errors, if they are available, by turning them into one-dimensional weights for each data point. To this end, we convert the photometric and astrometric measurement uncertainties into scalar values using the root sum of squares (RSS), as well as Gaussian error propagation (details in Appendix~\ref{appendix:SVR-weights}).

\subsubsection{Empirical isochrones and uncertainty regions}

\begin{figure}[t]
    \centering
    \resizebox{\hsize}{!}{\includegraphics{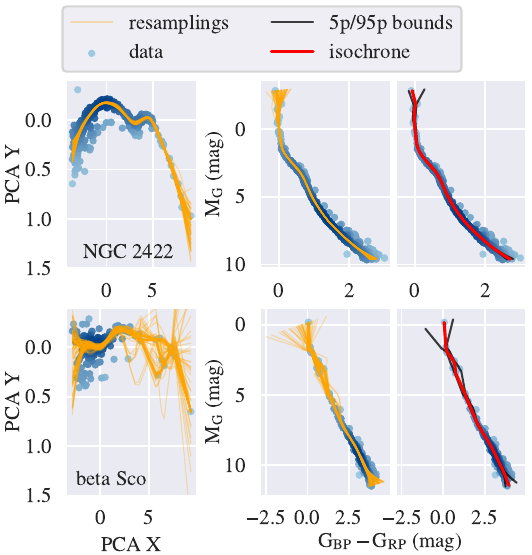}}
    \caption{Illustration of the bootstrapping and resampling process. The procedure is shown for the well-behaved cluster NGC~2422 (\emph{top row}), and the more scattered beta~Sco group (\emph{bottom row}). The \emph{orange lines} in the left column of the figure indicate 100 resampled SVR isochrones created via bootstrapping of the PCA cluster data. The center column depicts the same lines projected into the respective cluster CMD. The right column shows the median (\emph{red line}), 5$^{\text{th}}$ and 95$^{\text{th}}$ percentile boundaries (\emph{black lines}) calculated from the resampled set.} 
    \label{fig:03-bootstrapping}
\end{figure}

\begin{figure*}[ht]
        \centering
        \resizebox{\hsize}{!}{\includegraphics{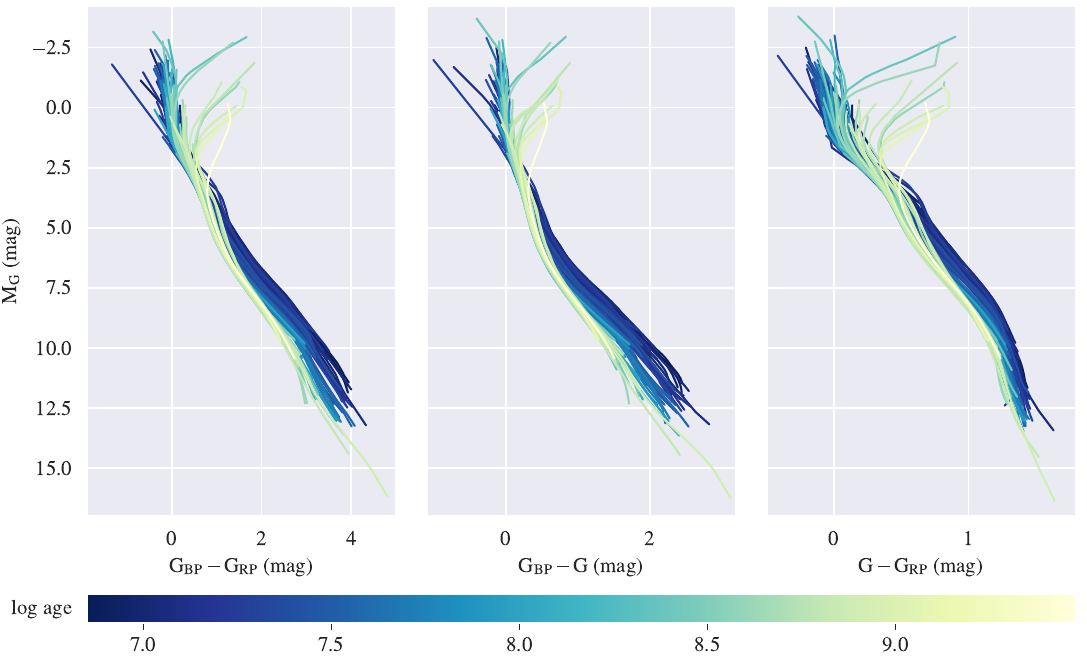}}
        \caption{Summary plot of all \emph{Gaia} empirical isochrones calculated for the 83 archive clusters, shown for the three different color indices $\mathrm{G}_{\mathrm{BP}}-\mathrm{G}_{\mathrm{RP}}$, $\mathrm{G}_{\mathrm{BP}}-\mathrm{G}$, and $\mathrm{G}-\mathrm{G}_{\mathrm{RP}}$. The color coding of the isochrones corresponds to the estimated cluster age found in the literature (see Table~\ref{tab:02-Cluster-specs} for details). An interactive version of the figure is also available \href{https://homepage.univie.ac.at/alena.kristina.rottensteiner/Empirical_isochrone_archive.html}{here}.}
        \label{fig:04-Summary_plot}
\end{figure*}

The last two steps of the algorithm workflow are designed to create a more reliable empirical isochrone than can be achieved by a single regression. Therefore, we bootstrap the PCA-transformed cluster data to create a slightly different input array for training a new SVR model with the predetermined hyperparameters. To get a robust result, we create $n_{\text{boot}} = 1000$ bootstrapped datasets for training SVR models and calculate the regression curves in PCA space by predicting the original PCA dataset. We then use the set of resampled curves to determine the empirical isochrone as the median of the values for each color index of the original data. We further compute the 5$^{\text{th}}$ and 95$^{\text{th}}$ percentile values, respectively. They correspond to an uncertainty region around the empirical isochrone and represent its reliability or trustworthiness locally along the data distribution in the CMD. The bootstrapping and the median or percentile array calculations are undertaken in PCA space before the results are inversely transformed into the original CMD space using the transformation matrix defined for the original cluster data.

The process is visualized in detail for two example clusters in Fig.~\ref{fig:03-bootstrapping}, using only $n_{\mathrm{boot}}=100$ bootstrapped curves for better visibility. For well-sampled clusters, such as the one in the \emph{top row} of the figure, the individual curves are generally in excellent agreement and largely overlapping. As a result, the uncertainty regions are very narrow over the whole populated sequence. On the other hand, if the cluster distribution is more scattered or exhibits sparsely populated regions, as in the case of the young beta~Sco group, those areas are characterized by larger uncertainty bounds, which tend to broaden toward the upper and lower limits of the empirical isochrone.

\section{Results}
\label{sec:Results}

We derived empirical isochrones for the three CMD color combinations of \emph{Gaia} DR2 and DR3 for all 83 clusters in the final source catalog (Sect.~\ref{sec:Data-final-cut}). The tables containing the results and the calculated uncertainty bounds are available online; an example of their contents can be seen in Appendix~\ref{Appendix:Tables}. 

The results based on the third \emph{Gaia} data release are displayed in Fig.~\ref{fig:04-Summary_plot} and color-coded according to the estimated cluster age found in the literature (Table~\ref{tab:02-Cluster-specs}), akin to the spatial distribution shown in Fig.~\ref{fig:02-Cluster-XYZ}. An interactive version of the summary figure, in which isochrones can be selected arbitrarily and studied in more detail, or their evolution may be visualized by successively adding them to the diagram using a slider corresponding to their estimated ages, is also available online and can be accessed via the link in the figure caption. Besides the summary plot, figures~\ref{fig:04-Summary_Matrix_1} and \ref{fig:04-Summary_Matrix_2} provide a separate view of the results for each cluster, using a $\mathrm{G}_{\mathrm{BP}}-\mathrm{G}_{\mathrm{RP}}$ color-magnitude diagram and the corresponding empirical isochrone and uncertainty bounds. The panels displaying the clusters are sorted in ascending order of their estimated literature ages.

From the individual cluster CMDs, it is apparent that the SVR-generated isochrones generally trace the respective stellar distributions very well. They even preserve specific details, such as the knee-like bend in the upper main sequence ($\mathrm{M}_{\mathrm{G}}\sim 4$ mag, $\mathrm{G}_{\mathrm{BP}}-\mathrm{G}_{\mathrm{RP}} \sim 1 - 1.5$ mag) characteristic for young stellar populations, in many cases, even though this part of the CMD is often sparsely populated. Regarding the lower main sequence region, the empirical isochrones pass through the most densely populated part of the stellar distribution, which we assume to be the most reliable indicator of the true distribution of the cluster stars, that they would assume without the scatter induced by observational errors. Another characteristic captured by the empirical isochrones is the arc toward the upper right of the CMD of evolved clusters. This feature is reminiscent of a main-sequence turn-off, albeit not as intricate as ones available from model isochrones due to the lack of observable sources at this evolutionary stage. Likewise, the empirical indicator of a main sequence turn-off is only present when our cluster selection includes observed members in the red giant region. This is why some evolved clusters with similar literature ages do not exhibit a turn-off in the empirical isochrones while those in adjacent panels of Fig.~\ref{fig:04-Summary_Matrix_2} do. This circumstance should also be considered when looking at the number of main sequence turn-offs visible in the summary figure.

Overall, Fig.~\ref{fig:04-Summary_plot} displays a generally good agreement between the distribution of the empirical isochrones and the cluster ages found in literature, as indicated by the mostly smooth color gradient visible between the isochrones depicting clusters that were determined to be young and those portraying ones that have been estimated to be older. Looking at the evolved clusters for which we do recover a main sequence turn-off indicator, we also report a good agreement between the model-determined ages and the empirically determined ones based on the evolution of the turn-off point positions, although a few disagreements are also apparent. For instance, the Stock~2 cluster has a lower empirical isochrone turn-off, which indicates a more evolved state than its literature age suggests. On the opposite end of the age range, it can be observed that the empirical isochrones corresponding to the youngest clusters in our sample successfully and unanimously capture the relative shift of the lower main sequence toward the upper right of the CMD.

Overall, the good agreement in the comparison between the literature cluster ages determined via model fitting and the purely observation-based empirical isochrones provides an important mutual verification for both approaches. Apart from possible systematic errors, model-dependent cluster ages from isochrone fitting do indeed often represent the stellar evolution of clusters very well. On the other hand, our archive of empirical isochrones proves to cover an extensive range of evolutionary phases of clusters. Both facts are apparent from the correlation between the cluster age from literature, indicated by their color, and the locations of the corresponding empirical isochrones in Fig.~\ref{fig:04-Summary_plot}. Regardless, this strong correlation breaks down for individual clusters. 

\subsection{Validation using Sco-Cen clusters}
\label{sec:Results-validation-ScoCen}

To assess the practical feasibility of using the empirical isochrone archive as an age-scaling ladder, we validate our results against the clusters of \citetalias{2022Ratzenbock_Sigma}. The dataset offers the best testing conditions of our source catalogs, as its populations all originate from the same star-forming region, and their ages have been determined homogeneously. Additionally, of the 16 considered groups, only beta~Sco has a reported mean extinction greater than zero ($A_V = 0.5$ mag). Thus, the validation is neither biased against different age estimation methods nor against differing general parameters between the various groups. Moreover, with estimated ages between circa 7 and 20 Myr, the CMD distributions of the cluster isochrones are expected to show a significant shift of the lower main sequence toward the upper right, which should make them clearly distinguishable from one another. At the same time, their low absolute age spread means that they would provide a very fine grid for a possible age-scaling of young, unknown stellar populations.

\begin{figure}[t]
    \resizebox{\hsize}{!}{\includegraphics{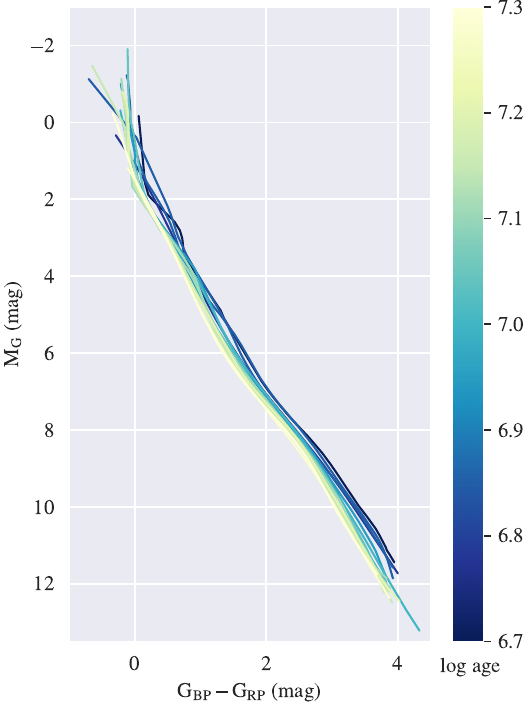}}
    \caption{Summary plot of empirical isochrones derived for the 16 clusters out of the 34 groups discovered in the Sco-Cen complex \citep{2022Ratzenbock_Sigma}, which are in the empirical isochrone archive. The lines are color-coded in accordance with the estimated ages published in \cite{2023Ratzenbock_Sco-Cen_ages}. The age sequence constitutes an almost perfect fit to the empirical isochrone shapes and positions within the CMD.}
    \label{fig:04-ScoCen-analysis}
\end{figure}

Figure~\ref{fig:04-ScoCen-analysis} displays the calculated empirical isochrones for the Sco-Cen groups, with the different colors again indicating their respective estimated ages \citep{2023Ratzenbock_Sco-Cen_ages}. As for Fig.~\ref{fig:04-Summary_plot}, a strong correlation between the locations, meaning the evolutionary phases, of the empirical isochrones and the associated ages of the populations is visible for the entire sample. By building up the plot sequentially, we found that in the region between $\sim 4$ and $11$ G magnitudes, where the uncertainty bounds of the empirical isochrones are the narrowest, there is even a perfect agreement between the isochrone position and age sequence. The trend also holds true in the upper main sequence ($< 2$ mag) for most of the groups, except for the sigma~Sco and the Antares clusters. A visual inspection of the CMDs of the two populations showed that their distributions exhibit a higher apparent scatter in their upper main sequences than the other Sco-Cen groups. As a result, the algorithm cannot derive a physically meaningful upper main sequence region for the two clusters.

From the favorable correlation, we conclude that the Sco-Cen empirical isochrones may indeed be used as an age-scaling grid or ladder for determining relative ages of other, unknown populations with similar ages. Given the excellent verification between literature and empirical isochrone positions in Fig.~\ref{fig:04-Summary_plot}, the principle can be extended to the whole archive, though extra care needs to be taken when comparing between clusters with ages determined by different authors, as those estimates might be biased again by their respective isochrone fitting or model choices.

\subsection{Isochrone blindspot as traced by empirical isochrones} 
\label{sec:Results-blindspot-II}

\begin{figure*}[t]
        \centering
        \resizebox{\hsize}{!}{\includegraphics{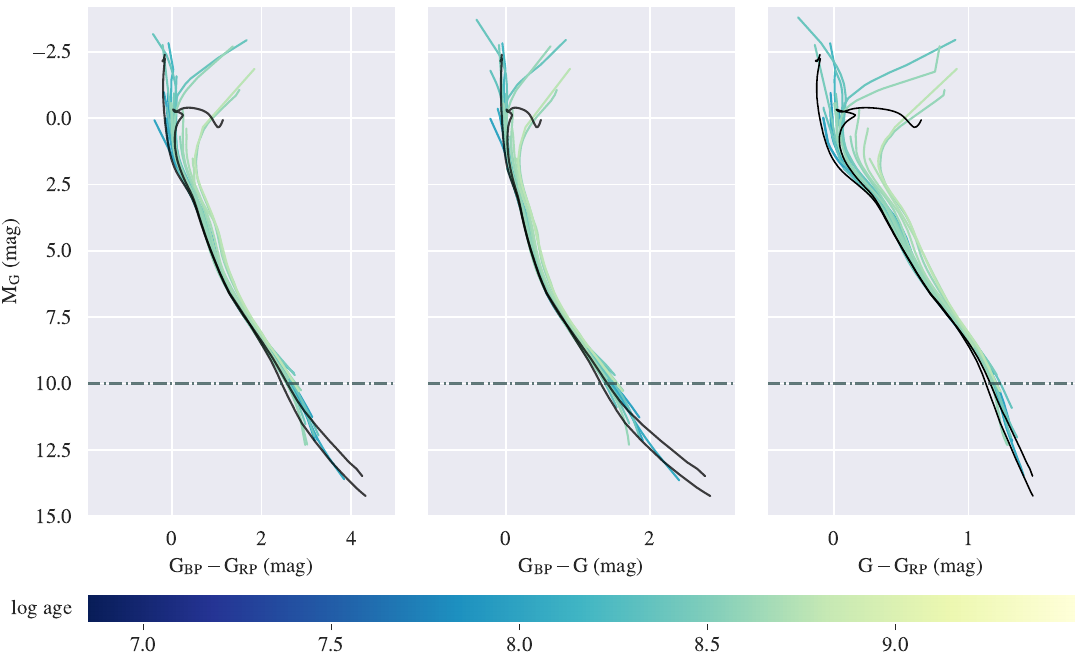}}
        \caption{Visualization of empirical isochrones compared with theoretical ones for the previously described isochrone blindspot ($\sim 100 - 500$ Myr). The \emph{black lines} correspond to model isochrones delineating the edges of the blindspot region, using PARSEC with default parameters. The \emph{colored lines} are all empirical isochrones belonging to clusters with estimated ages in this age range.}
        \label{fig:04-Blindspot_plot}
\end{figure*}

Another interesting comparison that can be made using the empirical isochrones of the archive concerns the isochrone blindspot between 100 and 500 Myr that was initially introduced in Fig.~\ref{fig:01-Blindspot}. The isochrones corresponding to the lower and upper age limits of said region are outlined in black in the panels of Fig \ref{fig:04-Blindspot_plot} for the different \emph{Gaia} color combinations, again using the PARSEC code with default parameters, no extinction and solar-like metallicity. Along with the theoretical isochrones, the figure shows all empirical isochrones of the archive clusters whose estimated ages coincide with the blindspot age range. 

It can be observed that the region where the theoretical isochrones overlap to form the isochrone blindspot is not as narrow when empirical isochrones are concerned. This broadening is most likely caused by the influence of the extinction and metallicity values for each cluster, thereby again highlighting the high dependence of the model-dependent isochrone fitting method on these two input parameters. Another reason for the observed broadening could be the uncertainty of the empirical isochrones around the main sequence turn-off locations of the clusters at the upper end of the blindspot age range. In cases where only a few select sources are observed in the red giant region, the empirical isochrones may divert from the main sequence earlier than theoretically predicted for the corresponding age. Finally, it is also possible that some of the oldest clusters in the figure may have their ages underestimated in literature, as their empirical main sequence turn-off points seem lower than theory would suggest. 

Nonetheless, we report a region between $\sim 5 - 9$ absolute G magnitudes, where the empirical isochrones emulate their theoretical counterparts in forming a very narrow section of overlapping curves -- in a reduced form, the blindspot remains visible, even using empirical isochrones.

Apart from the broadening of the upper main sequence, there is another critical feature mapped by the empirical isochrones in the plot, which concerns the lower main sequence (below the \emph{dash-dotted line}): Both the model isochrones and the empirical ones start to diverge from one another again at this point. However, the empirical curves systematically assume a different shape and curvature than the models, which is especially pronounced for the \emph{Gaia} $\mathrm{G}_{\mathrm{BP}}-\mathrm{G}_{\mathrm{RP}}$ and $\mathrm{G}_{\mathrm{BP}}-\mathrm{G}$ color indices. The color coding again shows an evolutionary sequence for the empirical lines. The discrepancy between models and observations is discussed further in Section~\ref{sec:Discussion-lower-MS}.

\subsection{Performance on different photometric systems}
\label{sec:Results-photometric-systems}

\begin{figure*}[t]
        \centering
        \resizebox{\hsize}{!}{\includegraphics{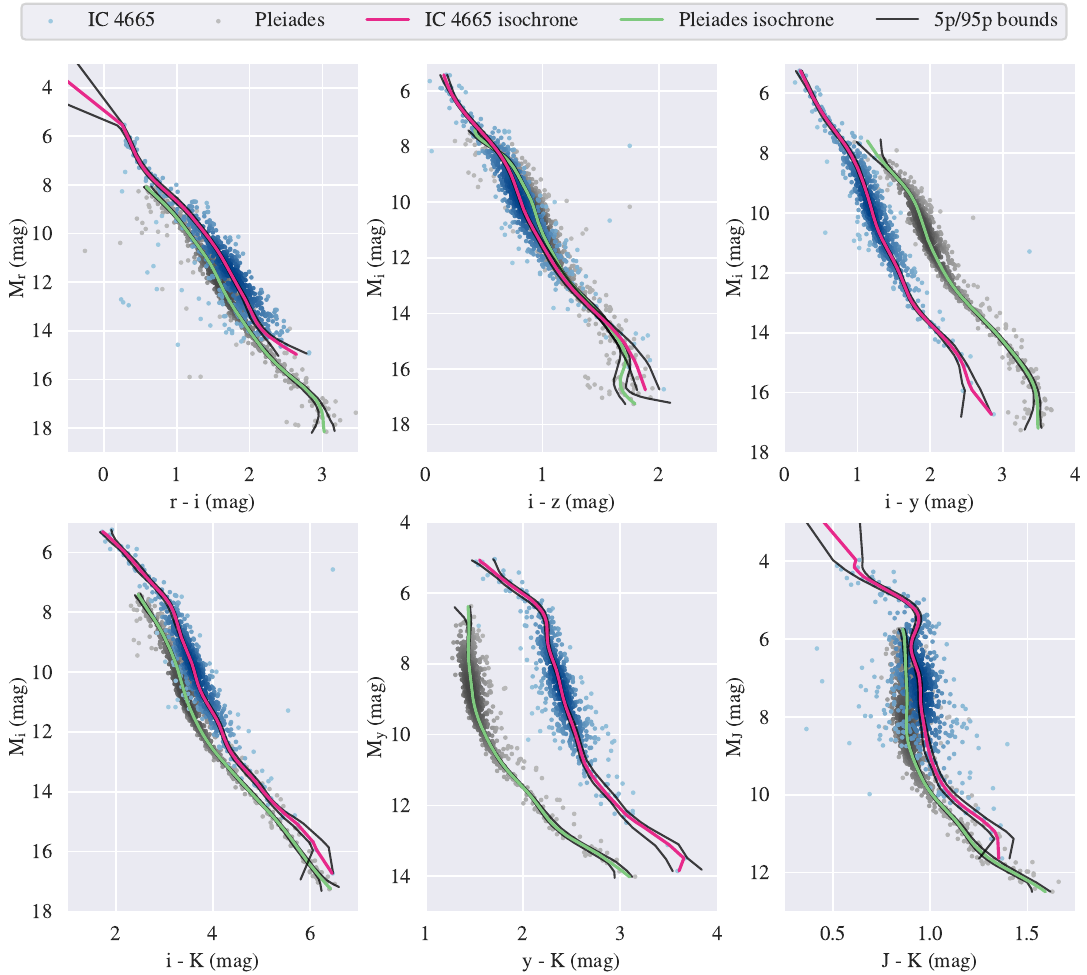}}
        \caption{Empirical isochrones calculated for the clusters IC~4665 and the Pleiades for six CMD variations using DANCe data.}
        \label{fig:04-Photometric-systems}
\end{figure*}

To gauge the aspired flexibility and adaptive properties of the PCA-SVR algorithm toward different CMD shapes and photometric systems other than \emph{Gaia}, we applied our method to the DANCe dataset of the two clusters IC~4665 and the Pleiades \citep{2018Pleiades_DANCe, 2019_IC4665_DANCe}. Figure~\ref{fig:04-Photometric-systems} displays the results for six different passband combinations. The isochrones calculated from these CMDs are also available online (see Appendix~\ref{Appendix:Tables} for details).

 As can be seen, the algorithm demonstrated an impressive performance on all tested CMDs, indicating great flexibility toward different kinds of input data. Even for the mostly vertical CMD and characteristic knee-shaped kink of the distribution of the $\mathrm{M}_{\mathrm{J}}$ vs. $\mathrm{J}-\mathrm{K}$ combination, reliable empirical isochrones that recover the distribution shape in detail were found. 

Compared to most of the results using \emph{Gaia} data, the edges of the uncertainty bounds and the isochrones are considerably less well-behaved for the new photometric systems, even describing abrupt turns in some cases. However, they penetrate the very faint regions close to the observation limit, which is why few sources are available for the fit. At the same time, the data scatter is considerably more prominent than in most \emph{Gaia} CMDs. In light of these complications, the performance of the algorithm is commendable. Even more, an adaptation of the hyperparameters to the new photometric systems was not necessary.

\section{Discussion}
\label{sec:Discussion}

In this section, we first address general quality aspects of using and interpreting our empirical isochrones and possible limitations of the algorithm before outlining concrete application scenarios of the archive as age-scaling ladder or age brackets. As most of the new information gained from empirical isochrones concerns the lower main sequence, we mainly focus on this region. Furthermore, the performance and information gain of empirical isochrones are compared with several popular models.

\subsection{Empirical isochrone quality}
\label{sec:Discussion-quality}

The quality of the extracted empirical isochrones is evaluated in a two-pronged manner: First, via comparison to the observed stellar distribution in the CMD and considering the shape and width of the uncertainty bounds and second, via the more general viewpoint regarding the sensitivity of the extracted curves on variations in the key observational parameters from which the CMD is created. 

Regarding the former, it can be seen from the cluster CMDs in Figs. \ref{fig:04-Summary_Matrix_1} and \ref{fig:04-Summary_Matrix_2} that the isochrone extraction with PCA and SVR yields accurate representations of the apparent shape of the cluster distributions. In general, there is little fluctuation between the various resampled isochrones, and the uncertainty regions indicated by \emph{black lines} are very narrow. The results show a successful hyperparameter tuning and the general good applicability of the machine learning procedure to the extraction task. The central parts of the cluster main sequences are exceptionally well-defined for almost all archive clusters. The only regions where the uncertainty bounds are larger are the edge regions.

Regarding the latter, we identified five key parameters that may impact the CMD of a stellar population itself, and by extension also the empirical isochrone: the measurement uncertainties in photometry and parallaxes, fraction of unresolved binaries, (differential) extinction, and field contamination. We performed a detailed exploration of the sensitivity of our proposed method to those parameters in Appendix \ref{appendix:Uncertainty-quantification}. To this end, we defined a parameter range holding three values for each key parameter and constructed a full factorial parameter grid consisting of 243 combinations. We then evaluated each grid point on three representative archive clusters by creating synthetic CMDs that incorporated the respective uncertainties in the key parameters at the given grid point and extracting empirical isochrones from them. We calculated an average deviation score between newly calculated and original isochrones using a nearest-neighbor distance metric. By heuristically determining thresholds for acceptable deviations between original and re-computed isochrones, we determined parameter ranges for reliable results based on our analysis (see Sect. \ref{appendix:param-ranges}). We briefly summarize our findings in the following paragraphs.

\subsubsection{Edge regions}

For a few clusters, the upper ends of the empirical isochrones start to diverge and flatten toward the left of the populated area instead of matching the brightest stars of the clusters. Prominent examples include the sigma~Sco, Antares, nu~Cen, UPK~422, and NGC~2516 clusters in Figs. \ref{fig:04-Summary_Matrix_1} and \ref{fig:04-Summary_Matrix_2}, respectively. This usually happens for a specific combination of CMD properties: Firstly, the upper main sequence only includes few and sparse sources, creating an exceptionally high gradient in population density between the upper and the lower mass end. Consequently, the bootstrapping process is even less likely to draw sources representing the upper main sequence than the general bias that the initial mass function imposes. Secondly, the brightest sources are associated with larger errors, usually in their parallax measurement, and therefore smaller weights in the regression. As a result, the machine learning algorithm tends to avoid those data points, but as it is designed to create a curve encompassing the entire dynamical range of the cluster, it creates the previously described artifacts. Often, these are already indicated by larger uncertainty bounds, but there are exceptions (see, e.g., UPK~422). When working with such isochrones, one should consider removing the empirical line's final entry to have a more reliable representation of the cluster.

\subsubsection{Measurement uncertainties in photometry and parallax}
\label{sec:Discussion-uncertainties}

Our sensitivity analysis showed, that neither photometric nor parallax measurement uncertainties have significant influence on the extracted isochrone. We purposefully used extreme assumptions for the errors in attributing the chosen uncertainties to every star in a cluster instead of drawing different values from a Gaussian distribution. Even so, we found that viewed individually, their respective impact on the isochronal shape is negligible. For the photometric uncertainty, we report an average deviation between old and re-computed isochrone of less than 1 \% for photometric uncertainties $\leq 0.03$ mag for each star. The same holds true for a fractional parallax error of 5 \% for each star. For a fractional parallax error of 10 \%, the average deviation between the original and a re-computed isochrone is around 2 \% using a $\mathrm{G}_{\mathrm{BP}}-\mathrm{G}_{\mathrm{RP}}$ passband combination in the CMD, ca. 1 \% using a $\mathrm{G}_{\mathrm{BP}}-\mathrm{G}$ CMD and below one percent for the $\mathrm{G}-\mathrm{G}_{\mathrm{RP}}$ CMD (Tab. \ref{tab:appendix-all-results}). Nevertheless, we again note that in case of fractional parallax errors larger than 10 \%, our approach to distance estimation via parallax inversion would not be applicable anymore.

When assuming the simultaneous presence of uncertainties in all parameters, we could extract reliable isochrones for more than 50 \% of the tested parameter combinations for parallax uncertainties up to 5 \% in the $\mathrm{G}_{\mathrm{BP}}-\mathrm{G}_{\mathrm{RP}}$ color combination. For the photometric uncertainties and other passband combinations, the fraction was typically lower than 50 \%.

\subsubsection{Unresolved binary sequence and field star contamination}
\label{sec:Discussion-binaries}

 Concerning field star contaminants and the unresolved binary sequence, our extraction method, in theory, should be robust against outliers as long as their number does not exceed that of (resolved) cluster members. For the archive clusters, this was only the case for the discarded RSG~7 cluster. We achieve this robustness by regulating the $\varepsilon$-tube of the SVR (Appendix~\ref{appendix:SVR-hyperparameters}). 

We verified this assumption in our sensitivity analysis in Appendix \ref{appendix:Uncertainty-quantification} by including both the unresolved binary fraction and field star contamination as parameters in our grid. When averaging over all three clusters and viewing the parameter influence in isolation, meaning with all other parameter influences set to zero (Tab. \ref{tab:appendix-all-results}), we verified that 30 \% binaries pose no problem to the reliability of the empirical isochrones in any of the passband combinations. On the other hand, the presence of 50 \% unresolved binary contamination exceed the reliability threshold, an expected outcome given the theoretical limitations of our method. As for the field fraction, contamination up to 50 \% has no impact on the reliability of the extracted isochrones, when acting in isolation.

We found that in the presence of other parameter uncertainties, disregarding photometric uncertainties, empirical isochrones for the $\mathrm{G}_{\mathrm{BP}}-\mathrm{G}_{\mathrm{RP}}$ passband combination could be produced for more than 50 \% of the grid points up to an unresolved binary fraction of 35 to 40 \% for the three test clusters. For the other passband combinations the thresholds are not reached for all clusters, thus not permitting a general statement. Interestingly, we find that artificially adding field contamination to the data can statistically have beneficial effects on intermediate age and old clusters, as they mostly add to the same locus as the cluster (lower) main sequence stars. Due to its dependence on the cluster age, the influence of field contamination on the isochrone reliability in the presence of other uncertainty factors cannot be simply generalized.

\subsubsection{Extinction} 

Observational data can be affected by interstellar extinction, which would shift the position of an empirical isochrone toward lower magnitudes and redder colors. Such mispositioning could potentially cause problems concerning the application of empirical isochrones as age-scaling ladders. However, extinction can be determined by using several sets of extra measurements, such as stellar spectra, inference methods such as \emph{Gaia} Apsis \citep{2023Gaia_APSIS}, or as a model parameter, for example via isochrone fitting. As a result, similarly to cluster ages, one can find a vast array of different extinction values for the archive clusters when consulting various works of literature \citep{2019Bossini, Cantat-Gaudin2020a,2021Dias}. Adjusting the empirical isochrones for extinction would therefore contradict the central focus of the isochrone archive of remaining model-independent and free of any priors, as we would have to place preference on one of the different approaches for extinction determination. 

\begin{figure}[ht]
        \centering
        \resizebox{\hsize}{!}{\includegraphics{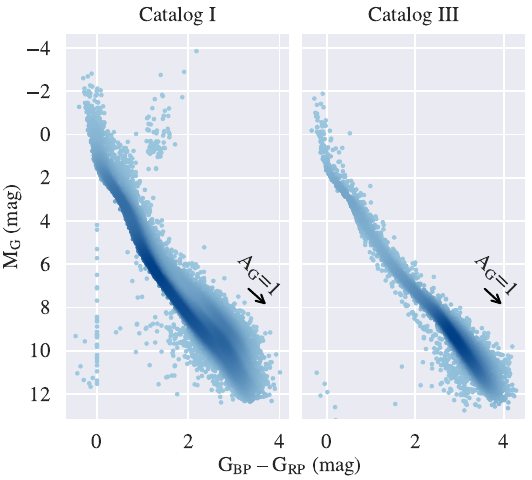}}
        \caption{Scope and direction of the extinction vector, visualized on \emph{Gaia} DR2 data of the \citetalias{Cantat-Gaudin2020a} and \citetalias{2022Ratzenbock_Sigma} selections using coefficients published by \cite{2019Wang}. For both cases, the extinction vector is almost parallel to the general orientation of the main sequence.}
        \label{fig:05-Extinction}
\end{figure}

Still, the effects of the extinction parameter on the isochrone quality in our specific use case need to be evaluated. As a first step, we calculated the extinction vector using the coefficients determined by \cite{2019Wang}. Since they only provide values for the second \emph{Gaia} data release, we use this data as well, as it is enough for a qualitative picture of the situation. Fig.~\ref{fig:05-Extinction} shows a CMD of all sources included in our selections of \citetalias{Cantat-Gaudin2020a} and \citetalias{2022Ratzenbock_Sigma}, respectively.\footnote{As in Fig.~\ref{fig:01-Blindspot-Pleiades}, the vertical line of data points in the CMD of \citetalias{Cantat-Gaudin2020a} only appears in their DR2 selection and vanishes when using DR3 data.} The extinction vector in the \emph{Gaia} DR2 G band is displayed alongside the data. It aligns well with the orientation of the bulk of the main sequence locus of stars in both test cases, meaning that any extinction present in the individual clusters would only result in a slight downward shift along the main sequence. We further note that according to literature \citep{2019Bossini, Cantat-Gaudin2020a, 2023Ratzenbock_Sco-Cen_ages}, the highest estimated extinction of a cluster in the archive corresponds to $A_{\mathrm{G}} = 0.6312$ mag ($A_V = 0.8$ mag) for UBC~17a.\footnote{\cite{2021Dias} report extinction values $A_V > 0.8$ mag for NGC~1662 ($A_V = 1.214$ mag), Alessi~20 ($A_V = 1.002$ mag) and Collinder~350 ($A_V = 0.981$ mag). However, these values are between a factor of 1.67 to 2.76 higher than those reported by both \cite{Cantat-Gaudin2020a,2019Bossini}.} Any shift would therefore only have a maximum length of around two-thirds of the indicated extinction vector.

We validated our assumption that the extinction values for the clusters in the archive are negligible in our sensitivity analysis in Appendix \ref{appendix:Uncertainty-quantification}. We created synthetic CMDs for three representative archive clusters, assuming a flat extinction level of either $A_{\mathrm{G}} = 0.25$ mag or $A_{\mathrm{G}} = 0.75$ mag, and calculated the deviation between the original and the newly computed isochrone and compared the results to empirically derived threshold values for each CMD passband combination and cluster, respectively. Viewed in isolation, we also found that for all but the $\mathrm{G}-\mathrm{G}_{\mathrm{RP}}$ CMD, an extinction value of $A_{\mathrm{G}}  = 0.25$ mag causes deviations well below the threshold value; however, this does not hold true for the larger extinction value of 0.75 mag (Tab. \ref{tab:appendix-all-results}). 

Given the presence of uncertainties in the other examined parameters, the median isochronal deviation lies within the accepted range for $A_{\mathrm{G}} \lesssim 0.3$ mag in the $\mathrm{G}_{\mathrm{BP}}-\mathrm{G}_{\mathrm{RP}}$ CMD. For the other tow combinations, only values well below 0.1 mag extinction should be considered in the presence of further significant parametric uncertainties. 

We also approximated the effects of differential extinction on the isochrone extraction method (see Appendix \ref{appendix:Differential}) by adding Gaussian noise to the flat extinction values and fixing the other investigated parameters to plausible values\footnote{Photometric uncertainty of 0.01 mag in all source, fractional parallax uncertainty of 1\% in all sources, 30 \% unresolved binary fraction and 25 \% field contamination fraction.}. Given these assumptions, the reliability threshold for a flat extinction assumption lies around 0.35 - 0.6 mag for the $\mathrm{G}_{\mathrm{BP}}-\mathrm{G}_{\mathrm{RP}}$ CMD and around 0.2 - 0.3 mag for the $\mathrm{G}_{\mathrm{BP}}-\mathrm{G}$ CMD. We found that for extinction levels $< 0.5$ mag and standard deviations $\sigma_{A_{\mathrm{G}}} \leq 1$ mag, the same thresholds apply as for the flat extinction level. 

We conclude, that for our archive clusters, we should be able to produce reliable isochrones in the $\mathrm{G}_{\mathrm{BP}}-\mathrm{G}_{\mathrm{RP}}$ passband combinations for almost all cases, even taking their estimated extinction into account. The third quantile and maximum of our reference extinction values are  $A_{\mathrm{G}} = 0.377$ mag and $A_{\mathrm{G}} = 0.6312$ mag, respectively. The former lies within the reliability threshold range defined in Appendix \ref{appendix:Differential}, while the latter is slightly above the upper limit. However, the actual uncertainties are likely to be below the assumptions of the simulations, which are overestimating the errors especially for the photometric and parallax uncertainties compared to a realistic distribution of uncertainties. As the sensitivity to extinction seems to scale with cluster age and/or distance, we searched the archive clusters for candidates with similar properties as the test case NGC~752, meaning $d > 400$ pc and an age of $> 1$ Gyr and checked whether their estimated extinction exceeds the reliability threshold for NGC~752 of 0.3 mag. We then did the same search for clusters similar to Blanco~1, that is $d > 200 pc$ and age $> 100$ Myr, with an extinction larger than the threshold value of 0.5 mag. None of the clusters in the archive corresponding to said properties have reported extinction values exceeding the respective threshold. The same applies to the $\mathrm{G}_{\mathrm{BP}}-\mathrm{G}$ CMD thresholds. For the $\mathrm{G}-\mathrm{G}_{\mathrm{RP}}$ passband combinations, we can only make the definitive statement that we can produce reliable empirical isochrones for flat extinction levels up to $A_{\mathrm{G}} = 0.25$ mag, disregarding the presence of uncertainties in other parameters.

It should be noted that users may subtract the extinction contribution from the empirical isochrones themselves in the case of uniform extinction at any time. In future works, we might also provide empirical isochrones considering the extinction parameters inferred by \emph{Gaia} Apsis.

\subsection{Lower main sequence analysis: Case studies}
\label{sec:Discussion-Case studies}

The lower main sequence of clusters is an important region for various reasons. First and foremost, low-mass stars are known to be the most numerous stars in the galaxy. Their properties have rendered them high-interest objects in the search for exoplanets \citep[see, e.g.,][]{2018Reiners}, which in turn makes estimations of their respective ages and masses paramount for furthering our understanding, especially regarding planetary formation and habitability research. Another vital aspect of why we should care for lower main sequence stars is that it is often an essential discriminating feature of younger populations ($\gtrsim 50 - 200$ Myr), as they often have no observable turn-off yet, and fall into the blindspot region. As Fig.~\ref{fig:04-Blindspot_plot} showed, the lower main sequence of such clusters can be empirically resolved and aid the correct age ordering of such populations.

\subsubsection{Theoretical vs. empirical isochrones down to the brown dwarf limit for IC~4665}
\label{sec:Discussion-lower-MS}

Modeling the dense and cool objects making up the faint end of (nearby) stellar population CMDs with stellar evolution codes involves complex physics. For instance, correct representations of nonideal effects within the stellar equations of state, as well as for nuclear reaction processes, and accurate stellar atmosphere models that account for the effects of molecular opacity \citep{1997Chabrier, 1998Baraffe} are needed. A particular obstacle to the successful modeling of the evolution of low-mass objects is posed by atmospheric convection or other stellar mixing processes such as overshooting \citep{2013Weiss}. Convection, at least in the mixing length theory formalism, reportedly still has some limitations, for instance, in mapping the thermal properties of convectively unstable atmospheres \citep{BHAC15}. Despite the overall reported good agreement between models and observations, especially in the optical bands, there is still often some visible disagreement regarding the lower main sequence of stars that appears over different models found in literature \citep[e.g.,][]{2014Chen, 2015Herczeg}.

\begin{figure}[t]
    \resizebox{\hsize}{!}{\includegraphics{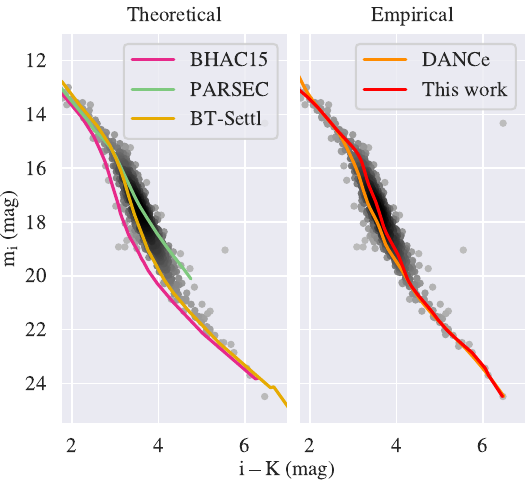}}
    \caption{Comparison between theoretical and empirical isochrone performances in the lower main sequence of the cluster IC~4665. \emph{Left:} Three theoretical isochrones corresponding to an age of 30 Myr and using solar metallicity, corrected regarding the estimated cluster extinction \citep{2019_IC4665_DANCe}. \emph{Right:} Empirical isochrones determined in this work (\emph{red line}) and by \cite{2019_IC4665_DANCe} (\emph{orange line}).}
    \label{fig:05-IC4665-vs-theory}
\end{figure}

The disagreement between theory and observations for the faint end of the main sequence has already been shown in Fig.~\ref{fig:04-Blindspot_plot} for the archive clusters. Now instead, we want to demonstrate the sensitivity of stellar evolution models to different input physics and how they compare to reality when applied specifically to the very low-mass end of a population. To this end, we display a $\mathrm{m}_{\mathrm{i}}$ vs. $\mathrm{i}-\mathrm{K}$ CMD of DANCe data of the IC~4665 cluster in Fig.~\ref{fig:05-IC4665-vs-theory}. We note that this plot displays apparent magnitudes instead of the usual absolute magnitudes on its $y-$axis. The left panel additionally displays model isochrones from BHAC15 \citep{BHAC15}, BT-Settl \citep{2014Allard}, and PARSEC\footnote{For this comparison, we chose the OBC bolometric correction instead of the default choice of YBC+Vega, as OBC matches the observations more closely.} \citep{2012PARSEC}, corresponding to the age ($30$ Myr) and extinction ($A_V = 0.72$ mag, transformed into the used bands using tables from \citealp{2019Wang}) values used in \cite{2019_IC4665_DANCe}. It is immediately obvious that neither of the three theoretical isochrones presents an accurate match for the data, but even more astonishing is the fact that they are in such stark disagreement with one another. The PARSEC isochrone does not penetrate the very faint magnitudes and has the least resemblance to the observations. The BHAC15 isochrone does not match the population very well, either. The best fit is achieved with the BT-Settl isochrone, but it still misses the sources at the faint end. Our findings support the age variations between different models already reported in the introduction in Sect.~\ref{sec:Ages-Ambiguity}. But even more, we emphasize that these isochrones are not only supposed to represent the same age but also identical metallicity and extinction values.

 In the right panel of the figure, the empirical isochrone derived by \cite{2019_IC4665_DANCe}, and the one calculated in this work are displayed. Both are much better fits than their theoretical counterparts for IC~4665. The differences between our and the DANCe isochrones are a consequence of the different extraction methods and the fact that the DANCe isochrone traces the lower left edge of the distribution, whereas we decided to derive the curve from the densest part, but the differences are less pronounced than for the model isochrones. In the areas with less scatter in the data, both empirical curves are in excellent agreement and accurately trace the sequence down to the faintest magnitudes.

The case study on IC~4665 shows that our observational data are good enough to map the observed distribution in detail, meaning that empirical isochrones could be important for constraining stellar evolution models. We anticipate that our results will provide reliable calibration data, specifically in the lower mass regime, where the physics gets increasingly complex, albeit they could also serve for calibration purposes along the rest of the main sequence.

\subsubsection{Age-bracketing of the Meingast~1 stream}
\label{sec:Discussion-MG1}

\begin{figure*}[t]
        \centering
        \resizebox{\hsize}{!}{\includegraphics{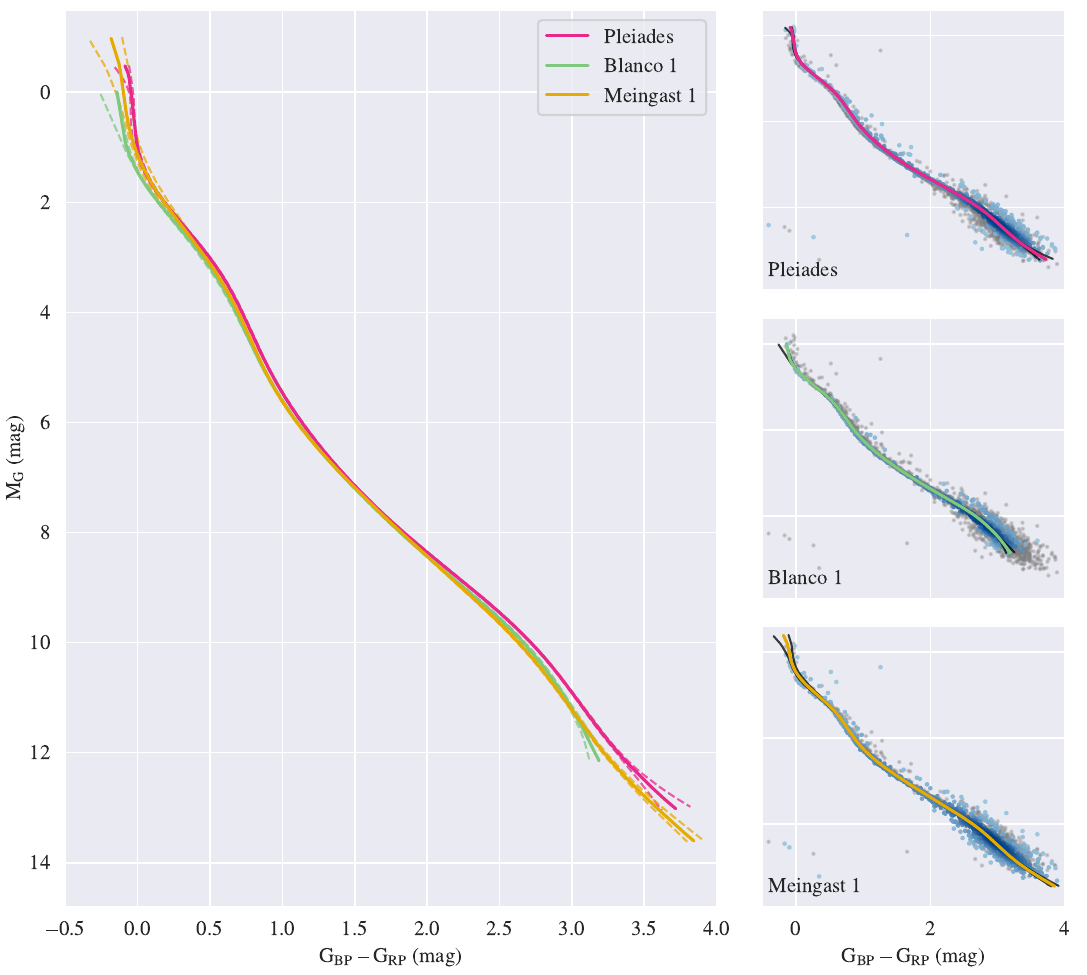}}
        \caption{Age bracketing of the Meingast~1 stream with empirical isochrones, whose corresponding cluster CMDs are shown in the right column of the figure. The gray points indicate the combined data of all three clusters, and the population connected to the plotted isochrone is overlaid in blue. By comparing the empirical isochrone of the stellar stream (\emph{yellow line}) with similar archive isochrones, we bracketed the stream and, in extension, its relative age, by the Pleiades (\emph{magenta line}) and by the Blanco~1 cluster (\emph{green line}).}
        \label{fig:05-MG1_brackets}
\end{figure*}

\begin{figure}[t]
    \resizebox{\hsize}{!}{\includegraphics{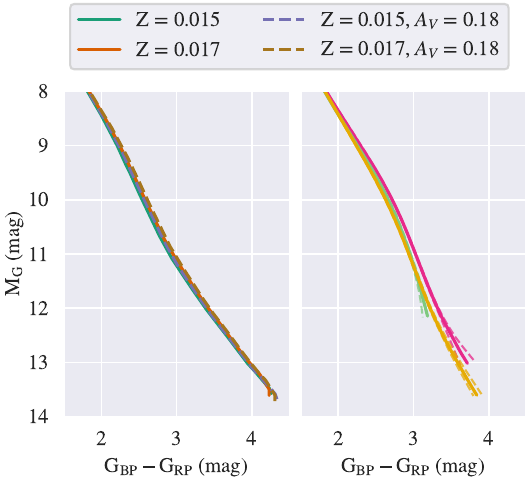}}
    \caption{Comparison of the difference between the empirical isochrones in the lower main sequence region with theoretical isochrones across a realistic extinction and metallicity range.\emph{Left:} PARSEC model isochrones with different possible extinction and metallicity values found for the Pleiades and the stream in literature. \emph{Right:} Zoom-in of the lower main sequence region of the age bracket of the Meingast~1 stream shown in Fig.~\ref{fig:05-MG1_brackets}.}
    \label{fig:05-MG1-Z-Av-check}
\end{figure}

We illustrate the idea behind the relative age determination using empirical isochrones on a type of stellar conglomerate discovered fairly recently -- the stellar stream Meingast~1 \citep{2019Meingast_ESSII}. As a relatively unfamiliar object with a well-defined CMD member selection and an estimated age right inside the blindspot region, it presents a particularly interesting case study for assessing the usefulness of the empirical isochrone archive. 
 
 The first discovered example of its class, the stream is a prime example of how easily isochronal ages can be distorted by nonmember stars at crucial points such as the turn-off region: Upon its discovery, the stream was estimated to be around 1 Gyr \citep{2019Meingast_ESSII}, due to the presence of a single evolved star right at the top of the member selection in the CMD. Subsequent works and age determinations using gyrochronology \citep{2019Curtis_Meingast1}, a lithium-rotation connection in low-mass members \citep{2020Arancibia-Silva}, isochrone fitting with an updated member selection \citep{2020ESSIV_Meingast1, 2020Roeser}, lithium abundance measurements \citep{2020Hawkins}, and empirical lithium equivalent widths \citep{2023Jeffries} have revised the original estimate and settled on an age around $110 - 120$ Myr. 
 
 However, apart from the discovery paper, all literature works have compared the stream to the Pleiades cluster, and almost all estimates even rest almost solely on this fact. For instance, its gyrochronological age was determined based on its similarity to one of only three benchmark clusters referenced as 120 Myr, 670 Myr, and 1 Gyr, respectively. Similarly, the age estimate from lithium abundance measurements for both the Pleiades and the stream has only very broad constraints of $\sim 120-200$ Myr and stems from comparing the stream data with measurements from the Pleiades, the Hyades, and the galactic disk. While the results indeed show that an apparent similarity exists between the two populations that can be traced by different indicators, a comparison between so few objects, though necessitated by lack of further observations, does not permit the determination of precise (relative) age limits. 

Once again, one has to fall back onto model-dependent approaches to get more precise age estimates. But for the Meingast~1 stream, even this strategy is hampered as it, along with the Pleiades, lies squarely inside the isochrone blindspot (Sect.~\ref{sec:Ages-blindspot}). This is also reflected in literature, where it is reported that PARSEC isochrones for the stream and the Pleiades show no significant differences between the 80 and 130 Myr \citep{2019Curtis_Meingast1}, or 100 and 150 Myr \citep{2020Roeser}. 

We, on the other hand, now have ample empirical information in the blindspot region, specifically in the lower main sequence, thanks to the empirical isochrone archive. Hence, we could select the ones best resembling the stream and determine relative ``age brackets'' for the stream, which are shown in Fig.~\ref{fig:05-MG1_brackets}: The upper bracket is provided by the Pleiades, whereas the Blanco~1 cluster forms the lower bracket. For our subsequent analysis, we focus exclusively on the regions where the uncertainty regions (\emph{dashed lines}) and the isochrones themselves are indistinguishable to minimize uncertainties brought on by the quality limitations of the empirical isochrones (Sect.~\ref{sec:Discussion-quality}). 

Similarly to the other literature works, we find a close resemblance between the Pleiades and the stream over almost the entire blindspot region. But we also see a clear shift of the empirical lower main sequences of the two populations starting around $\mathrm{M}_{\mathrm{G}} \approx 9$ mag and continuing until the broadening of the uncertainty bounds. Judging from only the lower main sequence, the Pleiades would therefore be slightly younger than Meingast~1. However, this is partly contested by literature (Table \ref{tab:05-Literature-ages}) and also not supported by the upper main sequence of the empirical isochrones. 

However, we find that along the lower main sequence the Meingast~1 stream isochrone matches the one of the Blanco~1 cluster almost exactly. The stream and Blanco~1 are both located below the galactic plane in the direction of the galactic south pole, indicating that their extinction should be negligible. In contrast, the Pleiades are often attributed a low but not zero extinction, although its value, when derived via isochrone fitting, is affected by the same uncertainties as the age estimates due to the strong parameter degeneracy. 

Another factor to consider in this difference between the isochrones is metallicity: Investigations of \cite{2020ESSIV_Meingast1} indicate a slight difference between the stream's and the Pleiades' metallicity values of $\Delta Z \simeq 0.002$. To determine whether the shift between the empirical isochrones of the two populations could be a result of either of the two parameters, we calculate PARSEC isochrones using the highest estimated extinction for the Pleiades found in our literature sources \citep[][$A_V = 0.18$ mag]{2021Dias}, and stellar metallicity values ranging from solar to $Z = 0.018$. The results of this qualitative analysis are displayed in Fig.~\ref{fig:05-MG1-Z-Av-check}. As can be seen from the comparison between the left and the right panel, the theoretical isochrones do not disperse enough to account for the apparent differences in the isochrones. In other words, the shift in the lower main sequence between the Pleiades and the stream cannot be explained solely by the combined effects of extinction and metallicity.

Judging from the results of the empirical lower main sequence analysis, we propose that the age of the Meingast~1 stream is between that of Blanco~1 and the Pleiades, with slightly more resemblance to the former population, than to the latter. Alas, comparing literature ages of those two populations yields a disagreement on which one is actually older and which is younger (see Table \ref{tab:05-Literature-ages}). Again, this highlights how flimsy model age determination can be. 

\begin{table}[t!]
    \caption{Different age estimates for the clusters bracketing the Meingast~1 stream. Since there is a general disagreement on which of the clusters is the older one, providing a relative age estimate for the stream is the best possibility.}
\begin{tabular*}{\linewidth}{@{\extracolsep{\fill}}lcc}
    \hline \hline
    Reference                           & Blanco~1  & Pleiades  \\
    \hline
    \multicolumn{3}{c}{\emph{Isochronal log(age)}}              \\
    \hline
    \citet{2019Bossini}                 & 7.975     & 7.973     \\
    \citet{2020b_Cantat-Gaudin_ages}    & 8.02      & 7.89      \\
    \citet{2021Dias}                    & 8.012     & 8.116     \\
    \hline
    \multicolumn{3}{c}{\emph{Empirical lithium EW log(age)}}    \\ 
    \hline
    \citet{2023Jeffries}                & 7.86      & 7.97      \\
    \hline

    \end{tabular*}

    \label{tab:05-Literature-ages}
\end{table}

Despite the disagreements in literature on which of the bracketing clusters actually corresponds to the lower and upper age limit, respectively, we can still provide a quite narrow relative age range of the stream. Given its position relative to the Pleiades in the lower main sequence, it seems as if it were a few Myr younger, even after factoring in possible extinction. From this perspective, it also seems as if the Blanco~1 cluster were indeed older than the Pleiades. This trend does seem reversed in the upper main sequence, but we again highlight that the regions with diverging uncertainty bounds should be interpreted with much more care than those with narrow uncertainty regions.

\section{Summary and conclusions}
\label{sec:Conclusion}

The new era of high-precision astrometric and photometric data provided in large quantities by all-sky surveys such as \emph{Gaia} allows us to find new synergies between astronomy and data science. In doing so, we were able to add new, purely observation-based information to the complex topic of stellar age determination. 

We developed a fast, flexible algorithm based on machine learning and statistical analysis tools that produces empirical isochrones for a multitude of CMD shapes and evolutionary stages of clusters. Its only inputs are the photometric and astrometric data needed to create any arbitrary color-magnitude diagram of a cluster or stellar population. By applying our algorithm to different recently published cluster member catalogs, we created an archive of empirical isochrones that includes 83 clusters within 500 pc representing an age range between 7 Myr and 3 Gyr. We draw the following conclusions from our results and subsequent analyses:

\begin{enumerate}
    \item When observing the general, larger trends, isochronal age determination is in good agreement with the empirical isochrone positions (Fig. \ref{fig:04-Summary_plot}).
    \item Validation against the Sco-Cen data of \citetalias{2022Ratzenbock_Sigma} (Fig. \ref{fig:04-ScoCen-analysis}) shows that it is possible to create a fine grid for relative age determinations and that for homogeneously determined groups and ages there exists a one-to-one correlation between the estimated absolute ages and the distribution of the empirical isochrones.
    \item The algorithm works for different photometric systems and CMD configurations (Fig. \ref{fig:04-Photometric-systems}), even though it was developed using only \emph{Gaia} data.
    \item The empirical isochrones are overall detailed and accurate CMD representations. High data scatter and sudden source drops are the limiting factors in the isochrone quality (Sect. \ref{sec:Discussion-quality}).
    \item A detailed sensitivity analysis concluded that, viewed in isolation, the empirical isochrones are reliable against at least a photometric measurement uncertainties of 0.03 mag, a fractional parallax uncertainties of 10 \%, a binary contamination of 30 \%, an extinction level of 0.25 mag $A_{\mathrm{G}}$\footnote{Except for the $\mathrm{G}-\mathrm{G}_{\mathrm{RP}}$ passband combination.}, and a field contamination fraction of 50 \%. A more detailed analysis of the effects of extinction concluded, that under the assumption of 0.01 mag photometric uncertainty, 1 \% parallax uncertainty, 30 \% unresolved binary fraction and 25 \% field contamination, empirical isochrones remain reliable given a mean extinction of 0.3 - 0.6 mag, depending on the cluster age, in the $\mathrm{G}_{\mathrm{BP}}-\mathrm{G}_{\mathrm{RP}}$ CMD, depending on the inspected cluster. For the $\mathrm{G}_{\mathrm{BP}}-\mathrm{G}$ CMD, this value range shifts to 0.2 - 0.3 mag. When considering combined uncertainty effects, the reliability may vary depending on the specific uncertainty values (Sect. \ref{appendix:Uncertainty-quantification}).
    \item Empirical isochrones add considerable information toward the hard-to-model lower main sequence. This is especially important in the isochrone blindspot region.
    \item There is still ample room for improvement when comparing model isochrones with observations, specifically at low stellar masses (Sect.~\ref{sec:Discussion-Case studies}). We are aware that modeling the physics of low-mass stars is particularly challenging and anticipate that our empirical isochrones could serve for calibration purposes in this regard (Fig. \ref{fig:05-IC4665-vs-theory}).
    \item The archive of isochrones can be used as age-scaling ladder for determining relative ages via empirical isochrone bracketing. We showed this in a case study of the newly found Meingast~1 stream, which is closely bracketed by not only the Pleiades cluster as established in the literature but also shares a lot of resemblance with the Blanco~1 cluster (Fig. \ref{fig:05-MG1_brackets}).
\end{enumerate}

In the future, we aim to expand the empirical isochrone archive toward an even larger sample of nearby open clusters and look for loci of empirical lines that can be connected to similar global attributes of open clusters.

\begin{acknowledgements}
The authors thank the referee Javier Olivares for providing very useful comments that helped to improve the quality of this work.The authors want to thank the research group in Vienna and especially João Alves, Josefa Großschedl, Núria Miret-Roig and Sebastian Ratzenböck, for their valuable insights, input and discussion to this work. Funded by the European Union (ERC, ISM-FLOW, 101055318). Views and opinions expressed are however those of the author(s) only and do not necessarily reflect those of the European Union or the European Research Council Executive Agency. Neither the European Union nor the granting authority can be held responsible for them. This work has made use of data from the European Space Agency (ESA) mission \emph{Gaia} (\url{https://www.cosmos.esa.int/gaia}), processed by the \emph{Gaia} Data Processing and Analysis Consortium (DPAC, \url{https://www.cosmos.esa.int/web/gaia/dpac/consortium}). Funding for the DPAC has been provided by national institutions, in particular the institutions participating in the \emph{Gaia} Multilateral Agreement.
This research made use of Astropy\url{http://www.astropy.org}, a community-developed core Python package for Astronomy \citep{astropy:2013, astropy:2018}.
We also acknowledge the use of Python \url{https://www.python.org}, along with the packages that were used in the data analysis of this work, including NumPy \citep{numpy}, scikit-learn \citep{scikit-learn}, Matplotlib \citep{matplotlib}, and Plotly \citep{plotly}.
This research has made use of the VizieR catalog access tool, CDS, Strasbourg, France (\citealp{vizier}; DOI: \href{http://dx.doi.org/10.26093/cds/vizier}{10.26093/cds/vizier}).
This research has made use of TOPCAT, an interactive graphical viewer and editor for tabular data \citep{Taylor2005}.
\end{acknowledgements}
\bibliography{main}
\clearpage
\begin{appendix}

\section{Cluster CMDs}
\label{Appendix:CMDs}

\begin{figure*}[ht]
        \centering
        \resizebox{\hsize}{!}{\includegraphics{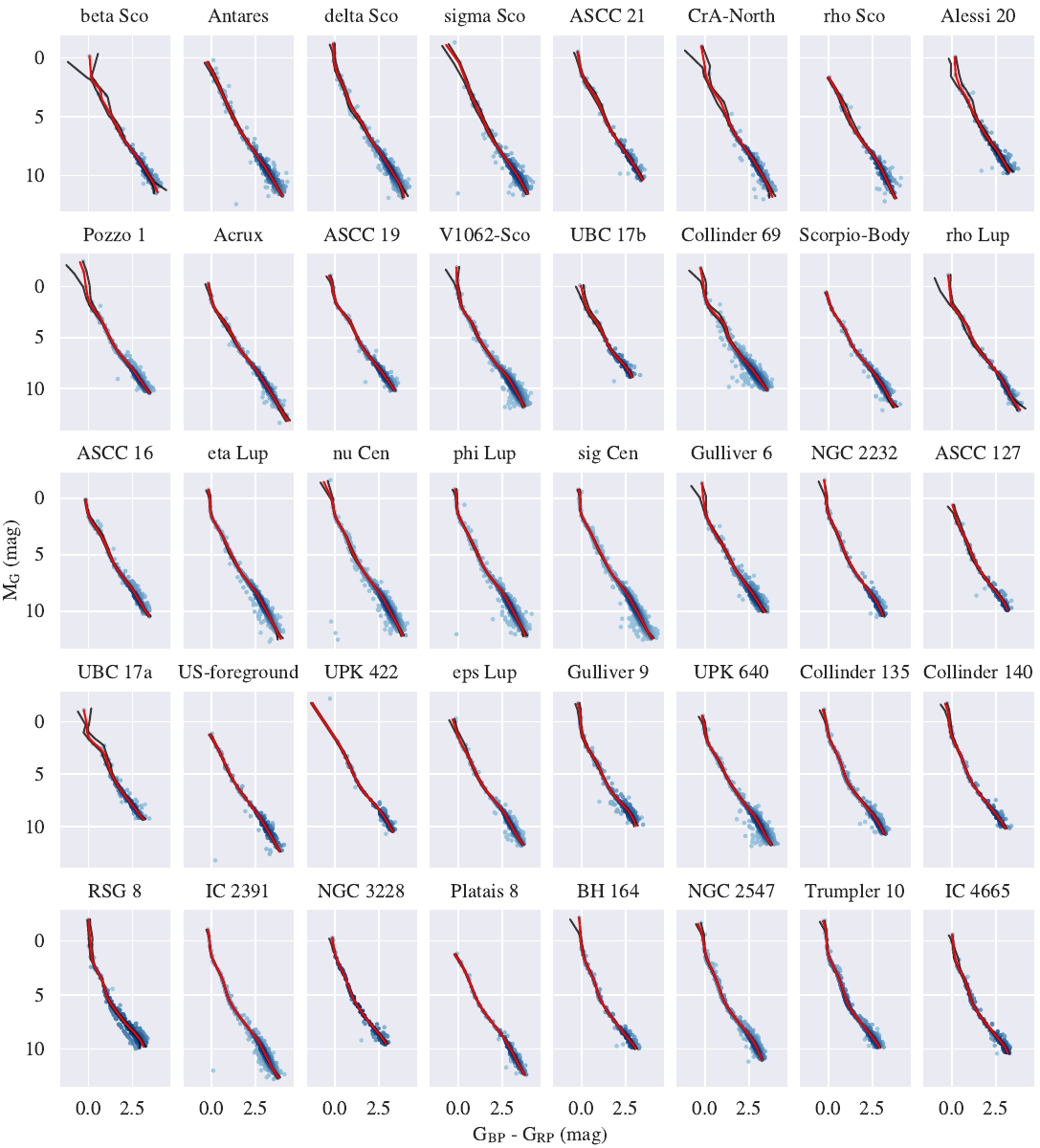}}
        \caption{CMDs of the individual archive clusters (1-40), along with their calculated empirical isochrones and uncertainty bounds. The clusters are ordered according to their literature ages in ascending fashion.}
        \label{fig:04-Summary_Matrix_1}
\end{figure*}

\begin{figure*}[ht]
        \centering
        \resizebox{\hsize}{!}{\includegraphics{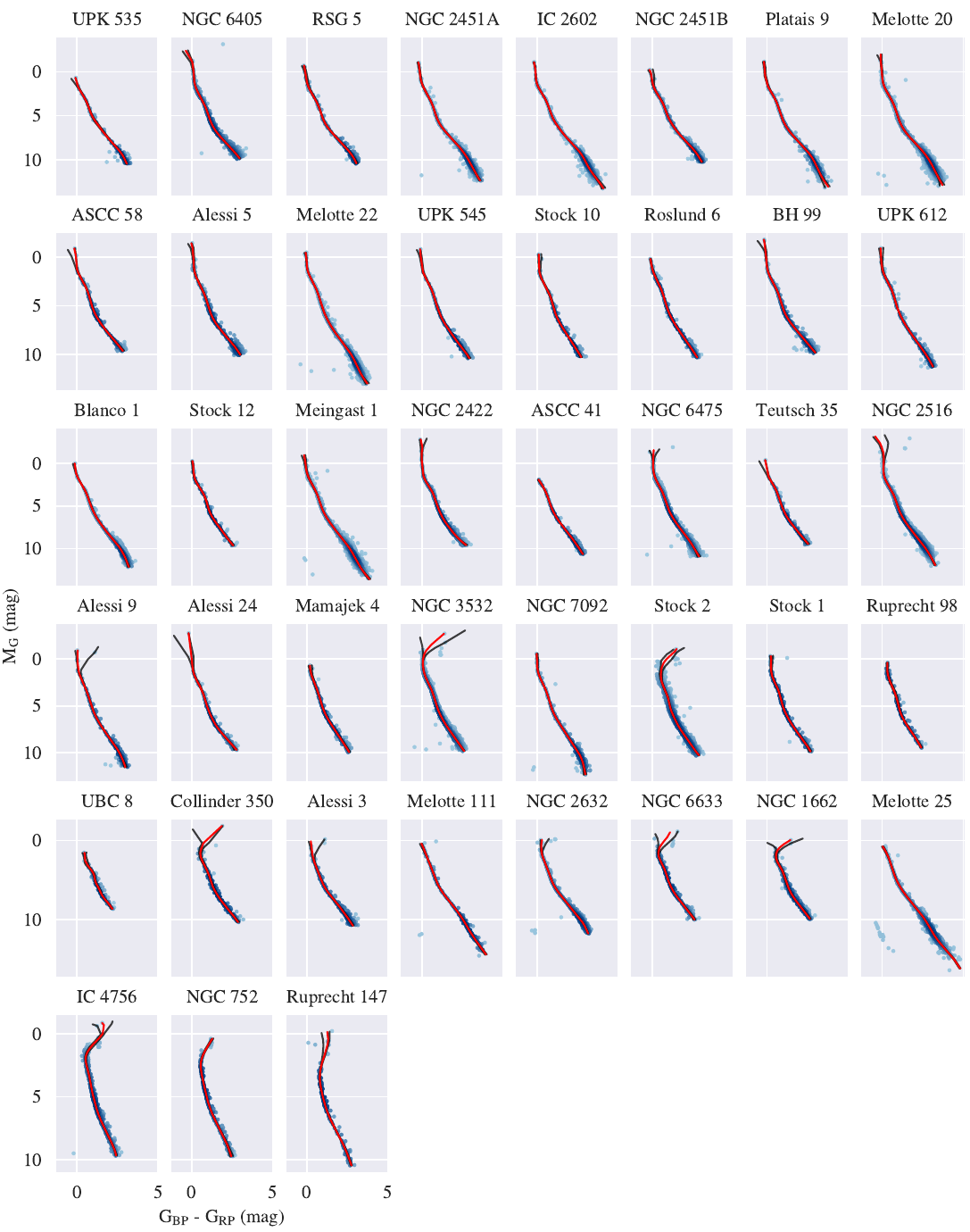}}
        \caption{Continuation of Fig.~\ref{fig:04-Summary_Matrix_1} for the remaining archive clusters (41-83).}
        \label{fig:04-Summary_Matrix_2}
\end{figure*}

To demonstrate the general quality of the empirical isochrones derived in this work, we showcase the results for all 83 clusters in our selection in the \emph{Gaia} $\mathrm{G}$ vs. $\mathrm{G}_{\mathrm{BP}}-\mathrm{G}_{\mathrm{RP}}$ CMDs in Figs. \ref{fig:04-Summary_Matrix_1} and \ref{fig:04-Summary_Matrix_2}. The clusters are arranged in ascending order based on their estimated ages (see Table~\ref{tab:02-Cluster-specs}). The empirical isochrone is shown in red, while the 5$^{\mathrm{th}}$ and 95$^{\mathrm{th}}$ percentile boundaries, calculated from 1000 bootstrapped curves, are represented by \emph{black lines}. It can be seen that while there is generally a good agreement between the literature ages and the evolution of the empirical isochrone shapes, there are also some obvious differences in the ordering, especially when inspecting rather evolved clusters with stars populating the giant region.

\newpage
\clearpage

\section{Empirical isochrone tables}
\label{Appendix:Tables}

In the following, we show an excerpt of one of the result tables for the empirical isochrones, as it may be found online. Table~\ref{tab:04-results} displays the first ten entries of the first archive cluster for the \emph{Gaia} DR3 $\mathrm{G}_{\mathrm{BP}}-\mathrm{G}_{\mathrm{RP}}$ passbands (the full table is available online). The same information is available for the $\mathrm{G}_{\mathrm{BP}}-\mathrm{G}_{\mathrm{RP}}$ and $\mathrm{G}_{\mathrm{BP}}-\mathrm{G}_{\mathrm{RP}}$ passband variations for all archive clusters. There are two distinct isochrones per passband combination available for the ten clusters appearing in both \citetalias{Cantat-Gaudin2020a} and \citetalias{meingast2020}, between we distinguish by assigning the empirical isochrones corresponding to the second catalog the suffix ``C2.'' Our adopted reference age estimates in this manuscript can be found in the last column of Table~\ref{tab:04-results}. Similar tables are available for the same passband combinations using \emph{Gaia} DR2 and for the DANCe CMDs shown in Fig.~\ref{fig:04-Photometric-systems}, respectively.

\begin{table*}
\caption{First ten table entries sorted by cluster name, showing only the columns concerning the $\mathrm{G}_{\mathrm{BP}}-\mathrm{G}_{\mathrm{RP}}$ combination of \emph{Gaia} DR3, as well as the reference age used in this work. The columns correspond to the lower (lb) and upper (ub) uncertainty bounds, as well as the empirical isochrone (iso) calculated for each cluster with $n_{\mathrm{boot}}=1000$.}
\begin{tabular*}{\linewidth}{@{\extracolsep{\fill}} c c c c c c c c c}
\hline\hline
Cluster   & BPRP\_lb\_x & BPRP\_lb\_y & BPRP\_iso\_x & BPRP\_iso\_y & BPRP\_ub\_x & BPRP\_ub\_y & \dots & ref\_age \\
\hline
ASCC\_127 & 0.0311      & 0.5926      & -0.0110            & 0.6027             & -0.3193     & 0.6772      & \dots   & 7.26     \\
ASCC\_127 & 0.0876      & 1.2667      & 0.0397             & 1.2782             & -0.0328     & 1.2957      & \dots   &   7.26     \\
ASCC\_127 & 0.1387      & 1.5687      & 0.0660             & 1.5863             & 0.0296      & 1.5951      & \dots   &   7.26     \\
ASCC\_127 & 0.2099      & 2.1380      & 0.1213             & 2.1594             & 0.0958      & 2.1655      & \dots   &   7.26     \\
ASCC\_127 & 0.2123      & 2.1846      & 0.1260             & 2.2055             & 0.1004      & 2.2116      & \dots   &   7.26     \\
ASCC\_127 & 0.2165      & 2.2862      & 0.1362             & 2.3056             & 0.1113      & 2.3116      & \dots   &   7.26     \\
ASCC\_127 & 0.2472      & 2.8702      & 0.2128             & 2.8785             & 0.1932      & 2.8832      & \dots   &   7.26     \\
ASCC\_127 & 0.2517      & 2.9404      & 0.2226             & 2.9475             & 0.2038      & 2.9520      & \dots   &   7.26     \\
ASCC\_127 & 0.2610      & 3.0639      & 0.2366             & 3.0698             & 0.2196      & 3.0739      & \dots   &   7.26     \\
ASCC\_127 & 0.2767      & 3.1764      & 0.2504             & 3.1828             & 0.2318      & 3.1873      & \dots   &   7.26     \\
\vdots    & \vdots      &\vdots       &\vdots              &\vdots              &\vdots &\vdots       &\dots   &\vdots      \\         
\hline
\end{tabular*}
\tablefoot{The full table will be made available at the CDS.}
\label{tab:04-results}
\end{table*}

\newpage
\clearpage

\section{Preprocessing the data with PCA}
\label{Appendix:PCA}

Although the concept of PCA was already developed more than a century ago, \citep[][]{1901PearsonPCA, 1933Hotelling}, its practical feasibility only started thriving with the emergence of computers. Since then, PCA has received recognition primarily as a dimensionality reduction method that unifies the ideas of increasing data interpretability and minimal information loss. Its working principle is based on creating new, uncorrelated variables, which successively maximize the variance from the original dataset, thereby retaining as much statistical information as possible. Consequently, PCA variables are always defined by the specific input dataset \citep{2016Jolliffe}. The basic principle of the method, as given for example in \cite{PCA_Bishop, 2016Jolliffe} can be applied to the problem at hand as follows:

We start with a set of $n$ observations, corresponding to the number of member stars $N_*$ observed for an arbitrary open cluster. For each of these stars, $p$ variables of interest are measured, which yields an $n \times p$ observation matrix $\mathbf{X}$ as our input data. In our case, only the color index and the absolute magnitude are needed for creating a CMD, meaning the formalism simplifies to $p=2$. We now seek linear combinations of the two columns of the observation matrix
\begin{align}
    \label{eq:03-Linear-combis} \sum_{i=1}^2 a_i \mathbf{x}_i & = \mathbf{X}\mathbf{a},
\end{align}
where $\mathbf{a} = \begin{pmatrix} a_1 & a_2 \end{pmatrix}^{\intercal}$ denotes a vector of constants that maximize the variance 
\begin{align}
    \label{eq:03-variance-linear-combis} \text{Var}(\mathbf{X}\mathbf{a}) &= \mathbf{a}^\intercal \mathbf{S} \mathbf{a}, \quad \text{with} \quad \mathbf{a}^\intercal \mathbf{a} = 1.
\end{align}
In the last line, $\mathbf{S}$ denotes the covariance of the observation matrix, and the equation is subjected to a unity restriction of the vector $\mathbf{a}$. Using this constraint is a common approach for achieving a well-defined analytical solution for PCA. The representation of the problem and its restriction in Eq. (\ref{eq:03-variance-linear-combis}) can now be rewritten in terms of a Lagrangian formulation $L(\mathbf{x}, \lambda) = f(\mathbf{x}) - \lambda g(\mathbf{x})$. Here, the function $f$ represents the function of the variance, $g$ denotes the imposed constraint, and $\lambda$ indicates a Lagrangian multiplier \citep[see][Appendix E for more details]{PCA_Bishop}. Thus, the original equation is transformed into its dual form and can be treated as a maximization problem
\begin{align}
    \label{eq:03-maximization} \max \left[ \mathbf{a}^\intercal\mathbf{S}\mathbf{a} - \lambda(\mathbf{a}^\intercal\mathbf{a} -1) \right].
\end{align}
Performing the derivation with respect to the vector $\mathbf{a}$ and setting the result to the zero vector yields
\begin{align}
    \label{eq:03-derivation} \mathbf{S}\mathbf{a} - \lambda\mathbf{a} = \mathbf{0} &\Leftrightarrow \mathbf{S}\mathbf{a} = \lambda \mathbf{a}.
\end{align}
From the new form of the equation, we see that $\mathbf{a}$ must be a unit-norm eigenvector with the corresponding eigenvalue given by $\lambda$. In PCA, one is generally interested in the largest eigenvalues, since they also correspond to the largest variance of the linear combinations defined by their respective eigenvectors. This can be shown easily using Eqs. (\ref{eq:03-variance-linear-combis}) and (\ref{eq:03-derivation})
\begin{align}
    \label{eq:03-eigenvalues=variances} \text{Var}(\mathbf{X}\mathbf{a}) &= \mathbf{a}^\intercal\mathbf{S}\mathbf{a} = \lambda\mathbf{a}^\intercal\mathbf{a} = \lambda.
\end{align}
Therefore, by setting $\mathbf{a}$ to the eigenvector corresponding to the largest eigenvalue, we define the first principal component. The covariance matrix $\mathbf{S}$ is always a real, symmetric matrix, which means that it has exactly $p$ real eigenvalues. In our case its two corresponding eigenvectors can be defined as an orthonormal vector set $\mathbf{a}_1^\intercal\mathbf{a}_2 = 0$. It can further be shown that the full set of eigenvectors of $\mathbf{S}$ are the solutions to obtaining the new, uncorrelated variables defined by the linear combinations seen in Eq. (\ref{eq:03-Linear-combis}), which successively maximize the variance \citep{2002Jolliffe}. Here, the uncorrelatedness is founded in the fact that the covariance of two such linear combinations $\mathbf{X}\mathbf{a}_1, \mathbf{X}\mathbf{a}_2$ given by the eigenvectors $\mathbf{a}_1$ and $\mathbf{a}_2$ are zero:
\begin{align}
    \label{eq:03-uncorrelatedness} \mathbf{a}_1^\intercal\mathbf{S}\mathbf{a}_2 & = \lambda\mathbf{a}_1^\intercal\mathbf{a}_2 = 0.
\end{align}
In other words, further principal components can be evaluated by iteratively choosing the next direction that maximizes the projected variance and is orthogonal to the ones already chosen as principal components \citep{PCA_Bishop}. The linear combinations of the observations matrix $\mathbf{X}$ and the eigenvectors $\mathbf{a}_k$ are what is commonly referred to as ``principal components.'' In its original form, PCA only describes linear transformations of data, thereby limiting its applicability. For this reason, various amendments to the method have been proposed to add a more sophisticated, nonlinear character to it \citep[see, e.g.,][]{PCA_Bishop}.

In the common use case of PCA for dimensionality reduction, one can further show that using a centered form of the observed variables $\mathbf{x}^*_j = x_{ij} - \Bar{x}_j$, PCA is equal to a Single Value Decomposition and use this fact for creating principal subspaces and retaining only the most important parameters \citep{2016Jolliffe}. However, in our case, we do not require an actual dimensionality reduction but instead use PCA to rotate the reference frame into the principal components spanned by the new variables defined in Eq. (\ref{eq:03-Linear-combis}). In doing so, we not only reduce the scatter in the data, but also preserve a functional dependency between the input variables for the support vector regression for clusters across all evolutionary stages and with all possible CMD morphologies (see Fig.~\ref{fig:03-PCA-components} in the main text).

\newpage
\clearpage

\section{Curve extraction with SVR}
\label{Appendix:SVR}

 SVR is used to extract empirical isochrones from the PCA-transformed observational data. It also represents the step in the algorithm where the most computational time is required; most of it can be attributed to the tuning of the hyperparameters. Furthermore, measurement errors can also be incorporated at this point in the workflow. In the following paragraphs, we first discuss the theory of the method before going into more detail about the hyperparameter influence on the shapes of the empirical isochrone, as well as the tuning process itself. Lastly, the calculation of the weight parameter, which we use to account for measurement errors during the regression, is discussed.

\subsection{The basic method}
\label{sec:Methods-SVR-basics}
 
Similar to other machine learning models, the basis for SVR is given by a regression problem regarding a function $y(\mathbf{x})$ of an $n$-dimensional observations vector $\mathbf{x}$, which can be written as
\begin{align*}
    y(\mathbf{x}) & = \mathbf{w}^\intercal \mathbf{\varphi}(\mathbf{x}) + b,
\end{align*}
with the variable $\varphi$ denoting a fixed feature space transformation of $\mathbf{x}$. To perform a regression, the weights $\mathbf{w}$ and the intercept $b$ are optimized by minimizing a regularized loss function
\begin{align}
    \label{eq:03-loss-function} L & = \frac{1}{2}\sum_{n=1}^N E(y_n,t_n) + \frac{\lambda}{2}\lVert \mathbf{w} \rVert ^2,
\end{align}
which describes the differences $(y_n - t_n)$, with $n=1,...N$, between the target values $t_n$ and the estimated values $y_n = y_n(\mathbf{x})$. There are many popular loss functions, such as least squares or Huber loss. However, SVR is founded on a function specifically designed to create sparse solutions. Its so-called $\varepsilon$-insensitive loss function takes the form
\begin{align}
    \label{eq:03-epsilon-loss-function} L_{\varepsilon} & = \begin{cases} 0 \quad \text{for} \quad |y_n- t_n| \leq \varepsilon\\
    |y_n - t_n| - \varepsilon \quad \text{otherwise,} \end{cases}
 \end{align}
meaning that it equals zero for all values that differ up to a threshold of $\varepsilon$ from their target values, the so-called $\varepsilon$-tube, whereas values exceeding this threshold are penalized. To describe the upper and lower margins of the loss function outside the $\varepsilon$-tube, ``Slack variables'' $\xi_n \geq 0$ and $\hat{\xi}_n \geq 0$, for all $n=1,...N$ are introduced:
\begin{align}
    \label{eq:03-upper-Slack} \xi_n & \geq t_n - y_n - \varepsilon \quad \text{"below" the $\varepsilon$-tube}\\
    \label{eq:03-lower-Slack} \hat{\xi}_n & \geq y_n - t_n - \varepsilon \quad \text{"above" the $\varepsilon$-tube}
\end{align}
Using the new variables, the primal problem of SVR can be written as follows:
\begin{align}
    \label{eq:03-Primal-problem} \min_{\mathbf{w},b,\xi,\hat{\xi}} & \left(C\sum_{n=1}^{N} \left(\xi_n+\hat{\xi}_n\right) + \frac{1}{2} \lVert \mathbf{w} \rVert ^2 \right).
\end{align}
Here, the parameter $C$ regulates the penalty assigned to values not included in the regression, although their loss does not equate to zero. The properties of the loss function of Eq. (\ref{eq:03-epsilon-loss-function}) are effectively contained within the Slack variables. To solve the problem under consideration of the restrictions imposed by Eqs. (\ref{eq:03-upper-Slack})-(\ref{eq:03-lower-Slack}) and $\xi_n, \hat{\xi}_n \geq 0$, it is customary to formulate the dual problem using Lagrange multipliers, similar to what was done for PCA in Sect. \ref{Appendix:PCA}, and subsequently optimize it. Following them leads to the formulation of an equation for predicted values
\begin{align}
\label{eq:03-SVR-final-form} y(\mathbf{x}) & = \sum_{n=1}^N \left(\alpha_n^{+} -\alpha_n^{-}\right)K(\mathbf{x},\mathbf{x}_n) +b.
\end{align}
The equation now includes the Lagrangian multipliers $\alpha_n^{+} \geq 0$ and $\alpha_n^{-} \geq 0$, which describe the two conditions of the previously used Slack variables, as well as a kernel function $K$. As a given data point can never be both above and below the regression curve, one of the $\alpha$ variables always equates to zero for the same index $n$. Support vectors are points that contribute to the predictions generated by Eq.~(\ref{eq:03-SVR-final-form}), meaning those for which either $\alpha_n^{+}$ or $\alpha_n^{-}$ is greater than zero. The coefficient $b$ can be either calculated using the Karush-Kuhn-Tucker conditions \citep[for details see, e.g.,][and references therein]{SVR_Bishop2006} or approximated numerically. As a kernel function, we choose the \texttt{scikit-learn} radial basis function kernel
\begin{align}
    \label{eq:03-rbf-kernel} K(\mathbf{x},\mathbf{x}') & = \exp{\left(-\gamma \lVert \mathbf{x}-\mathbf{x}' \rVert ^2\right)},
\end{align}
where $\gamma$ is conventionally tied to the variance $\sigma^2$ of the observation vector $\mathbf{x}$ via $\gamma =\frac{1}{2\sigma^2}$.

\subsection{Hyperparameter influence and tuning}
\label{appendix:SVR-hyperparameters}

Following the derivations of the previous section, one can observe that there are three variables that are not subjected to analytical optimization but need to be tuned separately according to the input data. The first of these, \texttt{epsilon}, stems from the definition of the loss function of Eq. (\ref{eq:03-loss-function}) and corresponds to the width of the tube around the regression curve, within which no penalty is assigned. The second hyperparameter is the penalty \texttt{C} found in Eq. (\ref{eq:03-Primal-problem}), which controls the amount of data that the regression curve may ignore. The third and final hyperparameter is given via the \texttt{gamma} variable appearing in Eq. (\ref{eq:03-rbf-kernel}) and is specific to the chosen kernel function. Its value determines the inverse radius of influence of each data point.

\begin{figure*}[ht]
        \centering
        \resizebox{\hsize}{!}{\includegraphics{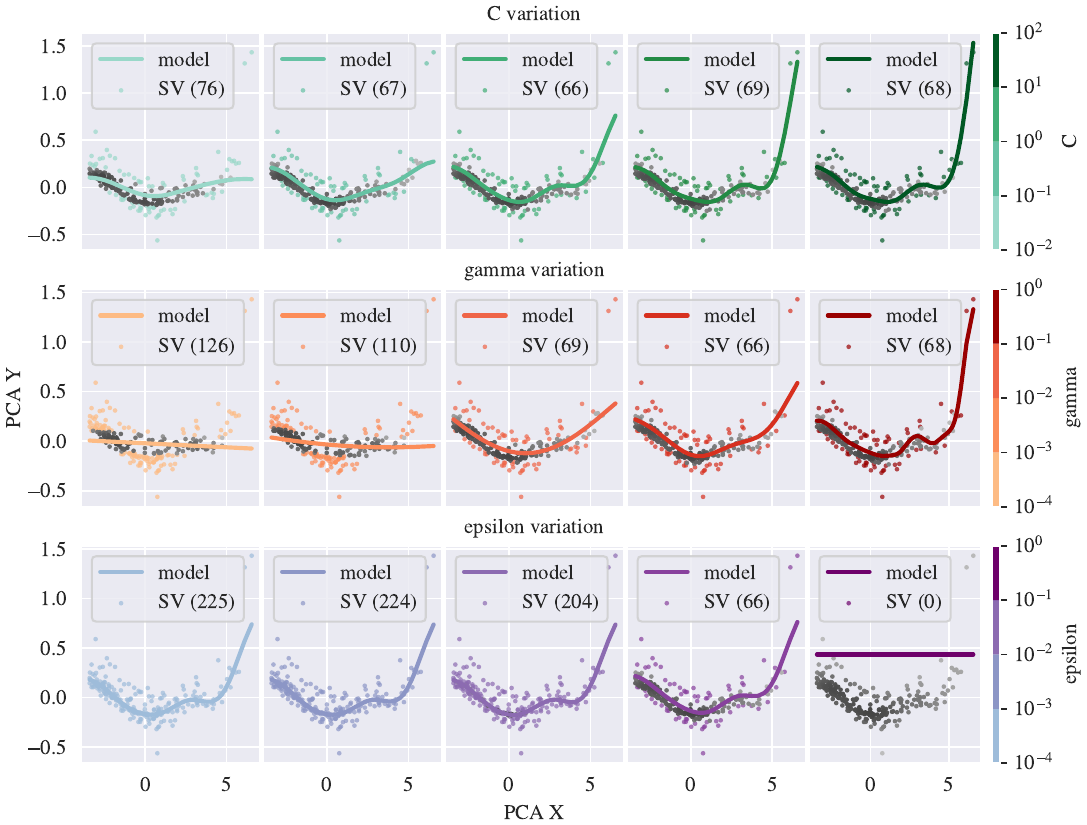}}
        \caption{Impact of different hyperparameter values on the shape of the SVR curve in PCA space on the example of NGC~1662. The data points used as support vectors are depicted as colored dots in each plot, whereas the other data are visible as black dots.}
        \label{fig:03-Hyperparameter-Matrix}
\end{figure*}

\begin{figure*}[ht]
        \centering
        \resizebox{\hsize}{!}{\includegraphics{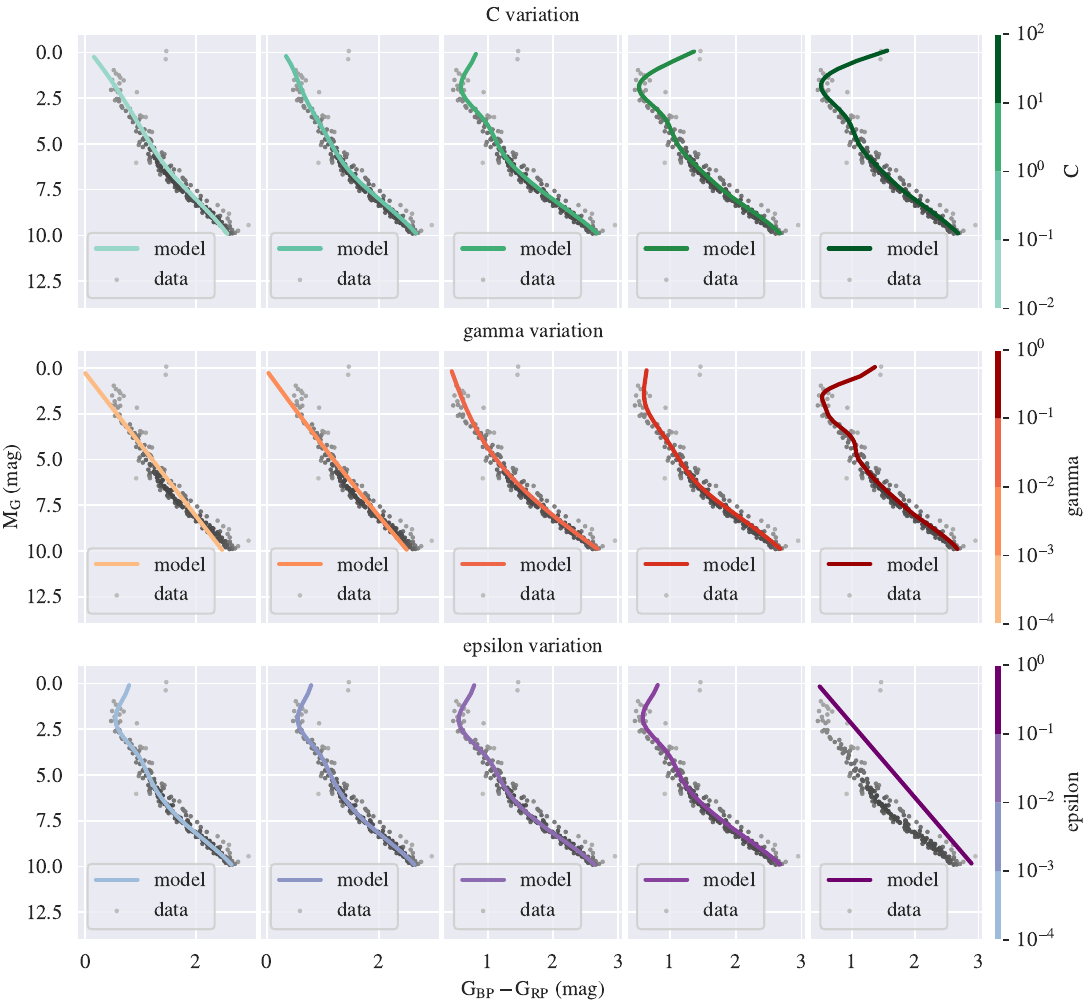}}
        \caption{Impact of different hyperparameter values on the shape of the empirical isochrone, along with the reverse-transformed CMD data of NGC~1662.}
        \label{fig:03-CMD-Hyperparameter-Matrix}
\end{figure*}

To visualize the effects of the different hyperparameters on the empirical isochrone shapes, Fig.~\ref{fig:03-Hyperparameter-Matrix} displays a parameter variation matrix of the three SVR hyperparameters on the example of the PCA-transformed data of the NGC~1662 cluster. Each row of the figure corresponds to the variation of one of the variables, while the remaining two parameters are kept fixed to the default values implemented in \texttt{sklearn} (C=1.0, \texttt{epsilon} = 0.1, \texttt{gamma} = 'scale' = $\frac{1}{n_{\text{features}}\sigma^2}$). In Fig.~\ref{fig:03-CMD-Hyperparameter-Matrix}, the same matrix is shown, but the data and regression curves have been transformed back into the CMD parameter space.

The panels in the first row of the figures illustrate that a larger value for \texttt{C} results in increasingly complex and longer regression curves, as missing data points with the regression curve is penalized more strongly. In an astrophysical context, this means that for high penalty parameters, more emphasis is placed on including the sparsely populated upper main sequence and main sequence turn-off regions than otherwise. In the second row, the \texttt{gamma} parameter is varied. As it is inversely tied to the influence radius of each data point, small values tend to result in a strong smoothing of the regression curve, while large values lead to overfitting. In Fig.~\ref{fig:03-Hyperparameter-Matrix}, it can be seen that for both the variations of \texttt{gamma} and \texttt{C}, the amount of support vectors identified by the model varies only slightly. This changes upon inspection of the last row of the figure, as the \texttt{epsilon} parameter mainly controls the number of support vectors used to build the regression model. For very small values of \texttt{epsilon}, all data are considered by the model, whereas for extremely large values no support vectors are defined at all. As shown in Fig.~\ref{fig:03-CMD-Hyperparameter-Matrix}, the \texttt{epsilon} parameter plays a vital role in the correct positioning of the empirical isochrone, meaning the ability of the algorithm to deal with outliers or strong scatter. Especially for populations with a strong unresolved binary sequence, as for example the selected example NGC~1662, the default value of the $\varepsilon$-tube implemented in \texttt{sklearn} is too lenient a choice, as the placement of the isochrone is then biased toward the binary sequence. Taken together, the strong changes that different values for the hyperparameters have on the shape and positioning of the empirical isochrone, particularly the sensitivity to unresolved binaries, demands that we perform a hyperparameter tuning.

\begin{figure*}[ht]
        \centering
        \resizebox{\hsize}{!}{\includegraphics{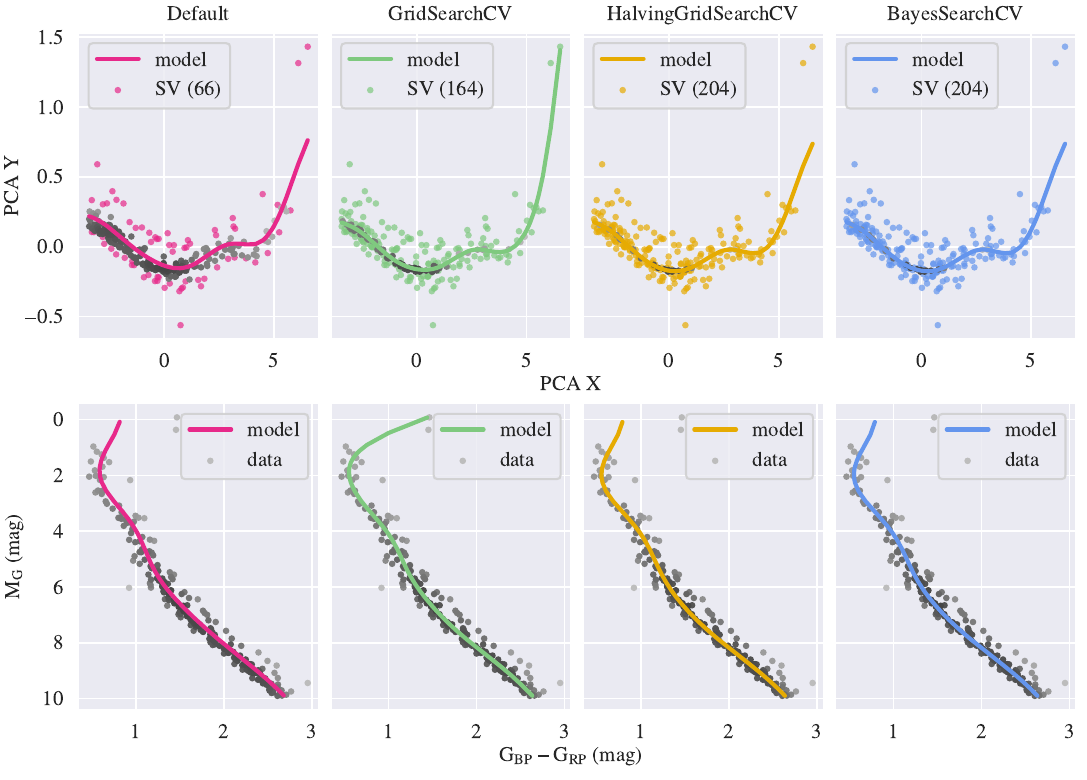}}
        \caption{Comparison of empirical isochrones produced with different hyperparameter optimization regimes. The equivocal values of the hyperparameters hint at a slight degeneracy between them, which also appears in the hyperparameter matrix depicted in Fig.~\ref{fig:03-Hyperparameter-Matrix} and Fig.~\ref{fig:03-CMD-Hyperparameter-Matrix}.}
        \label{fig:03-Tuning-effect}
\end{figure*}

The main drawback of hyperparameter tuning is that it needs to be performed individually for every cluster in the archive and may even be necessary when switching between different CMD filter combinations. However, the tuning process can be greatly sped up by using an automated parameter gridsearch implemented in \texttt{scikit-learn}. We use a 5-fold cross-validation on the scaled data and perform the gridsearch on a parameter grid ranging from \texttt{epsilon} $\in [0.01,0.0317]$ and \texttt{C} $\in [0.01, 100]$, with 20 logarithmically spaced step. We determined these bounds after running an exploratory parameter search on a significantly larger grid for a few test clusters. There are different gridsearch options available at \texttt{scikit-learn}. Figure~\ref{fig:03-Tuning-effect} displays the effects of using three different methods, namely (\texttt{GridSearchCV}, \texttt{HalvingGridSearchCV}, and \texttt{BayesSearchCV}) for the automated hyperparameter tuning, compared with the default SVR parameters, which are illustrated in the leftmost panel. Compared to the regression using default parameters, all three gridsearch methods produce superior results. However, likely due to a degeneracy between at least two of the hyperparameters, not all searches unambiguously reach the same optimized values. The \texttt{GridSearchCV} finds a larger penalty value and thus also traces the two most evolved sources in the CMD of NGC~1662 that are ignored by the other two regression curves. As indicated by the colored data points in the top row of the figure, the number of support vectors varies as well, with the \texttt{HalvingGridSearchCV} method determining the most support vectors and the \texttt{BayesSearchCV} relying on the fewest. Despite their different values for the hyperparameters, it is vital to note that for all well-populated areas of the CMD of NGC~1662, the empirical isochrones generated by the different gridsearch methods are all in agreement. The only significant differences occur at the very bright end of the distribution, where sources grow sparse in any case. In this region, the uncertainty of the empirical isochrones may grow quite large regardless of the employed search algorithm (see Sect. \ref{sec:Discussion-quality}). We therefore argue that all three methods are valid for our purposes and selected the computationally fastest and most stable one, namely the \texttt{GridSearchCV}.

\subsection{Weights}
\label{appendix:SVR-weights}

If measurement errors for a given cluster CMD are available, they are considered during the isochrone extraction in the form of scalar weights associated with each data data point in the diagram. The weight parameter is multiplied with the tuned penalty parameter of the SVR model to adjust the regression according to the observational errors.

\begin{figure}[ht]
    \resizebox{\hsize}{!}{\includegraphics{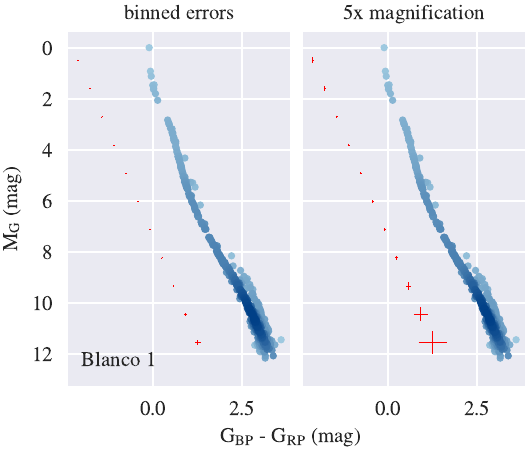}}
    \caption{Binned error data for the cluster CMD of Blanco~1. The right panel of the figure displays a five-fold magnification of the error bars.}
    \label{fig:03-errors}
\end{figure}

To transform the errors into weights, we employ a simple Gaussian error propagation formula. First, the error in the absolute magnitude $\Delta M$ is calculated by using Gaussian error propagation on the equation for the distance modulus
\begin{align*}
     \Delta M & = \sqrt{\left(\frac{5}{\log(10)\cdot \varpi} \cdot \Delta \varpi \right)^2 + \left(\Delta m\right)^2}, 
\end{align*}
where $\varpi$ denotes the parallax, $\Delta \varpi$ indicates its associated measurement uncertainty, and $\Delta m$ is the measurement error in apparent magnitude. The error in the color index is calculated as the RSS of the associated errors in apparent magnitudes
 \begin{align*}
     \Delta (m_X - m_Y) & = \sqrt{\left(\Delta m_X\right)^2 +\left(\Delta m_Y\right)^2},
\end{align*}
with the subscripts $X$ and $Y$ denoting the respectively chosen passbands for the CMD color axis. Subsequently, we combine the calculated errors by re-evaluating their root sum of squares and pass this value to the SVR model for each star during the model fitting and prediction process. We opt for this kind of error treatment as, especially considering \emph{Gaia} DR2 and DR3 data, the reported measurement uncertainties for sources within the applied distance cut of \SI{500}{pc} are commonly within a negligible mmag and milli-arcsecond regime \citep{Gaia_DR2_Summary_2018, Gaia_EDR3_2021}. We note, that the magnitude errors are assumed to be Gaussian in this case, although they are in truth Gaussian in the flux. As the magnitude uncertainties $\Delta m$ in the data catalogs used in this work are in the order of mmag, the log-norm can be approximated by a Gaussian. In case of larger magnitude errors, we recommend to verify the validity of the Gaussian approximation.\\

To test the validity of our approach, we calculated binned errors for all clusters of \citetalias{meingast2020}. Figure~\ref{fig:03-errors} displays the results for the Blanco~1 cluster. The left panel of the figure depicts errorbars created by dividing the cluster members into 12 equidistant bins, defined between the minimum and maximum absolute magnitudes of the cluster distribution, and calculating the mean errors in both variables for each bin via Gaussian error propagation. The errorbars have been shifted to the left side of the member distribution in the plot. In the right panel of the figure, the errors have been multiplied by a factor of 5 to enhance their visibility and to qualitatively highlight the trend of increasing measurement uncertainty toward the fainter end of the CMD distribution of the cluster. We studied plots as the one shown in Fig.~\ref{fig:03-errors} for all clusters of \citetalias{meingast2020} and conclude that the errors are minuscule compared to the general scatter in the cluster data and the approach of using simple Gaussian error propagation to turn them into weights seems justified.


\section{Quantification of the influence of measurement uncertainties}
\label{appendix:Uncertainty-quantification}

We systematically explored the methodological sensitivity of our data-driven extraction approach to uncertainties in key observational parameters. We identified the following factors as particularly influential to the empirical isochrone of a cluster CMD: measurement uncertainties in photometry and parallaxes, fraction of unresolved binaries, (differential) extinction, and field contamination. The goal of our analysis was to quantify the sensitivity of our method to each of these factors and to establish intervals where biases are negligible. 

We performed the analysis for three clusters from the archive, which we chose as representatives for the different ages, distances and member numbers of archive clusters. Their parameters are given in Tab. \ref{tab:appendix-Clusters}. We further treated each of the three possible \emph{Gaia} passband combinations as separate cases.

\begin{table}[h]
    \caption{Sample clusters for the sensitivity analysis of the empirical isochrones.}
    \begin{tabular*}{\linewidth}{@{\extracolsep{\fill}}cccc}
    \hline \hline
    Cluster   & Estimated age (Myr) & median distance (pc) & $N*$ \\
    \hline
    delta Sco & 6.8           & 142             & 430  \\
    Blanco 1  & 105           & 237             & 839  \\
    NGC 752   & 1170          & 440             & 225  \\
    \hline
    \end{tabular*}
    \label{tab:appendix-Clusters}
\end{table}

We also distinguish between two possible extinction scenarios: First, we assume a flat extinction level for all stars in a cluster, which is generally valid for our cluster selection, as they are older than $\approx 6$ Myr and are not notably reddened \citep[e.g.,][]{2020b_Cantat-Gaudin_ages}. We discuss this scenario in a full factorial parameter exploration in Sect. \ref{appendix:full-factorial} and define parameter ranges of minimal impact on the results in \ref{appendix:param-ranges}. The impact of differential extinction is estimated in Sect. \ref{appendix:Differential}.

\subsection{Full factorial analysis}
\label{appendix:full-factorial}

For each of the five identified key parameters, namely the photometric uncertainty $\Delta m$, parallax uncertainty fraction $f_{ \Delta \varpi}$, unresolved binary fraction $f_{\mathrm{bin}}$, constant extinction level $A_{\mathrm{G}}$, and field contamination fraction $f_{\mathrm{field}}$, we define a value range (see Tab. \ref{tab:appendix-grid}) and construct a parameter grid, consisting of a total of 243 combinations. At each grid point, we then
\begin{enumerate}
    \item generate synthetic CMD points from the original isochrone.
    \item add the parameter uncertainties in the order: $\Delta m$, $f_{\Delta \varpi}$, $f_{\mathrm{bin}}$,  $A_{\mathrm{G}}$, $f_{\mathrm{field}}$.
    \item re-compute an empirical isochrone from the new CMD and compare it to the original.
\end{enumerate}
For the purpose of our analysis, we added the photometric and parallax uncertainties to each star of the synthetic CMD, inducing the maximum possible perturbation of the data. This means, that error contributions are rather overestimated instead of underestimated. The specifics of the routine are outlined in the following paragraphs.

\subsubsection{Parameter grid}
The value ranges for the parameters of interest are listed in Tab. \ref{tab:appendix-grid}. The first value of a given parameter range is an idealized minimum value, whereas the maximum value represents an edge case given the chosen source data. The maximum value for the photometric uncertainty approximates the highest photometric errors typically found in the \emph{Gaia} DR3 catalog for faint sources within 500 pc. The chosen fractional parallax uncertainty approximately represents the 90$^{\mathrm{th}}$ percentile of fractional parallax errors in the \emph{Gaia} source catalog within 500 pc. It also corresponds to about the maximum uncertainty where distance approximation via inversion of the parallax is justified in our chosen distance cut. For the binary fraction, we include typical values from the literature  \citep[][30 - 50 \% expected binary fraction in simulations]{MiMO}. For the $A_{\mathrm{G}}$ range we consult the second  quantile of the reported cluster extinctions in the archive (Q1: 0.03 mag, Q2: 0.22 mag, Q3: 0.377 mag) as well as a value higher than the maximum reported extinction of the archive clusters (UBC 17a, $A_{\mathrm{G}} \approx$ 0.6312 mag). For the field fraction, we evenly cover the parameter range between 0 and 50 \%.

\begin{table}[h]

    \caption{Parameter grid for the quantification of the effects that measurement uncertainties can have on the extracted empirical isochrone. The grid is evaluated at each point, resulting in a total of 243 combinations.}
    \begin{tabular*}{\linewidth}{@{\extracolsep{\fill}}c|lll}
    \hline \hline
                            & minimum & intermediate & maximum \\
    \hline 
    $\Delta m$          & 0       & 0.01         & 0.03    \\
    $f_{\Delta \varpi}$     & 0       & 0.05         & 0.1     \\
    $f_{\mathrm{bin}}$      & 0       & 0.3          & 0.5     \\         
    $A_{\mathrm{G}}$        & 0       & 0.25         & 0.5     \\
    $f_{\mathrm{field}}$    & 0       & 0.25         & 0.5     \\
    \hline
    \end{tabular*}
    \label{tab:appendix-grid}
\end{table}

\subsubsection{Generation of synthetic CMD data}
We start from the discrete values of a calculated empirical cluster isochrone $I$ of a given cluster CMD, which correspond to the color index and absolute magnitude parameters listed in \ref{tab:04-results}, respectively. In order to include photometric and parallax uncertainties, the absolute magnitude is converted back into apparent magnitudes via the distance modulus, using the mean cluster distance as approximation. 

The photometric uncertainty $\Delta m$ impacts the color axis measurement as well as the absolute magnitude measurement of a given star.  Regarding the former, the only information available to us from the empirical isochrone is the scalar value of the color index, which we cannot resolve into the individual passbands without further assumptions. Therefore, we have to substitute the true photometric uncertainty of the individual passbands, which will generally not be identical, with one calculated from the color index value. Within this simplified scenario, we can distinguish between four cases:
\begin{enumerate}
    \item The uncertainty is added in both passbands.
    \item The uncertainty is subtracted in both passbands.
    \item The uncertainty is added to the first passband and subtracted from the second.
    \item The uncertainty is subtracted from the first passband and added to the second.
\end{enumerate}
Both the first and second case result in a net change of zero along the color axis. The third case causes an increase of twice the uncertainty value, while the fourth case yields a decrease of twice the uncertainty value. We randomly assign one of the four cases to each source in the CMD. It should be noted, that when calculating the photometric uncertainties with the information of the two passbands instead of the color index, the shift along the color axis will not always be symmetric as each color measurement is associated with an individual measurement error. However, we purposely chose a very high photometric uncertainty, which definitely accounts for the true expected scatter in a CMD even with the simplified scenario. To incorporate the simulated photometric uncertainty into the absolute magnitude measurements, the uncertainty $\Delta m$ is randomly either added to or subtracted from the apparent magnitude measurement for each star. The same is done for the parallax uncertainty, which is calculated using the mean cluster parallax as proxy, before the absolute magnitudes are calculated again. We note, that this way of simulating photometric and parallax uncertainties constitutes only a rough approximation. Nonetheless, combined with our choice of large uncertainty values, we generate a maximal source dispersion, which is what we are interested in for our sensitivity quantification.

Next, we add artificial unresolved binaries to the simulated CMD by increasing the absolute magnitude of a randomly drawn fraction $f_{\mathrm{bin}}$ of the CMD stars by -0.753 mag. This shift mainly assumes equal mass binaries, but also accounts for the reported shift of non-equal binaries on the binary sequence (see Discussion in Sect. \ref{sec:Methods-pitfalls}). By randomly drawing stars across the CMD, we disregard the mass-multiplicity relation. However, we argue that high-mass binaries, with a statistical probability of up to 80\% of having a binary are mostly located at the uppermost end of the main sequence in \emph{Gaia} CMDs, which is almost vertical. Hence, they are shifted almost only vertical in any case, justifying our assumptions.

Afterwards, we add a constant extinction level $A_{\mathrm{G}}$ to the CMD data in both absolute magnitude and color index. For the former, the extinction level is directly added. In case of the color index, the color excess is calculated by first multiplying the \emph{Gaia} DR3 extinction coefficients, which we approximated using \cite{2003Draine}, an extinction law of $R_V$ = 3.1 and the assuming a flat SED, with the extinction level, and then subtracting them.\\
Lastly, we randomly sample a contaminating field fraction $f_{\mathrm{field}}$ of stars from the full \emph{Gaia} DR3 catalog within 500 pc and add them to the cluster CMD.

\subsubsection{Re-computation and similarity evaluation of the isochrones}
From the new CMD that now encompasses all uncertainties at a given grid point, we re-compute an empirical isochrone $I^*$ using the procedure outlined in Section \ref{sec:Methods-algorithm} and $n_{boot} =100$ resamplings. We then compare it to the original isochrone $I$ by interpolating 100 equally spaced points $i$ for $I$ and $10^5$ points $i^*$ for $I^*$. For each $i$, we calculate the distances ${d}^{100}_i$ to the 100 nearest neighbors on $I^*$ and take the average $\Bar{d_i}$. Summing over all averaged distances and dividing the result by the number of interpolated points $n=100$ yields a final, scalar value $\Delta_{\mathrm{iso, avg}}$. It can be interpreted as the average deviation between the original empirical isochrone and the newly calculated one at any given point in units of magnitudes. We opt for this method instead of calculating Euclidean distances or the Area between the isochrones for two reasons: First, our technique ensures translational invariance, meaning it is insensitive to whether the isochrones cross each other. Second, by choosing a large number of points along $I^*$ versus a small number of points along $I$, we effectively only compare the dynamical range in the CMD covered by both curves.

The deviation score is calculated for each grid point. As the empirical isochrone calculation procedure subjected to small fluctuations due to the many resamplings, the deviation score should be calibrated against them. Thus, we  perform ten calculations of $\Delta_{\mathrm{iso, avg}}$ between the original isochrone and one recalculated with all investigated parameters set to zero, and use the median value as zero-point for the calibration of all deviation scores of the parameter grid.

\begin{figure}[h]
    \centering
    \resizebox{\hsize}{!}{\includegraphics{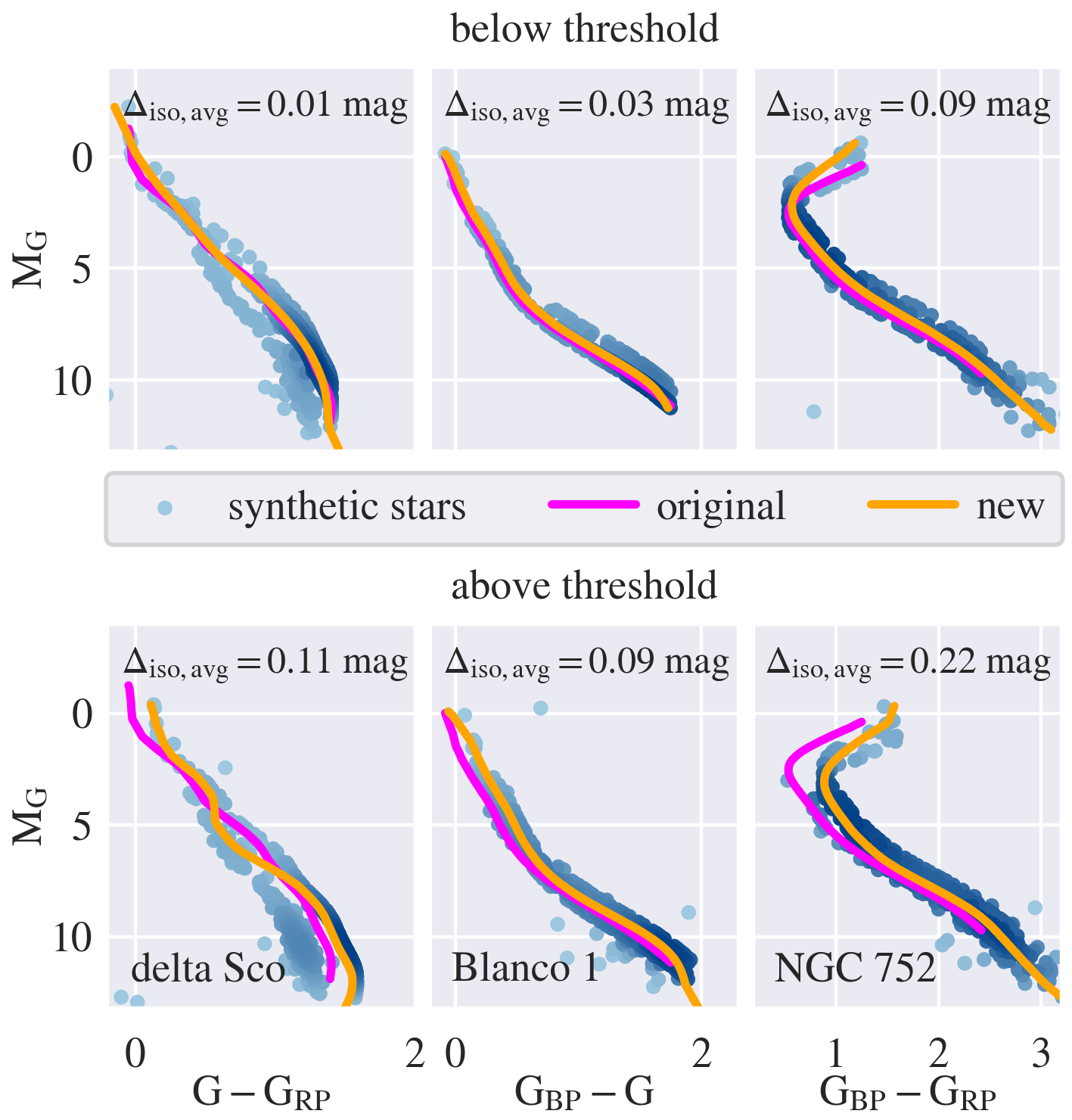}}
    \caption{Example synthetic CMDs for the three test clusters in different passband combinations. The top row shows the comparison between the original (\emph{magenta}) and the newly computed isochrone (\emph{orange}) for the chosen threshold values, whereas the bottom row shows the comparison for values exceeding the threshold.}
    \label{fig:appendix:good-vs-bad-deviation-scores}
\end{figure}

We then visually compare the recomputed isochrones to the original ones and determine a heuristic threshold $t_{\mathrm{good}}$, at which the recomputed isochrones still closely match the original ones, for each CMD type:
\begin{align*}
    & \mathrm{G}_{\mathrm{BP}}-\mathrm{G}_{\mathrm{RP}}: t_{\mathrm{good}} \leq 0.1~\text{mag}, \\
    & \mathrm{G}_{\mathrm{BP}}-\mathrm{G}: t_{\mathrm{good}} \leq 0.04~\text{mag}, \\
    & \mathrm{G}-\mathrm{G}_{\mathrm{RP}}: t_{\mathrm{good}} \leq 0.03~\text{mag}. 
\end{align*} 
To give a qualitative example of what CMDs below and above the set thresholds look like, Fig. \ref{fig:appendix:good-vs-bad-deviation-scores} shows six recomputed CMDs, two for each cluster and CMD type, respectively. The top row displays CMDs that we define as reliable, while the bottom row holds CMDs that surpass the threshold.

\subsubsection{Sensitivity analysis}

\begin{figure*}[h]
        \centering
        \resizebox{\hsize}{!}{\includegraphics{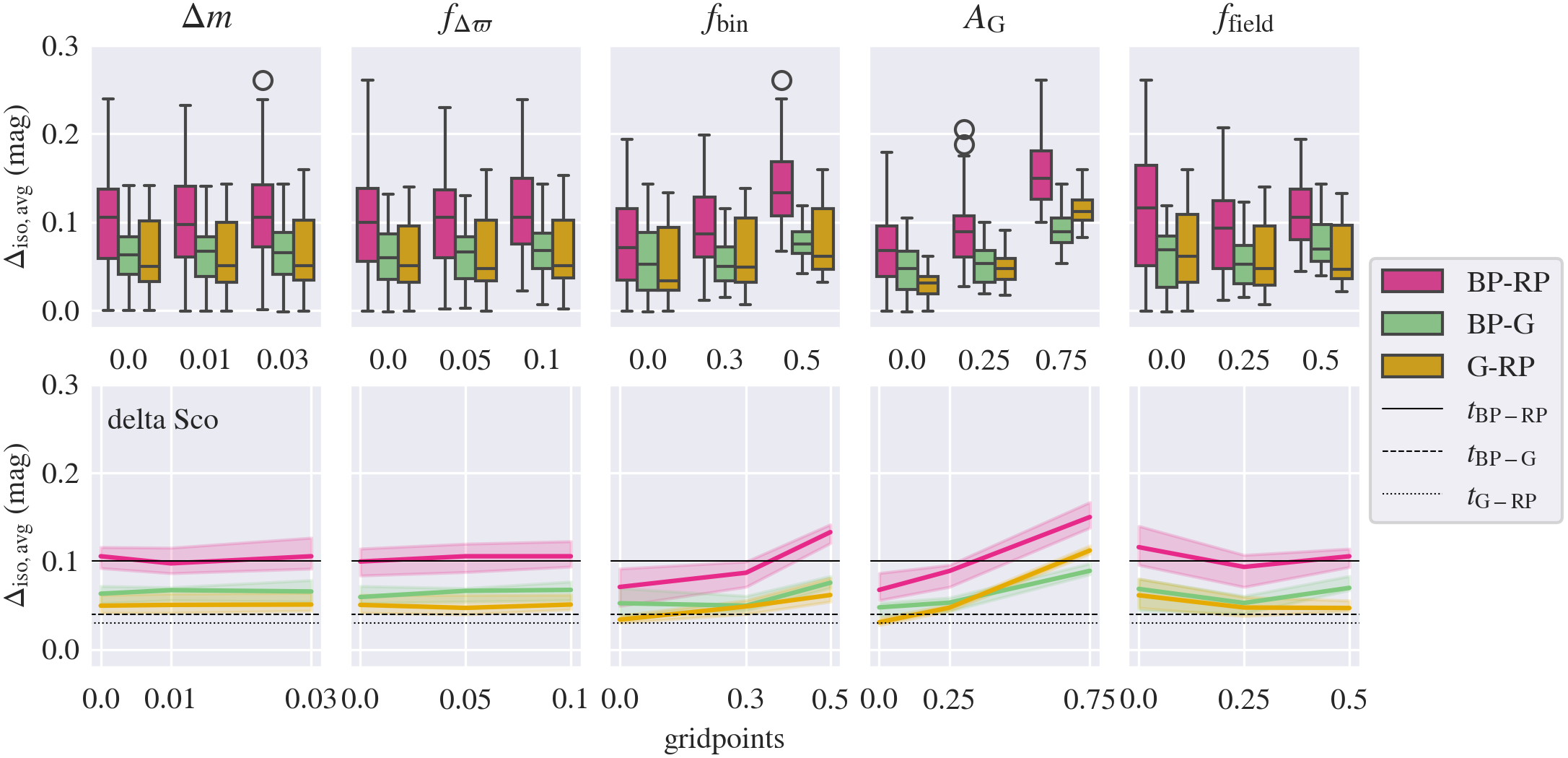}}
        \caption{Sensitivity analysis of the empirical isochronal shapes in dependence on the parameters $\Delta m$, $f_{\Delta \varpi}$, $f_{\mathrm{bin}}$, $A_{\mathrm{G}}$, and $f_{\mathrm{field}}$
         for the representative young cluster delta~Sco, with a median distance of $\sim 142$ pc and an estimated age of $\sim 7$ Myr \citep{2023Ratzenbock_Sco-Cen_ages}. \emph{Top row:} Boxplots illustrating the effects of the systematically varied factors on the averaged deviation score $\Delta_{\mathrm{iso, avg}}$, analyzed through a full factorial design. Each boxplot represents the deviation scores of all grid points containing the indicated value on the corresponding x-axis tick label. \emph{Bottom row:} Same as the top row, but this time only showing the evolution of the median $\Delta_{\mathrm{iso, avg}}$ at each grid point, along with the 95 \% confidence interval estimation. The color coding corresponds to the different \emph{Gaia} passband combinations. The black lines mark the empirical thresholds determined for each passband, respectively.}
        \label{fig:appendix-Boxplots-delta_sco}
\end{figure*}

\begin{figure*}[h]
        \centering
        \resizebox{\hsize}{!}{\includegraphics{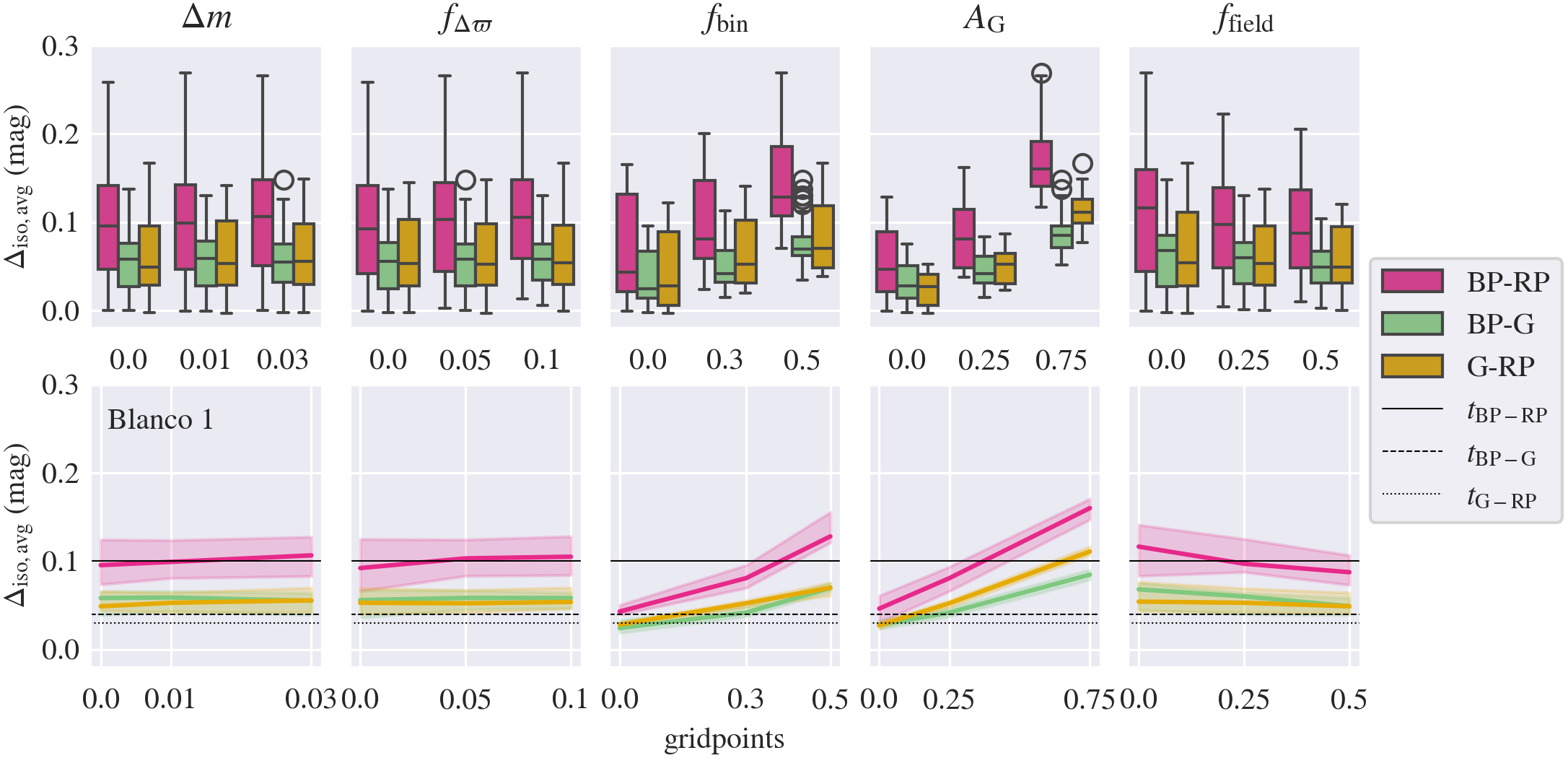}}
        \caption{Same as Fig. \ref{fig:appendix-Boxplots-delta_sco}, but for the representative intermediate age open cluster Blanco~1, with a median distance of $\sim 237$ pc and an estimated age between $\sim 94 - 105$ Myr \citep{2019Bossini, 2020b_Cantat-Gaudin_ages}.}
        \label{fig:appendix-Boxplots-Blanco1}
\end{figure*}

\begin{figure*}[h]
        \centering
        \resizebox{\hsize}{!}{\includegraphics{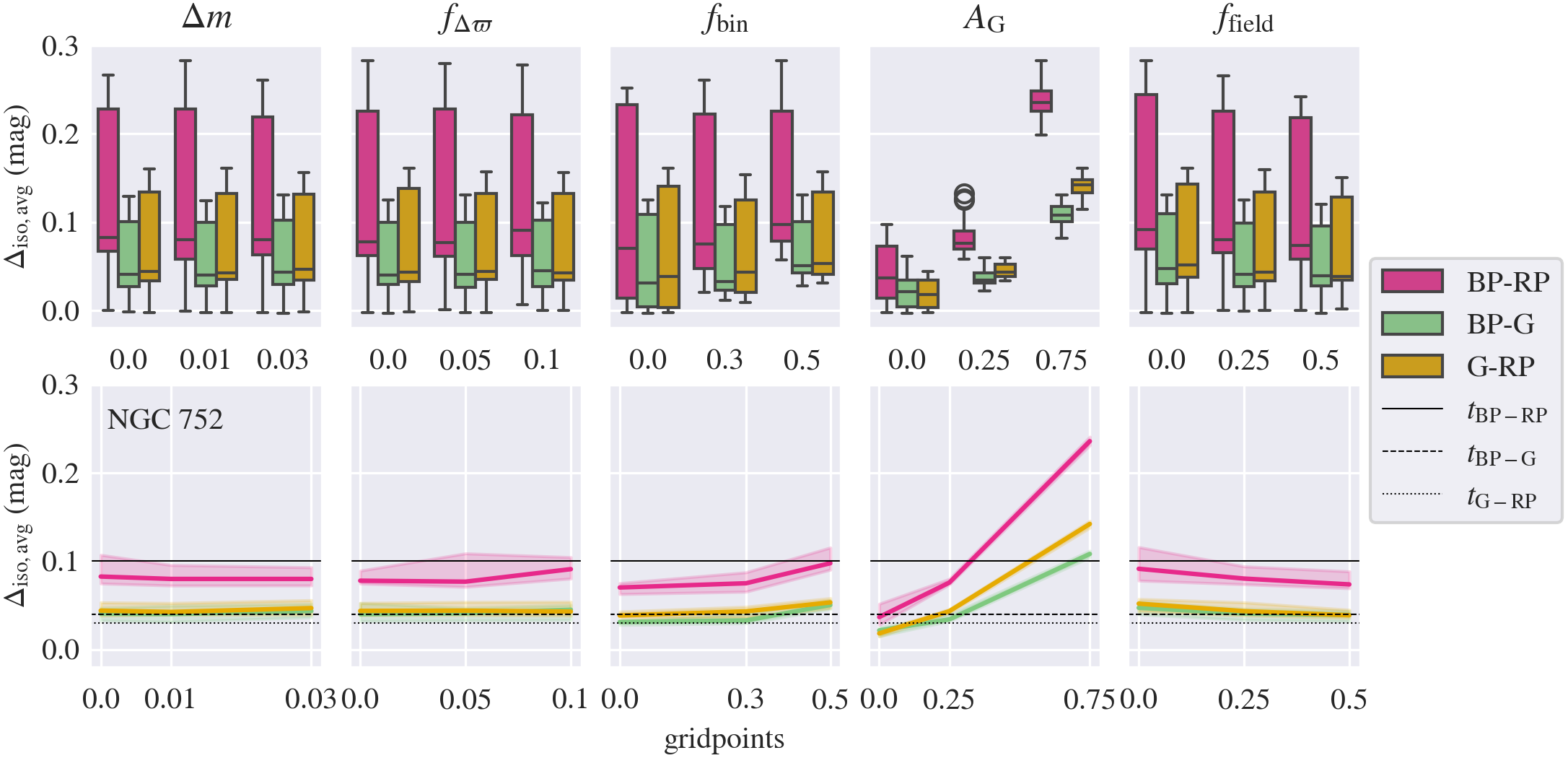}}
        \caption{Same as Fig. \ref{fig:appendix-Boxplots-delta_sco}, but for the representative old open cluster NGC~752, with a median distance of $\sim 440$ pc and an estimated age  between $\sim 1.17 - 1.52$ Gyr \citep{2020b_Cantat-Gaudin_ages, 2021Dias}.}
        \label{fig:appendix-Boxplots-NGC752}
\end{figure*}

To analyze how each investigated parameter affects the shape of the isochrone, we create boxplots of the deviation scores $\Delta_{\mathrm{iso, avg}}$ for each specific grid point of each parameter. These boxplots illustrate the range of isochronal deviation when uncertainties in all other investigated parameters are taken into account. We also create a lineplot showcasing only the median deviation scores at each grid point, to track the overall trends across the parameter ranges. The results are shown as boxplot-lineplot combinations for each of the three clusters in Figs. \ref{fig:appendix-Boxplots-delta_sco}-\ref{fig:appendix-Boxplots-NGC752}. The x-axis denotes the grid point values for the respective parameter, whereas the y-axis represents the average deviation scores. The color-coding corresponds to the different color combinations in the CMD color axes. The black lines indicate the upper thresholds for the respective CMD types.

The average deviations between the original and the recomputed isochrones range between 0 and 0.3 mag for all three inspected clusters. Outliers to the boxplots are marked by empty circles. Generally, the $\mathrm{G}_{\mathrm{BP}}-\mathrm{G}_{\mathrm{RP}}$ passband combination displays higher deviation scores across all grid points, while other combinations are less affected. We estimate that this discrepancy stems from the larger color range this CMD type occupies on the color axis. For all three clusters, the median deviation scores of the $\mathrm{G}_{\mathrm{BP}}-\mathrm{G}_{\mathrm{RP}}$ isochrones lie below the reliability threshold for at least two of the three investigated grid points. For the other two CMD passband combinations, the median deviation scores exceed their respective reliability thresholds for the two younger clusters in all but the lowest grid points for binary fraction and extinction. Only for NGC~752 do they adhere to the threshold values or drop below them for at least two of the three investigated grid points.

It can be observed that for all three example clusters, photometric uncertainty has a negligible influence on the isochronal shape. This outcome is expected because, although the values chosen for the parameter grid are relatively large given the generally very low photometric errors of \emph{Gaia} DR3, their overall impact on stellar positions in the CMD is minor compared to the magnitudes of the shifts induced by unresolved binaries or extinction. For delta Sco and Blanco~1, there exist slightly positive gradients in the median plots with increasing uncertainty values, whereas for NGC~752 the line is completely flat. For the former two clusters, the median deviation scores for the $\mathrm{G}_{\mathrm{BP}}-\mathrm{G}_{\mathrm{RP}}$ CMD lie very close to the empirical reliability threshold, whereas for the other two passband combinations they are always above the respective threshold boundary. For the oldest test cluster, the $\mathrm{G}_{\mathrm{BP}}-\mathrm{G}_{\mathrm{RP}}$ trendline is below the threshold over the entire photometric uncertainty parameter grid and the $\mathrm{G}_{\mathrm{BP}}-\mathrm{G}$ trendline is exactly at the threshold limit.

The plots indicate that, similar to the photometric uncertainty, the parallax uncertainty does not affect the deviation score, remaining roughly constant across the grid. There exists only a slight linear increase of the deviation score with increasing parallax uncertainty for all clusters. The  $\mathrm{G}_{\mathrm{BP}}-\mathrm{G}_{\mathrm{RP}}$ median line crosses the threshold after 5 \% fractional parallax errors in the cases of delta~Sco and Blanco~1.

Regarding the binary fraction, one can observe a small increase of the deviation score between zero and 30 \% unresolved binaries that does not push any of the median lines above the threshold for the $\mathrm{G}_{\mathrm{BP}}-\mathrm{G}_{\mathrm{RP}}$ CMD. From 30 \% to 50 \% simulated binary contamination, the increase is stronger and pushes past the threshold for delta Sco and Blanco~1. This is an expected result due to the nature of the SVR algorithm, which requires the amount of reliable data to be larger than the amount of outliers to correctly establish its support vectors. The influence of the contaminating binary sequence may however be alleviated by the simultaneous addition of field stars in cases where the field star locus coincides with the main sequence locus of the cluster, meaning for blindspot clusters and older. In those cases, the field stars actually bolster the main sequence population of the cluster and generate a better fit, which would explain both the smaller impact of the high binary fraction on the old NGC~752 cluster, and the slightly negative correlation between field fraction and deviation score apparent for NGC~752 and Blanco~1.

The extinction parameter now has the strongest influence on the isochronal shapes in all three cases, with the $\mathrm{G}_{\mathrm{BP}}-\mathrm{G}_{\mathrm{RP}}$ color combination being affected the most out of the three passband combinations.  One can also observe a stronger than linear increase in deviation scores with rising extinction levels. There appears to be a connection between increasing cluster age and/or distance and a stronger sensitivity to extinction. For instance, for delta~Sco all five key parameters produce roughly the same spread in deviation scores. For NGC~752 on the other hand, the uncertainties induced by extinction clearly dominate the overall isochronal deviation from its original shape.
 
 The field fraction shows an inverse relation between deviation score and higher field star content for the two older clusters, as well as a more complex pattern for the younger delta~Sco, whose main sequence is significantly above that of the blindspot region. There appears to be a minimum at 25 \% field contamination, the reasons for which are unclear.

\subsection{Parameter ranges of minimal deviation}
\label{appendix:param-ranges}

\begin{table}[h]
\caption{Isochronal deviation scores for all tested clusters and passband combinations, with all but one parameter set to zero, respectively. The values were rounded to the third decimal point. The bottom section lists the averaged isochronal deviation scores over all three clusters. Values exceeding the defined reliability thresholds are indicated in bold.}
\begin{tabular}{c c |lll}
\hline\hline
 &  & \multicolumn{3}{c}{$\Delta_{\mathrm{iso,   avg}}$ (mag)}   \\               
   Parameter       &    Value   & $\mathrm{G}_{\mathrm{BP}}-\mathrm{G}_{\mathrm{RP}}$ & $\mathrm{G}_{\mathrm{BP}}-\mathrm{G}$ & $\mathrm{G}-\mathrm{G}_{\mathrm{RP}}$ \\
\hline
\multicolumn{5}{c}{\emph{delta Sco}} \\
\hline
\multirow{2}{*}{$\Delta   m$}         & 0.01 & -0.001 & -0.001 & 0.000  \\
                                      & 0.03 & 0.001  & -0.002 & -0.001 \\
\multirow{2}{*}{$f_{\Delta \varpi}$}  & 0.05 & 0.002  & 0.002  & 0.002  \\
                                      & 0.1  & 0.031  & 0.020  & 0.001  \\
\multirow{2}{*}{$f_{\mathrm{bin}}$}   & 0.3  & 0.025  & 0.035  & 0.028  \\
                                      & 0.5  & 0.170  & 0.089  & 0.053  \\
\multirow{2}{*}{$A_{\mathrm{G}}$}     & 0.25 & 0.032  & 0.022  & 0.027  \\
                                      & 0.75 & 0.116  & 0.087  & 0.106  \\
\multirow{2}{*}{$f_{\mathrm{field}}$} & 0.25 & 0.020  & 0.023  & 0.021  \\
                                      & 0.5  & 0.056  & 0.054  & 0.034  \\
\hline
\multicolumn{5}{c}{\emph{Blanco 1}} \\
\hline
\multirow{2}{*}{$\Delta   m$}         & 0.01 & -0.001 & -0.001 & 0.000  \\
                                      & 0.03 & 0.000  & -0.003 & -0.003 \\
\multirow{2}{*}{$f_{\Delta \varpi}$}  & 0.05 & 0.003  & 0.000  & -0.003 \\
                                      & 0.1  & 0.018  & 0.010  & 0.004  \\
\multirow{2}{*}{$f_{\mathrm{bin}}$}   & 0.3  & 0.024  & 0.023  & 0.025  \\
                                      & 0.5  & 0.125  & 0.072  & 0.049  \\
\multirow{2}{*}{$A_{\mathrm{G}}$}     & 0.25 & 0.039  & 0.019  & 0.027  \\
                                      & 0.75 & 0.142  & 0.083  & 0.106  \\
\multirow{2}{*}{$f_{\mathrm{field}}$} & 0.25 & 0.015  & 0.017  & 0.003  \\
                                      & 0.5  & 0.022  & 0.020  & 0.006  \\
\hline
\multicolumn{5}{c}{\emph{NGC 752}} \\
\hline
\multirow{2}{*}{$\Delta   m$}         & 0.01 & -0.001 & -0.001 & -0.001 \\
                                      & 0.03 & -0.003 & -0.003 & -0.002 \\
\multirow{2}{*}{$f_{\Delta \varpi}$}  & 0.05 & 0.001  & -0.002 & -0.002 \\
                                      & 0.1  & 0.014  & 0.001  & 0.003  \\
\multirow{2}{*}{$f_{\mathrm{bin}}$}   & 0.3  & 0.037  & 0.022  & 0.015  \\
                                      & 0.5  & 0.091  & 0.040  & 0.042  \\
\multirow{2}{*}{$A_{\mathrm{G}}$}     & 0.25 & 0.075  & 0.032  & 0.044  \\
                                      & 0.75 & 0.252  & 0.124  & 0.160  \\
\multirow{2}{*}{$f_{\mathrm{field}}$} & 0.25 & 0.001  & 0.004  & 0.001  \\
                                      & 0.5  & 0.002  & 0.001  & 0.002  \\
\hline   
\hline
\multicolumn{5}{c}{\emph{Average}} \\
\hline
\multirow{2}{*}{$\Delta   m$}         & 0.01 & -0.001         & -0.001         & 0.000          \\
                                      & 0.03 & -0.001         & -0.003         & -0.002         \\
\multirow{2}{*}{$f_{\Delta \varpi}$}  & 0.05 & 0.002          & 0.000          & -0.001         \\
                                      & 0.1  & 0.021          & 0.010          & 0.003          \\
\multirow{2}{*}{$f_{\mathrm{bin}}$}   & 0.3  & 0.028          & 0.027          & 0.022          \\
                                      & 0.5  & \textbf{0.129} & \textbf{0.067} & \textbf{0.048} \\
\multirow{2}{*}{$A_{\mathrm{G}}$}     & 0.25 & 0.049          & 0.024          & \textbf{0.033} \\
                                      & 0.75 & \textbf{0.170} & \textbf{0.098} & \textbf{0.124} \\
\multirow{2}{*}{$f_{\mathrm{field}}$} & 0.25 & 0.012          & 0.015          & 0.008          \\
                                      & 0.5  & 0.027          & 0.025          & 0.014   
\\   
\hline
\end{tabular}
\tablefoot{The results listed here correspond to the values of a single evaluation of the entire parameter grid. Due to the resampling strategy of our extraction method, small changes in the values may occur between runs, but they were found to be on the order of mmag.}
    \label{tab:appendix-all-results}
\end{table}

Comparing the point of intersection between the median lines, or boxplots, and the threshold line for each parameter, we can analyze the reliability of the calculated empirical isochrones in the simultaneous presence of measurement uncertainties in all investigated parameters. For the $\mathrm{G}_{\mathrm{BP}}-\mathrm{G}_{\mathrm{RP}}$ color combination, we determine that we can produce reliable isochrones for more than 50 \% of our parameter combinations for parallax uncertainties up to 5 \%, binary fractions up to around $35 \%$ and extinction levels up to $\sim 0.3$ mag, globally across the representative clusters. Regarding the field fraction, all three clusters are below the threshold at 25 \% field contamination, but the behavior of the trend lines is age-dependent and hence cannot be easily generalized. The behavior of the photometric uncertainty median lines is too heterogeneous for generalization as well.
For the remaining two CMD color combinations, the thresholds are not unanimously met in any case.

In isolation, meaning with all other parameters set to zero, the deviation scores of the individual parameters are much smaller. The isolated values for each cluster, as well as for the average over all clusters, are listed in Tab. \ref{tab:appendix-all-results}. For the averages, values exceeding the defined reliability thresholds are marked in bold.

The results show that, viewed in isolation, empirical isochrones remain reliable for all intermediate parameter values from Tab. \ref{tab:appendix-grid}, except for the $\mathrm{G}-\mathrm{G}_{\mathrm{RP}}$ combination, which exceeds the set threshold already at an extinction of $A_{\mathrm{G}} = 0.25$ mag. Furthermore, regarding the photometric uncertainty, fractional parallax uncertainty and field contamination, even the set maxima of the parameter ranges induce changes below the reliability thresholds in all tested cases. The high binary fraction of 50 \% causes the deviation to exceed the reliability threshold in all cases, as expected due to the nature of the extraction algorithm. Similarly, the high extinction level of 0.75 $A_{\mathrm{G}}$ magnitudes, which strongly perturbs the original CMD, leads to unreliable results as well.

\subsection{Differential extinction}
\label{appendix:Differential}

\begin{figure}[h]
    \resizebox{\hsize}{!}{\includegraphics{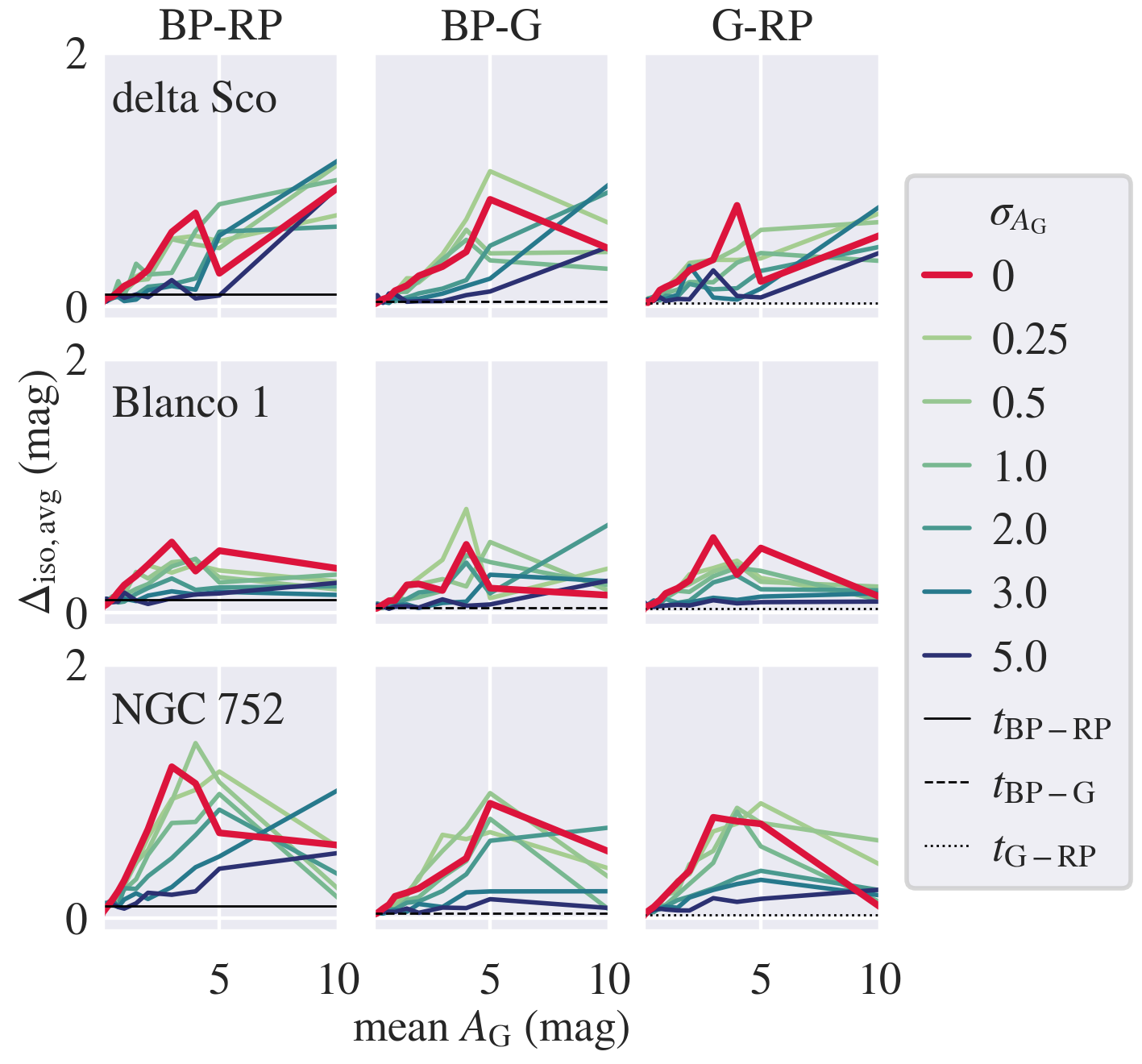}}
    \caption{Evolution of the isochrone deviation score in dependence of the mean extinction level, for different Gaussian noise additions with mean zero and standard deviations $\sigma_{A_{\mathrm{G}}}$. The evolution given a flat extinction level is indicated by the red line. The different rows and columns showcase the plots for the different example clusters and CMD passband combinations, respectively. The empirical reliability thresholds are indicated by the black lines in each panel.}
    \label{fig:appendix-diff-e}
\end{figure}

\begin{figure}[ht]
    \resizebox{\hsize}{!}{\includegraphics{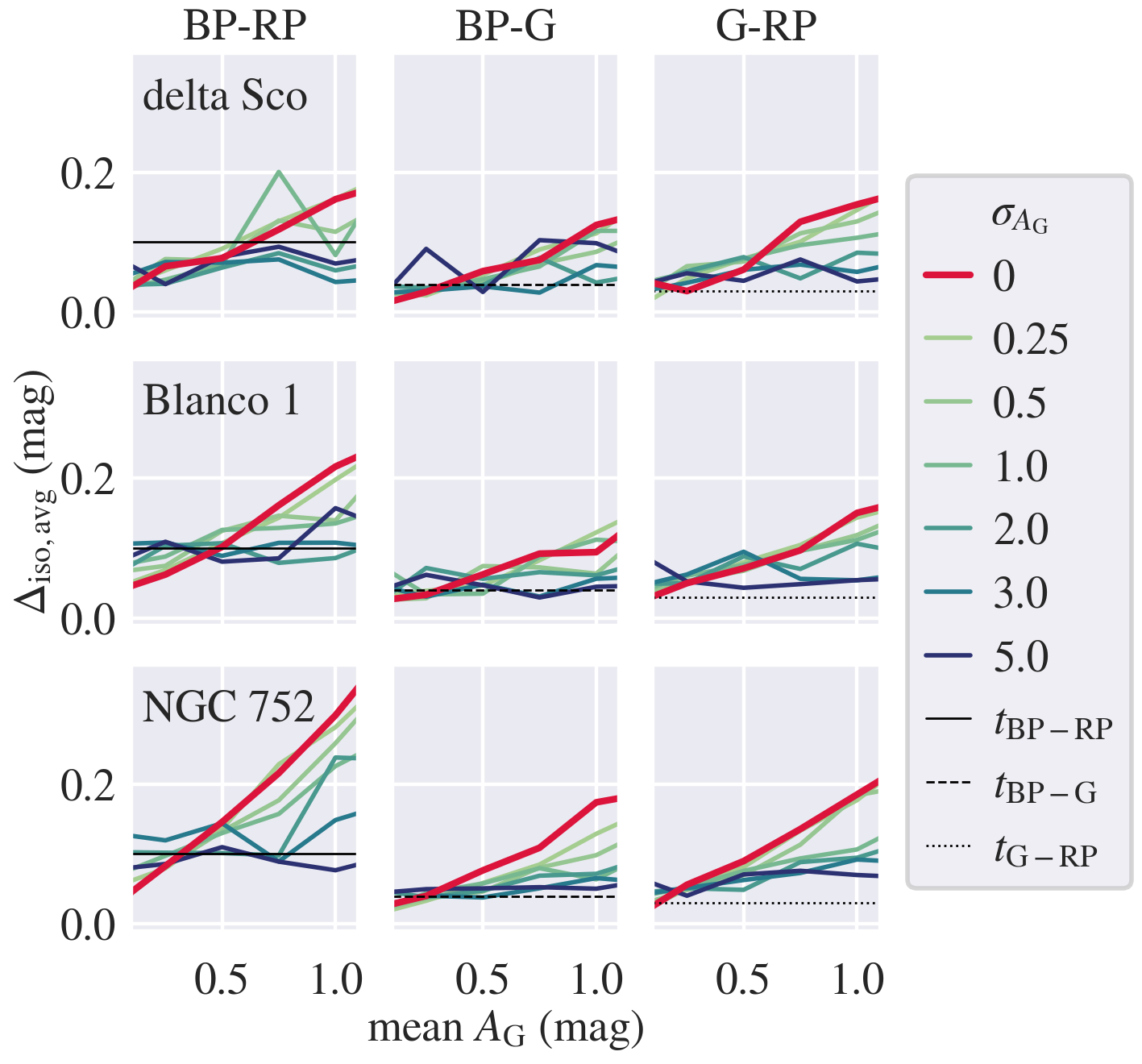}}
    \caption{Zoom-in of Fig. \ref{fig:appendix-diff-e}, showing the parameter space between a mean extinction level of 0.25 and 1 mag for all $\sigma_{A_{\mathrm{G}}}$ values.}
    \label{fig:appendix-diff-e-2}
\end{figure}

Under the assumption of differential extinctions, stars in a cluster present a distribution of extinction values instead of a constant value. The shape of this distribution is generally unknown as the dust distribution causing it can be arbitrary. It can also be caused by both dust inside the cluster and in the foreground along the line of sight to the observer. To model it for our purposes, we follow the simple approach described by \cite{2020_SPISEA}: We first assume a flat extinction level $A_{\mathrm{G}}$, as for the full factorial analysis. Then, we add a Gaussian noise component $\epsilon$ to the flat extinction level, which is drawn from a normal distribution with mean $\mu =  0$ and standard deviation $\sigma_{A_{\mathrm{G}}}$, yielding an individual extinction value $A^*_{\mathrm{G}}$ for each star. In case the new extinction value is negative, we set it to zero.
We again use the three representative clusters as test cases and analyze the different CMD passband combinations separately. Instead of evaluating a full grid, we fix the four already investigated parameters $\Delta m = 0.01 $ mag, $f_{\Delta \varpi} = 0.01$, $f_{\mathrm{bin}} = 0.3$, and $f_{\mathrm{field}} = 0.25$ to typical cluster values. For the flat extinction level, we define 10 grid points $A_{\mathrm{G}} \in [0.1, 0.25, 0.5, 0.75, 1, 1.5, 2, 3, 4, 5, 10]$ mag. For the standard deviation characterizing the Gaussian noise $\sigma_{A_{\mathrm{G}}}$, we test six different values $\sigma_{A_{\mathrm{G}}} \in [0.25, 0.5, 1, 2, 3, 5]$. 
We follow the same workflow as in Sect. \ref{appendix:full-factorial}, with the exception that we now calculate two empirical isochrones: One for the case of a flat extinction level for all stars in the synthetic CMD, as in the case of the full factorial grid, and the second for the simulated differential extinction case. The deviation scores with regard to the original isochrone are calculated for both curves as before. 
Figure \ref{fig:appendix-diff-e} shows the result over the entire grid, while Fig. \ref{fig:appendix-diff-e-2} displays a zoom-in into the region between 0 and 1 mean $A_{\mathrm{G}}$ magnitudes.

The general trend shows, that the curve corresponding to a flat extinction level (\emph{red line}) often produces higher deviation scores than the addition of Gaussian noise. Furthermore, due to the Gaussian approximation of differential extinction and the requirement that all extinctions are $\geq 0$, the influence of larger $\sigma_{A_{\mathrm{G}}}$ values only comes into full effect at sufficiently large extinction levels. One can observe spikes in $\Delta_{\mathrm{iso, avg}}$ over the flat extinction level curve around $\sim 3$ mag  and a general increase after $\sim 5$ mag, but in general, the flat extinction level curve encompasses the differential extinction distributions well up until 5 mag. As seen already in the full factorial sensitivity analysis, the older and farther away NGC~752 cluster is most strongly affected by extinction than its younger and nearer counterparts, especially when looking at the zoom-in plots. There, one can also discern that for extinctions $< 0.5$, the flat extinction level assumption is generally lower than the differential extinction assumptions. For extinction levels between 0.1 and 1 magnitude, the flat extinction curve reaches the threshold around 0.3 - 0.6 mag for the  $\mathrm{G}_{\mathrm{BP}}-\mathrm{G}_{\mathrm{RP}}$ CMD, depending on the inspected cluster. The threshold is met around 0.2 - 0.3 mag for the  $\mathrm{G}_{\mathrm{BP}}-\mathrm{G}$ CMD. For the  $\mathrm{G}-\mathrm{G}_{\mathrm{RP}}$ CMD, the threshold is never reached.
The lines denoting the differential extinction assumptions show versatile behaviors. However, for cases with $\sigma_{A_{\mathrm{G}}} \leq 1$, the same statements as for the flat extinction level trendline hold true. Therefore, we conclude that for moderate differential extinction, we can produce reliable isochrones up to the threshold imposed by the flat extinction level in any case.

\end{appendix}
\end{document}